%% file: ijmperev.tex
\documentclass[twoside]{article} 
 
\usepackage{ijmpe1} 
\usepackage{graphicx} 
\usepackage{clpsref}

\catcode`\@=11 
\long\def\@makefntext#1{ 
\protect\noindent \hbox to 3.2pt {\hskip-.9pt 
$^{{\eightrm\@thefnmark}}$\hfil}#1\hfill}       
 
\def\thefootnote{\fnsymbol{footnote}} 
\def\@makefnmark{\hbox to 0pt{$^{\@thefnmark}$\hss}}    
 
\def\ps@myheadings{\let\@mkboth\@gobbletwo 
\def\@oddhead{\hbox{} 
\rightmark\hfil\eightrm\thepage} 
\def\@oddfoot{}\def\@evenhead{\eightrm\thepage\hfil 
\leftmark\hbox{}}\def\@evenfoot{} 
\def\sectionmark##1{}\def\subsectionmark##1{}} 
 
\def\qed{\hbox{${\vcenter{\vbox{            
   \hrule height 0.4pt\hbox{\vrule width 0.4pt height 6pt 
   \kern5pt\vrule width 0.4pt}\hrule height 0.4pt}}}$}} 
 
\renewcommand{\thefootnote}{\fnsymbol{footnote}}    
 
\def\bsc{{\sc a\kern-6.4pt\sc a\kern-6.4pt\sc a}}   
\def\bflatex{\bf L\kern-.30em\raise.3ex\hbox{\bsc}\kern-.14em 
T\kern-.1667em\lower.7ex\hbox{E}\kern-.125em X} 
 
\renewcommand{\thefigure}{\thesection.\arabic{figure}} 
\renewcommand{\thetable}{\thesection.\arabic{table}} 
\newcounter{dummyv} 
 
\newcommand{\lsim}{\mathrel{\rlap{\lower4pt\hbox{\hskip0pt$\sim$}} 
\raise1pt\hbox{$<$}}}           
\newcommand{\gsim}{\mathrel{\rlap{\lower4pt\hbox{\hskip0pt$\sim$}} 
\raise1pt\hbox{$>$}}}           
 
\newcommand{\sfrac}[2]{\mbox{\footnotesize $\frac{#1}{#2}$}} 
 
\begin{document} 
 
\runninghead{DSEs: A tool for Hadron Physics}{DSEs: A tool for Hadron 
Physics} 
 
\normalsize\textlineskip 
\thispagestyle{empty} 
\setcounter{page}{1} 
\setcounter{section}{0} 
 
\copyrightheading{}         
 
\vspace*{0.88truein} 
 
\fpage{1} 
\centerline{\bf DYSON-SCHWINGER EQUATIONS: A TOOL FOR HADRON PHYSICS} 
\vspace*{0.37truein} 
\centerline{\footnotesize PIETER MARIS} 
\vspace*{0.015truein} \centerline{\footnotesize\it Department of Physics, 
North Carolina State University} 
\baselineskip=10pt \centerline{\footnotesize\it Raleigh, NC 27695-8202, 
USA}\vspace*{10pt} \centerline{\footnotesize and} 
\vspace*{10pt} 
\centerline{\footnotesize CRAIG D.\ ROBERTS} 
\vspace*{0.015truein} 
\centerline{\footnotesize\it Physics Division, Argonne National Laboratory} 
\baselineskip=10pt 
\centerline{\footnotesize\it Argonne, IL 60439-4843, USA} 
\vspace*{0.225truein} 
\publisher{(received date)}{(revised date)} 
 
\vspace*{0.21truein} %
\abstracts{Dyson-Schwinger equations furnish a Poincar\'e covariant framework
within which to study hadrons.  A particular feature is the existence of a
nonperturbative, symmetry preserving truncation that enables the proof of
exact results.  The gap equation reveals that dynamical chiral symmetry
breaking is tied to the long-range behaviour of the strong interaction, which
is thereby constrained by observables, and the pion is precisely understood,
and seen to exist simultaneously as a Goldstone mode and a bound state of
strongly dressed quarks. The systematic error associated with the simplest
truncation has been quantified, and it underpins a one-parameter model
efficacious in describing an extensive body of mesonic phenomena.  Incipient
applications to baryons have brought successes and encountered challenges
familiar from early studies of mesons, and promise a covariant field theory
upon which to base an understanding of contemporary large momentum transfer
data.\\[0.5ex]}
{{\it Keywords}: \raisebox{2ex}{\parbox[t]{32em}{\flushleft 
Bethe-Salpeter Equation, Confinement; Dynamical Chiral Symmetry Breaking; 
Dyson-Schwinger Equations; Electroweak and Strong Form Factors; Faddeev 
Equation; Hadron Physics; QCD Modelling}}\\[0.5ex]}%
 
\textheight=7.8truein 
 
\setcounter{footnote}{0} 
\renewcommand{\thefootnote}{\alph{footnote}} 
 
 
\bigskip 
 
\noindent 
{\bf Contents}\medskip 
 
\hspace*{0.8em}\parbox{32em}{%
\noindent 1.\ Introduction \dotfill \pageref{introlabel} 
\smallskip 
 
\noindent 2.\ Dyson-Schwinger Equations \dotfill \pageref{sect2label} \smallskip 
 
\noindent 3.\  Foundation for a Description of Mesons \dotfill 
\pageref{sect3label} 
\smallskip 
 
\noindent 4.\ \textit{Ab Initio} Calculation of Meson Properties \dotfill 
\pageref{sect3pmlabel} \smallskip 
 
\noindent 5.\ On Baryons \dotfill \pageref{sect4label} \smallskip 
 
\noindent 6.\ Epilogue \dotfill \pageref{epiloguelabel} \smallskip 
 
\noindent \hspace*{0.8em} References \dotfill \pageref{Rlatticeproc} \smallskip } 
 
\pagebreak
 
 
\input sect1.tex

\input sect2.tex

\input sect3.tex

\input sect4.tex

\input sect5.tex

\input sect6.tex

\nonumsection{Acknowledgments} 
\noindent 
In preparing this article we benefited from conversations and correspondence
with
J.C.R.\ Bloch,
S.J.\ Brodsky,  
F.\ Coester, 
M.B.\ Hecht, 
G.A.\ Miller
and P.C.\ Tandy. 
This work was supported by the US Department of Energy (DOE), Nuclear Physics 
Division, under contract no.\ W-31-109-ENG-38, and DOE grant nos.\ 
DE-FG02-96-ER-40947, DE-FG02-97-ER-41048; and benefited from the resources of 
the US National Energy Research Scientific Computing Center. 
 
\nonumsection{References} 
 
\end{document}

%% file: sect1.tex
 
\vspace*{1pt}\textlineskip  
\section{Introduction}      
\addtocounter{section}{1} 
\vspace*{-0.5pt} 
\noindent 
%
%
A central\label{introlabel} goal of contemporary nuclear physics is to 
understand the properties of hadrons in terms of the elementary excitations in 
quantum chromodynamics (QCD): quarks, gluons and ghosts, because this is a 
crucial step in validating the theory.  Here ``elementary excitations'' means 
those quantum fields in terms of which QCD's Lagrangian is naturally expressed; 
for example, the analogues in quantum electrodynamics (QED) are the electron 
and photon.  Just as positronium does not appear in the QED Lagrangian but is 
understood as a relativistic bound state in the theory, hadrons do not appear 
in QCD's Lagrangian: they are supposed to be bound states of quarks and gluons. 
It is an observational fact that hitherto only colourless hadrons have been 
observed and that hadrons cannot be ``ionised;'' i.e., unlike positronium, for 
which the addition of a few ($\sim 7$) electron-volts will separate the 
electron and positron, no amount of energy available presently is sufficient to 
break a hadron into separated coloured constituents.  To understand this unique 
feature of confinement is one of the most significant challenges in physics. 
 
The defining problems of hadron physics can now be stated: specify and solve 
the quantum field theoretical bound state problem whose solution is the hadron 
spectrum; calculate the interactions of these hadrons between themselves and 
with electroweak probes; and use these interactions as incisive tools with 
which to identify the origins of confinement and elucidate its effects on, and 
expression in, observables.  These problems are essentially nonperturbative. 
 
In considering hadron physics it is natural to think of numerical simulations 
of lattice-QCD, which over the last thirty years has become an independent 
branch of high-energy physics with numerous large-scale collaborations.  An 
essential aspect of lattice-QCD is the enumerable, finite grid used to 
represent the spacetime continuum, which means that a comparison between 
simulations and experiment necessarily involves extrapolations.  In modern 
studies the extrapolation to the continuum limit; i.e., to small spacing 
between the lattice sites, is reliably handled.  However, that is not true of 
the extrapolation to infinite volume.  Furthermore, most simulations continue 
to employ the so-called quenched approximation, which introduces an \textit{a 
priori} unquantifiable systematic error.  Significant resources are being 
expended to overcome these difficulties, by developing improved algorithms, and 
by the obvious expedient of acquiring faster computers with more memory. 
Lattice-QCD's current status can be appraised from the proceedings of any of 
the major lattice conferences, e.g., Refs.\ [\ref{Rlatticeproc}].  There are 
naturally successes and problems, and we leave their description to 
practitioners, e.g., Ref.\ [\ref{Rlatticesumm}]. 
 
In describing many aspects of hadron physics, light-front concepts and 
techniques are useful, as may be seen, e.g., from Ref.\ [\ref{Rlightcone}]. The 
application, exploration and improvement of these methods, too, forms an 
identifiable subfield but one that is not insular: much is being gained by 
capitalising on the opportunities that exist to inform other nonperturbative 
approaches and models, and the feedback that brings.  This framework has seen 
numerous recent successes, e.g., a renormalisation group approach to the 
light-front Hamiltonian,\cite{rjperry} applications\cite{transverse} of 
light-front Hamiltonian techniques to the transverse lattice formulation of 
QCD, and refinements of discretised light-front quantisation.\cite{hiller}  A 
perspective on the approach's status and future is presented in Ref.\ 
[\ref{Rsjbrodsky}]. 
 
Reviewing this material it is readily apparent that drawing a connection
between QCD and hadron observables is difficult, and that is why modelling
remains a keystone of hadron physics.  Constituent-quark-like models
currently provide a peerless description of the baryon spectrum and baryon
decays,\cite{capstick,plessas} correlating a wealth of data via few
parameters.  Furthermore, while such models do not simultaneously give a
satisfactory description of low-lying light-quark mesons, for which an
accurate representation of dynamical chiral symmetry breaking (DCSB) is
essential, they can be adapted and applied to guide the search for exotic and
hybrid \linebreak had\-rons\footnote{In fact, these quantum mechanical models
must be adapted because exotic hadrons are, by definition, those states which
cannot be constructed in the simplest versions of the model, with only
constituent-quark degrees of freedom.} \ and
glueballs.\cite{swanson,page,swanson2} When focusing on properties of the
baryon octet and decuplet, such as elastic and transition form factors at
small to moderate momentum transfer, mean field models of baryon structure
ameliorate some of the deficiencies of constituent-quark models and provide
additional insights.\cite{tonyCBM,Birse:cx,reinhardTS,thomasweise} At higher
momentum transfers a proper expression of Poincar\'e covariance becomes
important in models of such processes.\cite{plessas,polyzou}
 
Contemporary Dyson-Schwinger equation (DSE) studies complement these 
approaches.  Modern comparisons with and predictions of experimental data can 
properly be said to rest on model assumptions but they can be tested within the 
framework and also via comparison with lattice-QCD simulations, and the 
predictions are very good.  Furthermore, progress in understanding the intimate 
connection between symmetries and truncation schemes has enabled the proof of 
exact results. Herein we review recent phenomenological applications and the 
foundation of their success.  There are naturally challenges, which we shall 
also highlight.

%% file: sect2.tex
 
\vspace*{1pt}\textlineskip  
\section{Dyson-Schwinger Equations}     
\addtocounter{section}{1} 
\vspace*{-0.5pt} 
\noindent 
The best\label{sect2label} known DSE is the simplest: the \textit{Dyson} or 
\textit{gap} equation, which describes how the propagation of a fermion is 
modified by its interactions with the medium being traversed.  In QCD that 
equation is:\footnote{We employ a Euclidean metric throughout, with: 
$\{\gamma_\mu,\gamma_\nu\} = 2\delta_{\mu\nu}$; $\gamma_\mu^\dagger = 
\gamma_\mu$; and $a \cdot b = \sum_{i=1}^4 a_i b_i$.} 
\begin{equation} 
\label{gendse} S^{-1}(p) = Z_2 \,(i\gamma\cdot p + m_{\rm bare}) +\, Z_1 
\int^\Lambda_q \, g^2 D_{\mu\nu}(p-q) \frac{\lambda^a}{2}\gamma_\mu S(q) 
\Gamma^a_\nu(q;p) \,; 
\end{equation} 
i.e., the renormalised DSE for the dressed-quark propagator.  In Eq.\ 
(\ref{gendse}), $D_{\mu\nu}(k)$ is the renormalised dressed-gluon propagator, 
$\Gamma^a_\nu(q;p)$ is the renormalised dressed-quark-gluon vertex, $m_{\rm 
bare}$ is the $\Lambda$-dependent current-quark bare mass that appears in the 
Lagrangian and $\int^\Lambda_q := \int^\Lambda d^4 q/(2\pi)^4$ represents a 
\textit{translationally-invariant} regularisation of the integral, with 
$\Lambda$ the regularisation mass-scale.\footnote{It is only with a 
translationally invariant regularisation scheme that Ward-Takahashi identities 
can be preserved, something that is crucial to ensuring vector and axial-vector 
current conservation.  The final stage of any calculation is to take the limit 
$\Lambda \to \infty$.} \ In addition, $Z_1(\zeta^2,\Lambda^2)$ and 
$Z_2(\zeta^2,\Lambda^2)$ are the quark-gluon-vertex and quark wave function 
renormalisation constants, which depend on the renormalisation point, $\zeta$, 
and the regularisation mass-scale, as does the mass renormalisation constant 
\begin{equation} 
\label{Zmass} Z_m(\zeta^2,\Lambda^2) = 
Z_4(\zeta^2,\Lambda^2)/Z_2(\zeta^2,\Lambda^2) . 
\end{equation} 
The renormalised mass is 
$m(\zeta) := m_{\rm bare}(\Lambda)/Z_m(\zeta^2,\Lambda^2)$. 
When $\zeta$ is very large the right-hand-side (r.h.s.) can be evaluated in 
perturbation theory whereby one finds 
\begin{equation} 
m(\zeta) = 
\frac{\hat m}{(\ln \zeta/\Lambda_{\rm QCD})^{\gamma_m}\!\!\!\!\!\!\!} 
\;\;\;\;\; , 
\end{equation} 
with $\gamma_m = 12/(33 - 2 N_f)$, where $N_f$ is the number of current-quark 
flavours that contribute actively to the running coupling, and $\Lambda_{\rm 
QCD}$ and $\hat m$ are renormalisation group invariants: $\hat m$ is the 
renormalisation-group-invariant current-quark mass. 
 
The solution of Eq.~(\ref{gendse}) is the dressed-quark propagator, which 
takes the form 
\begin{equation} 
 S^{-1} (p) =  i \gamma\cdot p \, A(p^2,\zeta^2) + B(p^2,\zeta^2) \\ 
 = \frac{1}{Z(p^2,\zeta^2)}\left[ i\gamma\cdot p + M(p^2)\right] 
\label{sinvp} 
\end{equation} 
and is obtained by solving the gap equation subject to the renormalisation 
condition that at some large spacelike $\zeta^2$ 
\begin{equation} 
\label{renormS} \left.S^{-1}(p)\right|_{p^2=\zeta^2} = i\gamma\cdot p + 
m(\zeta)\,. 
\end{equation} 
 
The gap equation illustrates the features and flaws of each DSE.  It is a
nonlinear integral equation for $S(p)$ and hence can yield much-needed
nonperturbative information.  However, the kernel involves the two-point
function $D_{\mu\nu}(k)$ and the three-point function $\Gamma^a_\nu(q;p)$.
The equation is therefore coupled to the DSEs these functions satisfy. Those
equations in turn involve higher $n$-point functions and hence the DSEs are a
tower of coupled integral equations with a tractable problem obtained only
once a truncation scheme is specified.  It is unsurprising that the best
known truncation scheme is the weak coupling expansion, which reproduces
every diagram in perturbation theory.  This scheme is systematic and valuable
in the analysis of large momentum transfer phenomena because QCD is
asymptotically free but it precludes any possibility of obtaining
nonperturbative information, which we identified as a key feature of the
DSEs.
 
In spite of the problem with truncation, gap equations have long been used 
effectively in obtaining nonperturbative information about many-body systems 
as, e.g., in the Nambu-Gorkov formalism for superconductivity.\cite{mattuck} 
The positive outcomes have been achieved through a simple expedient of 
employing the most rudimentary truncation, e.g., Hartree or Hartree-Fock, and 
comparing the results with observations.  Naturally, agreement under these 
circumstances is not an unambiguous indication that the contributions omitted 
are small nor that the model expressed in the truncation is sound. However, it 
does justify further study, and an accumulation of good results is grounds for 
a concerted attempt to substantiate a reinterpretation of the truncation as the 
first term in a systematic and reliable approximation. 
 
\subsection{Nonperturbative truncation} 
\addtocounter{subsection}{1} 
\label{sec:nonptruncation} 
\noindent 
To explain why Eq.\ (\ref{gendse}) is called a gap equation we consider the 
chiral limit, which is readily defined\cite{mrt98} because QCD exhibits 
asymptotic freedom and implemented by employing\cite{mr97} 
\begin{equation} 
\label{chirallimit} Z_2(\zeta^2,\Lambda^2) \, 
m_{\rm bare}(\Lambda) \equiv 0\,, \;\; \Lambda \gg \zeta\,. 
\end{equation} 
An equivalent statement is that one obtains the chiral limit when the
renorma\-li\-sa\-tion-point-invariant current-quark mass vanishes; i.e.,
$\hat m = 0$.  In this case the theory is chirally symmetric, and a
perturbative evaluation of the dressed-quark propagator from Eq.\
(\ref{gendse}) gives
\begin{equation} 
\label{Bpert0} 
B_{\rm pert}^0(p^2) = m \left( 1 - \frac{\alpha}{\pi} \ln 
\left[\frac{p^2}{m^2} \right] + \ldots \right) \stackrel{m\to 0}{\equiv} 0\,; 
\end{equation} 
viz., the perturbative mass function is identically zero in the chiral limit. 
It follows that there is no gap between the top level in the quark's filled 
negative-energy Dirac sea and the lowest positive energy level. 
 
However, suppose that one had at hand a truncation scheme other than 
perturbation theory and that subject to this scheme Eq.\ (\ref{gendse}) 
possessed a chiral limit solution $B^0(p^2)\not\equiv 0$.  Then interactions 
between the quark and the virtual quanta populating the ground state would 
have nonperturbatively generated a mass gap.  The appearance of such a gap 
breaks the theory's chiral symmetry.  This shows that the gap equation can be 
an important tool for studying DCSB, and it has long been used to explore 
this phenomenon in both QED and QCD.\cite{cdragw} 
 
The gap equation's kernel is formed from a product of the dressed-gluon 
propagator and dres\-sed-quark-gluon vertex.  However, in proposing and 
developing a truncation scheme it is insufficient to focus only on this 
kernel.  We have observed that the gap equation can be a tool for studying 
DCSB but it is only useful if the truncation itself does not destroy the 
chiral symmetry. 
 
Chiral symmetry is expressed via the axial-vector Ward-Takahashi 
identity:\label{chiralsymmetry} 
\begin{equation} 
P_\mu \, \Gamma_{5\mu}(k;P)  =  S^{-1}(k_+)\, i\gamma_5 + i\gamma_5 
\,S^{-1}(k_-)\,,\; k_\pm = k\pm  P/2 , \label{avwti} 
\end{equation} 
wherein $\Gamma_{5\mu}(k;P)$ is the dressed axial-vector vertex.  This 
three-point function satisfies an inhomogeneous Bethe-Salpeter equation 
(BSE): 
\begin{equation} 
\label{avbse} 
\left[\Gamma_{5\mu}(k;P)\right]_{tu} 
 =  Z_2 \left[\gamma_5\gamma_\mu\right]_{tu} + \int^\Lambda_q 
[S(q_+) \Gamma_{5\mu}(q;P) S(q_-)]_{sr} K_{tu}^{rs}(q,k;P)\,, 
\end{equation} 
in which $K(q,k;P)$ is the fully-amputated quark-antiquark scattering kernel, 
and the colour-, Dirac- and flavour-matrix structure of the elements in the 
equation is denoted by the indices $r,s,t,u$.  The Ward-Takahashi identity, 
Eq.~(\ref{avwti}), entails that an intimate relation exists between the kernel 
in the gap equation and that in the BSE. (This is another example of the 
coupling between DSEs.) Therefore an understanding of chiral symmetry and its 
dynamical breaking can only be obtained with a truncation scheme that preserves 
this relation, and hence guarantees Eq.~(\ref{avwti}) without a 
\textit{fine-tuning} of model-dependent parameters. 
 
\subsubsection{Rainbow-ladder truncation} 
\addtocounter{subsubsection}{1} 
\noindent 
At least one such scheme exists.\cite{truncscheme} Its leading-order term is
the so-called re\-nor\-ma\-li\-sa\-tion-group-improved rainbow-ladder
truncation, whose analogue in the many body problem is an Hartree-Fock
truncation of the one-body (Dyson) equation combined with a consistent
ladder-truncation of the related two-body (Bethe-Salpeter) equation.  To
understand the origin of this leading-order term observe that the
dressed-ladder truncation of the quark-antiquark scattering kernel is
expressed in Eq.\ (\ref{avbse}) via
\begin{eqnarray} 
\nonumber \lefteqn{ [ L(q,k;P)]_{t u}^{t^\prime u^\prime } \, 
[\Gamma_{5\mu}(q;P)]_{u^\prime t^\prime }  :=[S(q_+) \Gamma_{5\mu}(q;P) 
S(q_-)]_{sr} K_{tu}^{rs}(q,k;P)}\\ 
\nonumber & =& \!\!\!\! - g^2(\zeta^2)\, D_{\rho\sigma}(k-q) \, 
\left[\rule{0mm}{0.7\baselineskip} \Gamma^a_\rho(k_+,q_+)\,S(q_+) 
\right]_{tt^\prime} \left[\rule{0mm}{0.7\baselineskip} 
S(q_-)\,\Gamma^a_\sigma(q_-,k_-) \right]_{u^\prime u}\, 
[\Gamma_{5\mu}(q;P)]_{t^\prime u^\prime} \\ 
\end{eqnarray} 
wherein we have only made explicit the renormalisation point dependence of the 
coupling.  One can exploit multiplicative renormalisability and asymptotic 
freedom to establish that on the kinematic domain for which $Q^2:=(k-q)^2 \sim 
k^2\sim q^2$ is large and spacelike 
\begin{equation} 
\label{LqkPuv} 
[L(q,k;P)]_{t u}^{t^\prime u^\prime } 
= - 
4 \pi \alpha(Q^2) \, D_{\rho\sigma}^{\rm free}(Q)\, 
\left[\rule{0mm}{0.7\baselineskip} 
        \frac{\lambda^a}{2}\gamma_\rho \,S^{\rm free}(q_+)\right]_{tt^\prime} 
\left[\rule{0mm}{0.7\baselineskip}S^{\rm free}(q_-)\, 
        \frac{\lambda^a}{2}\gamma_\sigma\right]_{u^\prime u}, 
\end{equation} 
where $\alpha(Q^2)$ is the strong running coupling and, e.g., $S^{\rm free}$ is 
the free quark propagator.  It follows that on this kinematic domain the 
r.h.s.\ of Eq.\ (\ref{LqkPuv}) describes the leading contribution to the 
complete quark-antiquark scattering kernel, $K_{tu}^{rs}(q,k;P)$, with all 
other contributions suppressed by at least one additional power of $1/Q^2$. 
 
The renormalisation-group-improved ladder-truncation supposes that 
\begin{equation} 
\label{ladder} 
K_{tu}^{rs}(q,k;P) = - 4 \pi \alpha(Q^2) D_{\rho\sigma}^{\rm free}(Q)\, 
\left[\rule{0mm}{0.7\baselineskip} 
        \frac{\lambda^a}{2}\gamma_\rho \right]_{ts} 
\left[\rule{0mm}{0.7\baselineskip} 
        \frac{\lambda^a}{2}\gamma_\sigma\right]_{r u}
\end{equation} 
is also a good approximation on the infrared domain and is thus an assumption 
about the long-range ($Q^2 \lsim 1\,$ GeV$^2$) behaviour of the interaction. 
Combining Eq.\ (\ref{ladder}) with the requirement that Eq.\ (\ref{avwti}) be 
automatically satisfied leads to the renormalisation-group-improved 
rainbow-truncation of the gap equation: 
\begin{equation} 
\label{rainbowdse} S^{-1}(p) = Z_2 \,(i\gamma\cdot p + m_{\rm bare}) + 
\int^\Lambda_q 4 \pi \alpha(Q^2) D_{\mu\nu}^{\rm free} (p-q) 
\frac{\lambda^a}{2}\gamma_\mu \, S(q) \, \frac{\lambda^a}{2} \gamma^a_\nu\,. 
\end{equation} 
 
As will become apparent, the rainbow-ladder truncation provides the foundation 
for an explanation of a wide range of hadronic phenomena.  There are some 
notable exceptions; e.g., the description of the scalar meson sector is 
unconvincing, and it is natural to seek the reason.  The recognition that this 
truncation is the first term in a systematic procedure has provided an 
answer.\cite{truncscheme} 
 
\subsection{Systematic procedure} 
\addtocounter{subsection}{1} 
\noindent 
The truncation scheme of Ref.~[\ref{Rtruncscheme}] is a dressed-loop expansion 
of the dressed-quark-gluon vertices that appear in the half-amputated 
dressed-quark-anti\-quark scattering matrix: $S^2 K$, a 
re\-nor\-ma\-li\-sa\-tion-group in\-va\-ri\-ant.\cite{detmold} All $n$-point 
functions involved thereafter in connecting two particular quark-gluon vertices 
are \textit{fully dressed}. The effect of this truncation in the gap equation, 
Eq.~(\ref{gendse}), is realised through the following representation of the 
dressed-quark-gluon vertex, $i \Gamma_\mu^a = \frac{i}{2}\lambda^a\,\Gamma_\mu 
= l^a \Gamma_\mu$: 
\begin{eqnarray} 
\nonumber 
\lefteqn{Z_1 \Gamma_\mu(k,p)   =   \gamma_\mu +  \frac{1}{2 N_c} 
\int_\ell^\Lambda\! g^2 D_{\rho\sigma}(p-\ell) 
\gamma_\rho S(\ell+k-p) \gamma_\mu S(\ell) 
\gamma_\sigma}\\ 
\nonumber &+ & \frac{N_c}{2}\int_\ell^\Lambda\! g^2\, 
D_{\sigma^\prime \sigma}(\ell) \, D_{\tau^\prime\tau}(\ell+k-p)\, 
\gamma_{\tau^\prime} \, S(p-\ell)\, 
\gamma_{\sigma^\prime}\, 
\Gamma^{3g}_{\sigma\tau\mu}(\ell,-k,k-p) + [\ldots]\,. \\ 
\label{vtxexpand} 
\end{eqnarray} 
Here $\Gamma^{3g}$ is the dressed-three-gluon vertex and it is readily apparent 
that the lowest order contribution to each term written explicitly is O$(g^2)$. 
The ellipsis represents terms whose leading contribution is O$(g^4)$; viz., the 
crossed-box and two-rung dressed-gluon ladder diagrams, and also terms of 
higher leading-order. 
 
This expansion of $S^2 K$, with its implications for other $n$-point functions, 
yields an ordered truncation of the DSEs that guarantees, term-by-term, the 
preservation of vector and axial-vector Ward-Takahashi identities, a feature 
that has been exploited\cite{mrt98,marisAdelaide,mishaSVY} to prove Goldstone's 
theorem and other exact results in QCD.  It is readily seen that inserting Eq.\ 
(\ref{vtxexpand}) into Eq.\ (\ref{gendse}) provides the rule by which the 
rainbow-ladder truncation can be systematically improved. 
 
\subsubsection{Planar vertex} 
\begin{figure}[t] 
 
\centerline{\includegraphics[height=9em]{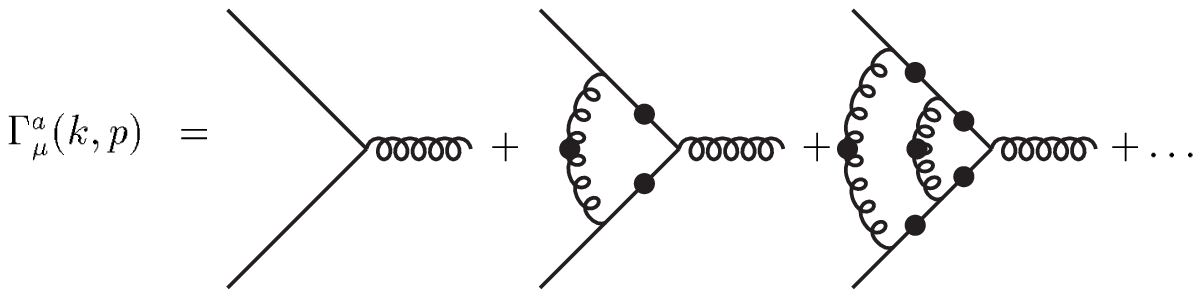}} 
\fcaption{\label{Gamma_inf} Integral equation for a planar dressed-quark-gluon 
vertex obtained by neglecting contributions associated with explicit gluon 
self-interactions.  Solid circles indicate fully dressed propagators.  The 
vertices are not dressed.  (Adapted from Ref.\ [\protect\ref{Rdetmold}].)} 
 
\end{figure}

\addtocounter{subsubsection}{1} 
\noindent 
The effect of the complete vertex in Eq.\ (\ref{vtxexpand}) on the solutions of 
the gap equation is unknown.  However, insights have been drawn from a 
study\cite{detmold} of a more modest problem obtained by retaining only the sum 
of dressed-gluon ladders; i.e., the vertex depicted in Fig.~\ref{Gamma_inf}. 
The elucidation is particularly transparent when one employs the following 
choice for the dressed-gluon line in the figure:\cite{mn83} 
\begin{equation} 
\label{mnmodel} {\cal D}_{\mu\nu}(k):= g^2 \, D_{\mu\nu}(k) = 
\left(\delta_{\mu\nu} - \frac{k_\mu k_\nu}{k^2}\right) (2\pi)^4\, {\cal G}^2 \, 
\delta^4(k)\,, 
\end{equation} 
which defines an ultraviolet finite model so that the regularisation mass-scale 
can be removed to infinity and the renormalisation constants set equal to 
one.\footnote{The constant ${\cal G}$ sets the model's mass-scale and using 
${\cal G}=1$ simply means that all mass-dimensioned quantities are measured in 
units of ${\cal G}$.} \  This model has many positive features in common with 
the class of renormalisation-group-improved rainbow-ladder models and its 
particular momentum-dependence works to advantage in reducing integral 
equations to algebraic equations with similar qualitative features.  There is 
naturally a drawback: the simple momentum dependence also leads to some 
model-dependent artefacts, but they are easily identified and hence not cause 
for concern.

The general form of the dressed-quark gluon vertex involves twelve distinct 
scalar form factors but using Eq.~(\ref{mnmodel}) only $\Gamma_\mu(p) 
:=\Gamma_\mu(p,p)$ contributes to the gap equation. This considerably 
simplifies the analysis since, in general, 
\begin{equation} 
\Gamma_\mu(p)  =  \alpha_1(p^2)\, \gamma_\mu 
+ \,  \alpha_2(p^2)\, \gamma\cdot p\,p_\mu 
- \, \alpha_3(p^2)\, i \,p_\mu 
+ \alpha_4(p^2) \, i \gamma_\mu \,\gamma\cdot p\,. \label{vtxmn} 
\end{equation} 
The summation depicted in Fig.~\ref{Gamma_inf} is expressed via 
\begin{equation} 
\label{vtxalgebraic} \Gamma_\mu(p) = \gamma_\mu + \frac{1}{8}\,\gamma_\rho\, 
S(p)\, \Gamma_\mu(p)\, S(p)\, \gamma_\rho\,, 
\end{equation} 
and inserting Eq.~(\ref{vtxmn}) into Eq.~(\ref{vtxalgebraic}) one finds $ 
\alpha_4\equiv 0$, and hence the solution 
simplifies: 
\begin{equation} 
\label{vtxinfty} \Gamma_\mu(p) = \alpha_1(p^2)\, \gamma_\mu + \, 
\alpha_2(p^2)\, \gamma\cdot p\,p_\mu - \, \alpha_3(p^2)\, i \,p_\mu\,. 
\end{equation} 
The three surviving functions are those which are most important in the 
dressed-quark-photon vertex.\cite{pieterpion} 
 
\begin{figure}[t] 
 
\centerline{\includegraphics[height=12em]{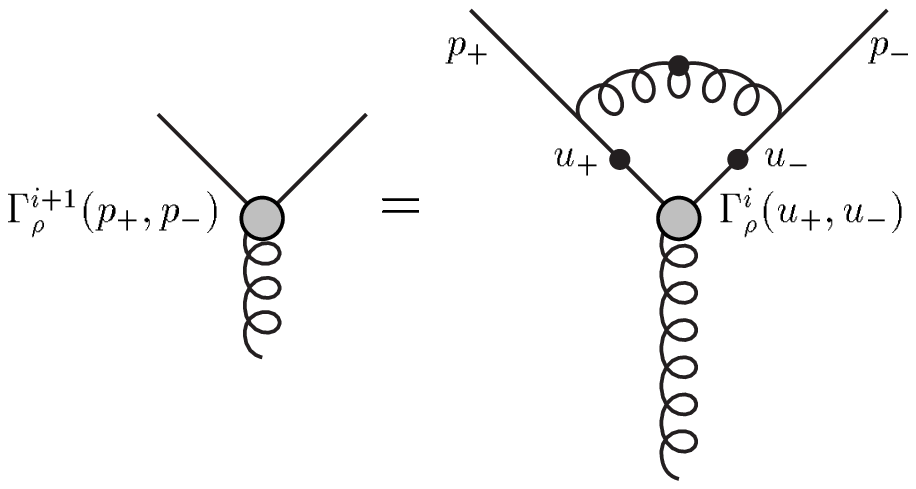}} 
\fcaption{\label{gamma_recursion} Recursion relation for the iterates in the 
fully-resummed dressed-gluon-ladder vertex: filled circles denote a
fully-dressed propagator or vertex.  Using Eq.~(\protect\ref{mnmodel}),
$p_+=p_-$.  (Adapted from Ref.\ [\protect\ref{Rdetmold}].)}
 
\end{figure} 
 
One can re-express this vertex as \begin{eqnarray} 
\Gamma_\mu(p) = \sum_{i=0}^\infty\,\Gamma_\mu^i(p) = 
\sum_{i=0}^\infty\, \left[ \alpha^i_1(p^2)\, \gamma_\mu + \, 
\alpha^i_2(p^2)\, \gamma\cdot p\,p_\mu - \, \alpha^i_3(p^2)\, i 
\,p_\mu\right], \label{vtxi} 
\end{eqnarray} 
where the superscript enumerates the order of the iterate: $\Gamma_\mu^{i=0}$ 
is the bare vertex, 
\begin{equation} 
\label{alpha0} 
\begin{array}{cc} 
\alpha_1^0 = 1\,,\; & \alpha_2^0=0=\alpha_3^0\,; 
\end{array} 
\end{equation} 
$\Gamma_\mu^{i=1}$ is the result of inserting this into the r.h.s.\ of 
Eq~(\ref{vtxalgebraic}) to obtain the one-rung dressed-gluon correction; 
$\Gamma_\mu^{i=2}$ is the result of inserting $\Gamma_\mu^{i=1}$, and is 
therefore the two-rung dressed-gluon correction; etc.  A key 
observation\cite{detmold} is that each iterate is related to its precursor via 
the simple recursion relation depicted in Fig.~\ref{gamma_recursion} and, 
substituting Eq.~(\ref{vtxi}), this recursion yields ($s=p^2$) 
\begin{equation} 
\label{matrixrecurs} \mbox{\boldmath $\alpha$}^{i+1}(s):= 
\left( 
\begin{array}{l} 
\rule{0ex}{2.5ex}\alpha_1^{i+1}(s) \\\rule{0ex}{2.5ex} \alpha_2^{i+1}(s) \\ 
\rule{0ex}{2.5ex}\alpha_3^{i+1}(s) 
\end{array}\right) 
= {\cal O}(s;A,B)\, \mbox{\boldmath $\alpha$}^{i}(s)\,, 
\end{equation} 
\begin{equation} 
{\cal O}(s;A,B) = \frac{1}{4} \,\frac{1}{\Delta^2} 
\left( 
\begin{array}{ccc} 
\rule{0ex}{2.5ex} - \Delta & 0 & 0\\ 
\rule{0ex}{2.5ex} 2  A^2 & s A^2 - B^2 & 2 A B \\ 
\rule{0ex}{2.5ex} 4 A B & 4 s A B & 2 (B^2 - s A^2) 
\end{array} \right), 
\end{equation} 
$\Delta = s A^2(s) + B^2(s)$.  It follows that 
\begin{equation} 
\label{boldalpha} 
\mbox{\boldmath $\alpha$} 
= \left(\sum_{i=1}^\infty\, {\cal O}^i\right) \, \mbox{\boldmath $\alpha$}^0 
= \frac{1}{1 - {\cal O}} \,\mbox{\boldmath $\alpha$}^0\, 
\end{equation} 
and hence, using Eq.\ (\ref{alpha0}), 
\begin{eqnarray} 
\nonumber 
\alpha_1 & = & \frac{4\, \Delta}{1 + 4\,\Delta}\,,\\ 
 \alpha_2 & = & \frac{- \,8 \,A^2} { 
1 + 2\,( B^2 - s\,A^2) - 8\,\Delta^2 } \, 
\frac{1 + 2\,\Delta}{1 + 4\,\Delta} \,, \label{alpharesults}\\ 
\nonumber \alpha_3 & = & 
\frac{- 8 \,A B}{ 1 + 2 (B^2 - s\, A^2) - 8\,\Delta^2 }\,. 
\end{eqnarray} 
 
The recursion relation thus leads to a closed form for the gluon-ladder-dressed 
quark-gluon vertex in Fig.~\ref{Gamma_inf}; viz., Eqs.~(\ref{vtxinfty}), 
(\ref{alpharesults}).  Its momentum-dependence is determined by that of the 
dressed-quark propagator, which is obtained by solving the gap equation, itself 
constructed with this vertex.  Using Eq.~(\ref{mnmodel}), that gap equation is 
\begin{equation} 
S^{-1}(p) = 
 i \gamma\cdot p + m + \gamma_\mu \, S(p) \,\Gamma_\mu(p) \\ 
\label{gapmodel} 
\end{equation} 
and substituting Eq.~(\ref{vtxinfty}) gives 
\begin{eqnarray} 
A(s) & = & 1 +\frac{1}{s A^2 + B^2} \left[ A \, (2 \alpha_1 - s 
\alpha_2) - B \,\alpha_3\right]\,,\label{Afull}\\ 
B(s) & = & m+ \frac{1}{s A^2 + B^2} \left[ 
B \, ( 4 \alpha_1 + s \alpha_2) - s A \,\alpha_3 \right]\,. \label{Bfull} 
\end{eqnarray} 
Equations (\ref{Afull}), (\ref{Bfull}), completed using
Eqs.~(\ref{alpharesults}), form a closed algebraic system.  It can easily be
solved numerically, and that yields simultaneously the complete
gluon-ladder-dressed vertex and the propagator for a quark fully dressed via
gluons coupling through this nonperturbative vertex.  Furthermore, it is
apparent that in the chiral limit, $m=0$, a realisation of chiral symmetry in
the Wigner-Weyl mode, which is expressed via the $B\equiv 0$ solution of the
gap equation, is always admissible.  This is the solution anticipated in Eq.\
(\ref{Bpert0}).
 
\begin{figure}[t] 
 
\centerline{\includegraphics[height=16em]{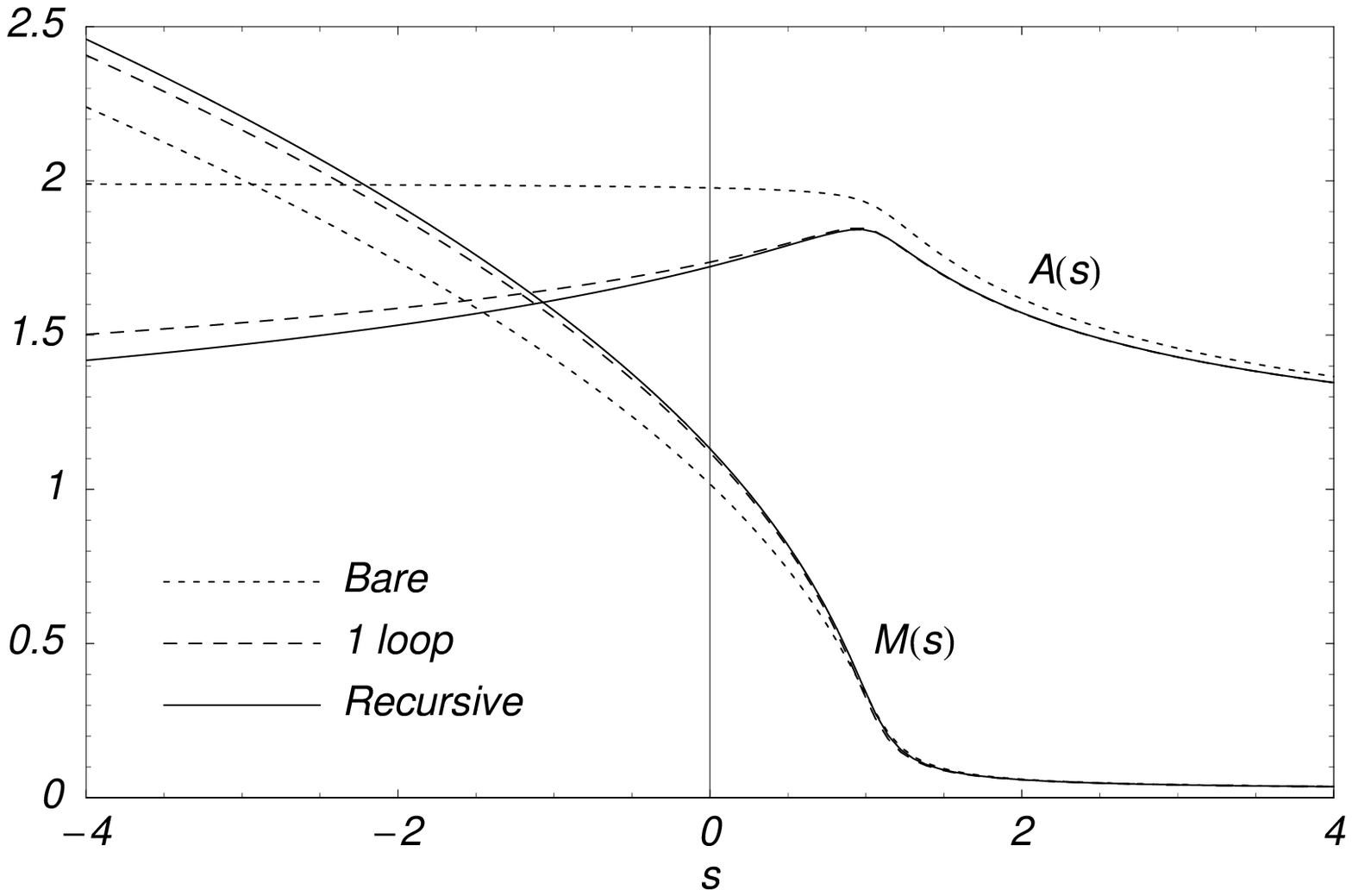}} 
\vspace*{7pt} 
\fcaption{\label{AM0plot} $A(s)$, $M(s)$ obtained from Eqs.\ 
(\protect\ref{alpharesults}), (\protect\ref{Afull}), (\protect\ref{Bfull}) 
with $m=0.023$: solid line.  All dimensioned quantities are expressed in 
units of ${\cal G}$ in Eq.~(\protect\ref{mnmodel}).  For comparison, the 
results obtained with the zeroth-order vertex: dotted line; and the one-loop 
vertex: dashed line, are also plotted.  (Adapted from Ref.\ 
[\protect\ref{Rdetmold}].)} 
 
\end{figure} 
 
\begin{figure}[t] 
 
\centerline{\includegraphics[height=16em]{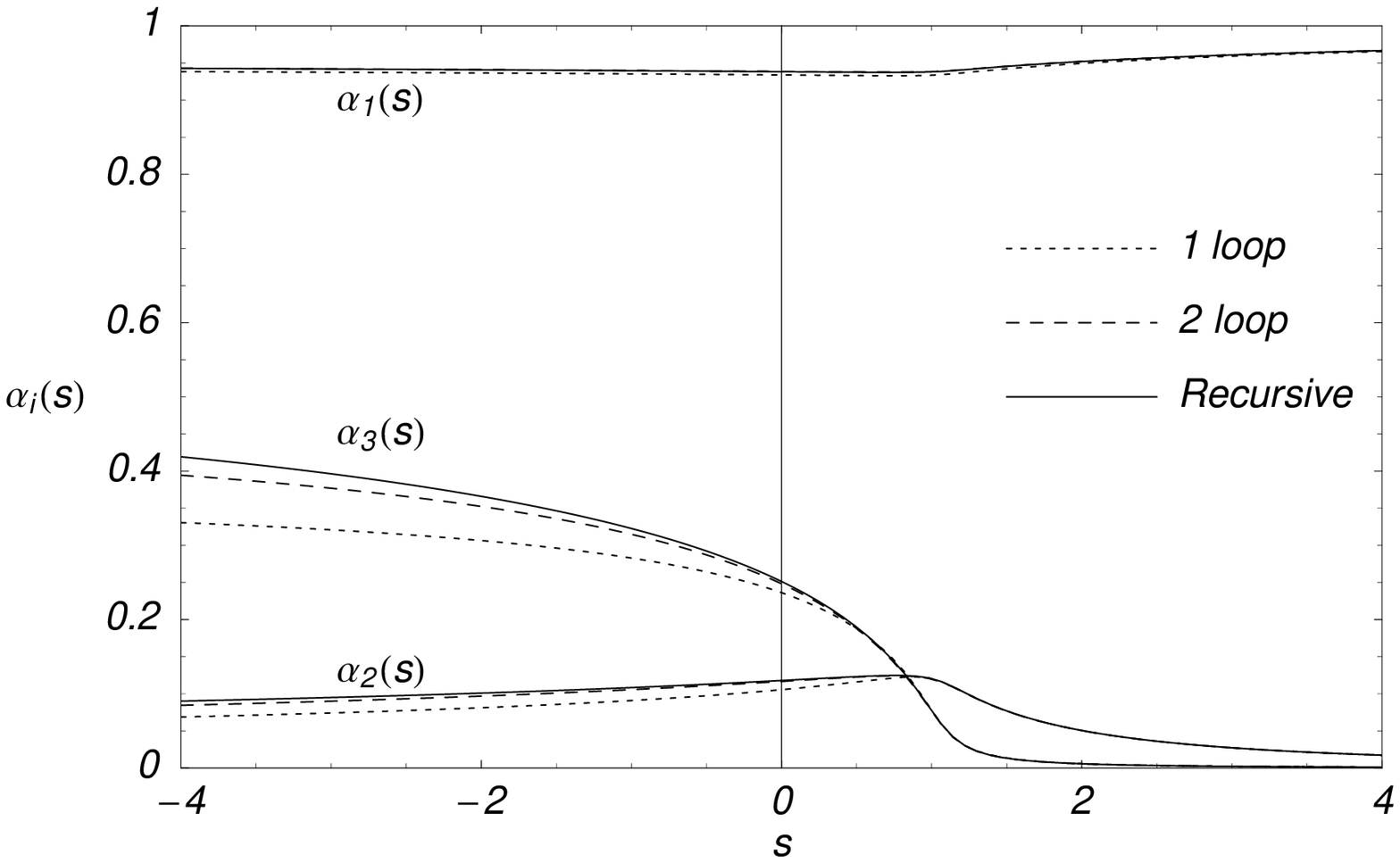}} 
\vspace*{7pt} 
\fcaption{\label{alphamplot} $\alpha_i$, $i=1,2,3$, calculated from Eqs.\ 
(\protect\ref{alpharesults}), (\protect\ref{Afull}), (\protect\ref{Bfull}) 
with $m=0.023$.  These functions calculated at one-loop (dotted line) and 
two-loop (dashed line) are also plotted for comparison.  (Adapted from Ref.\ 
[\protect\ref{Rdetmold}].)} 
 
\end{figure}

The chiral limit gap equation also admits a Nambu-Goldstone mode solution
whose $p^2\simeq 0$ properties are unambiguously related to those of the
$m\neq 0$ solution, a feature also evident in QCD.\cite{mishaSVY} A complete
solution of Eq.\ (\ref{gapmodel}) is available numerically, and results for
the dressed-quark propagator and gluon-ladder-dressed vertex are depicted in
Figs.~\ref{AM0plot}, \ref{alphamplot}.  It is readily seen that the complete
resummation of dressed-gluon ladders gives a dressed-quark propagator that is
little different from that obtained with the one-loop-corrected vertex; and
there is no material difference from the result obtained using the
zeroth-order vertex.  Similar observations apply to the vertex itself.  Of
course, there is a qualitative difference between the zeroth-order vertex and
the one-loop-corrected result: $\alpha_{2,3}\neq 0$ in the latter case.
However, once that effect is seeded, the higher-loop corrections do little.
The scale of these modest effects can be quantified by a comparison between
the values of $M(s=0)=B(0)/A(0)$ calculated using vertices dressed at
different orders:
\begin{equation} 
\begin{array}{l|llll} 
\sum_{i=0,N}\Gamma_\mu^i & N=0 & N=1 & N=2 & N=\infty\\[1.5ex]\hline 
M(0) & \rule{0em}{2ex} 1 & 1.105 &  1.115 & 1.117 
\end{array} 
\end{equation} 
The rainbow truncation of the gap equation is accurate to within 12\% and 
adding just one gluon ladder gives 1\% accuracy.  It is now important to couple 
this with an understanding of how the vertex resummation affects the 
Bethe-Salpeter kernel. 
 
\subsubsection{Vertex-consistent Bethe-Salpeter kernel} 
\begin{figure}[t] 
 
\centerline{\includegraphics[height=7em]{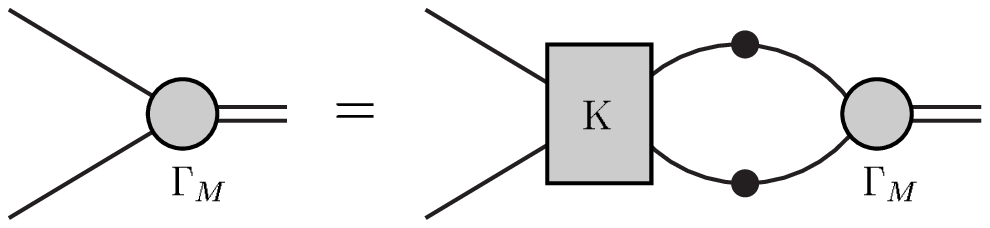}} 
\vspace*{7pt} 
\fcaption{\label{BSEpic} Homogeneous BSE, Eq.\ (\protect\ref{bsegen}).  Filled 
circles: dressed propagators or vertices; $K$ is the dressed-quark-antiquark 
scattering kernel.  A systematic truncation of $S^2 K$ is the key to 
preserving Ward-Takahashi identities.\protect\cite{truncscheme,herman} 
(Adapted from Ref.\ [\protect\ref{Rdetmold}].)} 
 
\end{figure}

\setcounter{dummyv}{\value{subsubsection}} 
\label{sec:BSK} 
\addtocounter{subsubsection}{1} 
\noindent 
The renormalised homogeneous BSE for the quark-antiquark channel denoted by $M$ 
can be expressed 
\begin{equation} 
\label{bsegen} 
    [\Gamma_M(k;P)]_{tu} =\int_q^\Lambda\! [\chi_M(q;P)]_{sr} 
    [ K(k,q;P)]_{tu}^{rs}\, 
\end{equation} 
where: $\Gamma_M(k;P)$ is the meson's Bethe-Salpeter amplitude, $k$ is the 
relative momentum of the quark-antiquark pair, $P$ is their total momentum; and 
\begin{equation} 
\label{chiM} 
\chi_M(k;P) = S(k_+)\, \Gamma_M(k;P) \,S(k_-)\,. 
\end{equation} 
Equation (\ref{bsegen}), depicted in Fig.\ \ref{BSEpic}, describes the 
residue at a pole in the solution of an inhomogeneous BSE; e.g., the lowest 
mass pole solution of Eq.\ (\ref{avbse}) is identified with the pion.  (NB.\ 
The normalisation of a Bethe-Salpeter amplitude is fixed by requiring that 
the bound state contribute with unit residue to the fully-amputated 
quark-antiquark scattering amplitude: $M = K + K (SS) K + 
\ldots$, see, e.g., Ref.\ [\protect\ref{Rllewellyn}].) 
 
On p.\ \pageref{chiralsymmetry} we observed that the automatic preservation 
of Ward-Takahashi identities in those channels related to strong interaction 
observables requires a conspiracy between the dressed-quark-gluon vertex and 
the Bethe-Salpeter kernel.\cite{truncscheme,herman} A systematic procedure 
for building that kernel follows\cite{detmold} from the 
observation\cite{herman} that the gap equation can be expressed via 
\begin{equation} 
\frac{\delta \Gamma[S] }{\delta S} = 0 \,, 
\end{equation} 
where $\Gamma[S]$ is a Cornwall-Jackiw-Tomboulis-like effective action.  The 
Bethe-Salpeter kernel is then obtained via an additional functional 
derivative: 
\begin{equation} 
K_{tu}^{rs} = - \frac{\delta \Sigma_{tu}}{\delta S_{rs}}\,. \label{KCJT} 
\end{equation} 
 
With the recursive vertex depicted in Fig.\ \ref{Gamma_inf}, the $n$-th order 
contribution to the kernel is obtained from the $n$-loop contribution to the 
self energy: 
\begin{equation} 
    \Sigma^n(p)= - \int_q^\Lambda\! {\cal D}_{\mu\nu}(p-q) \, l^a \gamma_\mu \, 
    S(q) l^a \,\Gamma_{\nu}^n(q,p). 
\end{equation} 
Since $\Gamma_\mu(p,q)$ itself depends on $S$ then Eq.~(\ref{KCJT}) yields the 
Bethe-Salpeter kernel as a sum of two terms and hence Eq.\ (\ref{bsegen}) 
assumes the form 
\begin{equation} 
\Gamma_M(k;P)  =  \int_q^\Lambda 
{\cal D}_{\mu\nu}(k - q)\,l^a \gamma_\mu \left[\rule{0em}{3ex} \chi_M(q;P) \, 
l^a\, 
\Gamma_\nu(q_-,k_-) + S(q_+) \, \Lambda_{M \nu}^{a}(q,k;P)\right], 
\label{genbsenL1} 
\end{equation} 
where we have used the mnemonic 
\begin{equation} 
\Lambda_{M \nu}^{a}(q,k;P) = \sum_{n=0}^\infty \Lambda_{M \nu}^{a;n}(q,k;P)\,. 
\label{Lambdatotal} 
\end{equation} 
While ${\cal D}_{\mu\nu}$ also depends on $S$ because of quark vacuum
polarisation diagrams, the additional term arising from the derivative of
${\cal D}_{\mu\nu}$ does not contribute to the Bethe-Salpeter kernel for
flavour nonsinglet systems, which are our current focus, and hence is
neglected for simplicity.
 
\begin{figure}[t] 
\centerline{\includegraphics[height=7em]{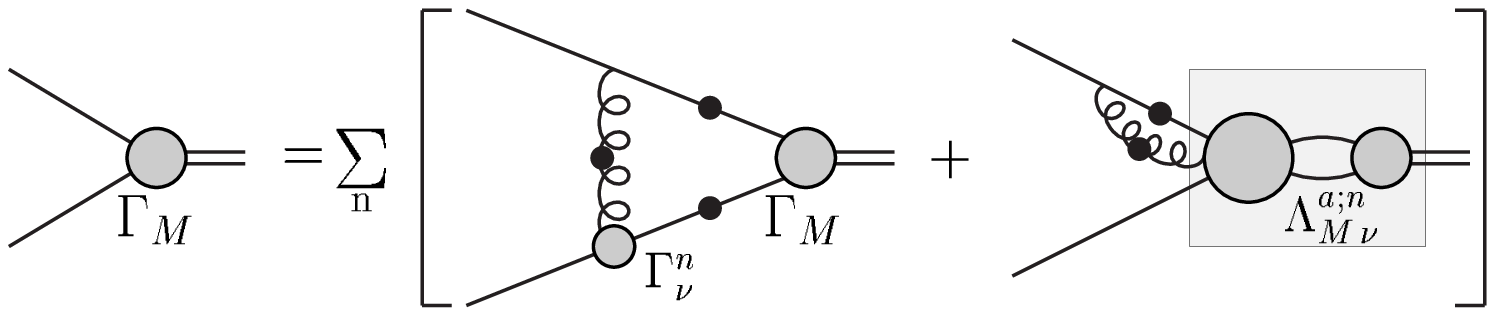}} 
\centerline{\includegraphics[height=7em]{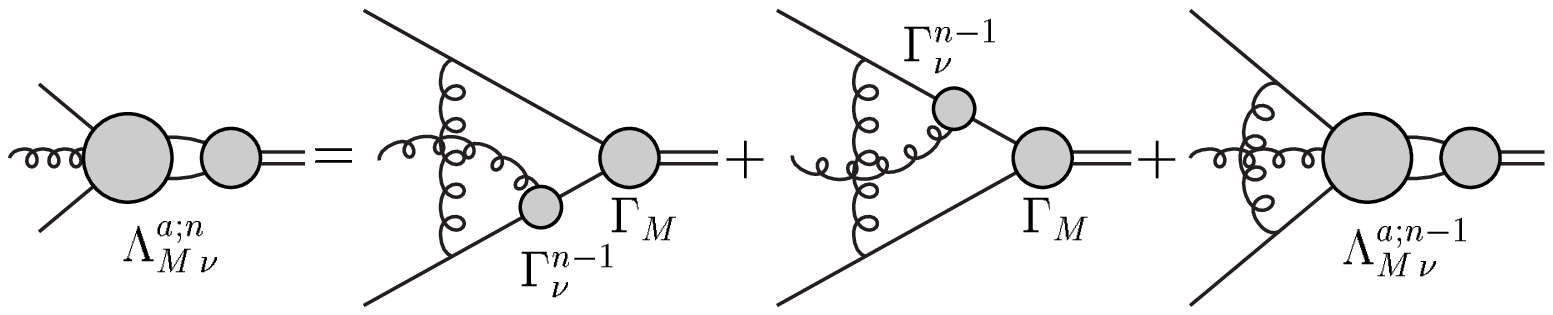}} 
\vspace*{7pt} \fcaption{\label{BSE2} Upper panel: BSE, Eq.\ 
(\protect\ref{genbsenL1}), which is valid whenever $\Gamma_\mu$ can be obtained 
via a recursion relation.  Lower panel: Recursion relation for $\Lambda_{M 
\nu}^{a;n}$, Eq.\ (\protect\ref{Lambdarecursion}). (Adapted from Ref.\ 
[\protect\ref{Rdetmold}].)} 
\end{figure} 
 
Equation~(\ref{genbsenL1}) is depicted in Fig.~\ref{BSE2}.  The first term is 
instantly available once one has an explicit form for $\Gamma^{n}_{\nu}$ and 
the second term, identified by the shaded box in Fig.\ \ref{BSE2}, can be 
obtained via an inhomogeneous recursion relation\cite{detmold} 
\begin{eqnarray} 
\nonumber\lefteqn{\Lambda^{a;n}_{M \nu}(\ell,k;P)}\\ 
\nonumber &= & \int_q^\Lambda\! {\cal 
D}_{\rho\sigma}(\ell-q)\,l^b \gamma_\rho\, \chi_M(q;P)\,  l^a 
\Gamma_\nu^{n-1}(q_-,q_-+k-\ell) 
\, S(q_-+k-\ell)\, l^b \gamma_\sigma\\ 
\nonumber & +&\int_{q}^\Lambda\! {\cal D}_{\rho\sigma}(k -q)\, l^b 
\gamma_\rho\, 
S(q_+ + \ell - k)\, l^a \Gamma_\nu^{n-1}(q_+ + \ell - k,q_+)\, \chi_M(q;P) 
\,l^b \gamma_\sigma \\ 
\nonumber &+ & 
\int_{q^\prime}^\Lambda\! {\cal D}_{\rho\sigma}(\ell -q^\prime) l^b 
\gamma_\rho\, S(q^\prime_+)\, \Lambda^{a;n-1}_{M \nu}(q^\prime,q^\prime 
+k - \ell;P)\, S(q_-^\prime+k-\ell) \, l^b \gamma_\sigma 
\,. \\ 
\label{Lambdarecursion} 
\end{eqnarray} 
This equation is also depicted in Fig.\ \ref{BSE2}.  Combining the two
figures, it is apparent that to form the Bethe-Salpeter kernel the free gluon
line is attached to the upper dressed-quark line.  It follows that the first
term in Eq.\ (\ref{Lambdarecursion}) invariably generates crossed gluon
lines; i.e., nonplanar contributions to the kernel.  The character of the
vertex-consistent Bethe-Salpeter kernel is now clear: it consists of
countably many contributions, a subclass of which are crossed-ladder diagrams
and hence nonplanar.  Only the rainbow gap equation, obtained with $i=0$ in
Eq.\ (\ref{vtxi}), yields a planar vertex-consistent Bethe-Salpeter kernel,
namely the ladder kernel of Eq.\ (\ref{ladder}).  In this case alone is the
number of diagrams in the dressed-vertex and kernel identical.  Otherwise
there are always more terms in the kernel.
 
\subsubsection{Solutions for the $\pi$- and $\rho$-mesons} 
\noindent 
Section 2.2.2 recapitulates on a general procedure that provides the 
vertex-consistent channel-projected Bethe-Salpeter kernel once 
$\Gamma^{n}_{\nu}$ and the propagator functions: $A$, $B$, are known.  That 
kernel must be constructed independently for each channel because, e.g., 
$\Lambda_{M \nu}^{a}$ depends on $\chi_M(q;P)$.  As with the study of the 
vertex in Sec.\ 2.2.1, an elucidation of the resulting BSEs' features is 
simplified by using the model of Eq.\ (\ref{mnmodel}), for then the 
Bethe-Salpeter kernels are finite matrices [cf.\ $(1 - {\cal O})^{-1}$ in 
Eq.\ (\ref{boldalpha})] and the homogeneous BSEs are merely linear, coupled 
algebraic equations. 
 
Reference [\ref{Rdetmold}] describes the solution of the coupled gap and
Bethe-Salpeter equations for the $\pi$- and $\rho$-mesons in detail.  Herein
we focus on the results, which are summarised in Table \ref{masses}.  It is
evident that, irrespective of the order of the truncation; viz., the number
of dressed gluon rungs in the quark-gluon vertex, the pion is massless in the
chiral limit.  This is in spite of the fact that the pion is composed of
heavy dressed-quarks, as is evident in the calculated scale of the
dynamically generated dressed-quark mass function: see Fig.\ \ref{AM0plot},
$M(0) \approx {\cal G} \approx 0.5\,$GeV.  These observations emphasise that
the masslessness of the $\pi$ is a model-independent consequence of
consistency between the Bethe-Salpeter kernel and the kernel in the gap
equation. Furthermore, the bulk of the $\rho$-$\pi$ mass splitting is present
for $m=0$ and with the simplest ($n=0$; i.e., rainbow-ladder) kernel, which
demonstrates that this mass difference is driven by the DCSB mechanism: it is
not the result of a carefully tuned chromo-hyperfine interaction.  Finally,
the quantitative effect of improving on the rainbow-ladder truncation; i.e.,
including more dressed-gluon rungs in the gap equation's kernel and
consistently improving the kernel in the Bethe-Salpeter equation, is a 10\%
correction to the vector meson mass.  Simply including the first correction
(viz., retaining the first two diagrams in Fig.\
\ref{Gamma_inf}) yields a vector meson mass that differs from the fully 
resummed result by $\lsim 1$\%.  The rainbow-ladder truncation is clearly 
accurate in these channels. 
 
\begin{table}[t] 
\tcaption{\label{masses} Calculated $\pi$ and $\rho$ meson masses, in GeV, 
quoted with ${\cal G}= 0.48\,{\rm GeV}$, in which case $m=0.023\, {\cal G} = 
11\,$MeV.  $n$ is the number of dressed-gluon rungs retained in the planar 
vertex, see Fig.~\protect\ref{Gamma_inf}, and hence the order of the 
vertex-consistent Bethe-Salpeter kernel: the rapid convergence of the kernel 
is apparent from the tabulated results.  (Adapted from Ref.\ 
[\protect\ref{Rdetmold}].)\smallskip} 
%
\centerline{\smalllineskip 
\begin{tabular}{l c c c c} 
 & $M_H^{n=0}$ & $M_H^{n=1}$ & $M_H^{n=2}$ & $M_H^{n=\infty}$\\[1ex]\hline 
$\pi$, $m=0$ & 0 & 0 & 0 & 0\\ 
$\pi$, $m=0.011$ & 0.152 & 0.152 & 0.152 & 0.152\\\hline 
$\rho$, $m=0$ & 0.678 & 0.745 & 0.754 & 0.754\\ 
$\rho$, $m=0.011$ & 0.695 & 0.762 & 0.770 & 0.770 \\\hline 
\end{tabular}} 
\end{table} 
 
\subsubsection{Summary} 
\addtocounter{subsubsection}{1} 
\noindent 
It is now evident that a Ward-Takahashi identity preserving Bethe-Salpeter 
kernel can always be calculated explicitly from a dressed-quark-gluon vertex 
whose diagrammatic content is enumerable.\footnote{That is not true if one 
employs an \textit{Ansatz} for the dressed-quark-gluon vertex whose 
diagrammatic content cannot be made explicit; e.g., that of Ref.\ 
[\ref{Rcp90}].}  Furthermore, in all but the simplest case, namely, the 
rainbow-ladder truncation, that kernel is nonplanar. 
 
While we described results obtained with a rudimentary interaction model in 
order to make the construction transparent, the procedure is completely 
general.  However, the algebraic simplicity of the analysis is naturally 
peculiar to the model.  With a more realistic interaction, the gap and vertex 
equations yield a system of twelve coupled integral equations.  The 
Bethe-Salpeter kernel for any given channel then follows as the solution of a 
determined integral equation. 
 
We have reviewed those points in the construction of Refs.\ 
[\ref{Rtruncscheme},\ref{Rdetmold}] that bear upon the fidelity of the 
rainbow-ladder truncation in the gap equation, and in the vector and flavour 
non-singlet pseudoscalar channels.  The error is small. In modelling it is 
therefore justified to fit one's parameters to physical observables at this 
level in these channels and then make predictions for other phenomena involving 
vector and pseudoscalar bound states in the expectation they will be reliable. 
That approach has been successful, as we shall illustrate. 
 
By identifying the rainbow-ladder truncation as the lowest order in a
systematic scheme, the procedure also provides a means of anticipating the
channels in which that truncation must fail.  The scalar mesons are an
example.  Parametrisations of the rainbow-ladder truncation, fitted as
described above, give masses for scalar mesons that are too
high.\cite{jain,conradsep,jacquesscalar,pmdiquark} That was thought to be a
problem.  However, we now know this had to happen because cancellations that
occur between higher order terms in the pseudoscalar and vector channels,
thereby reducing the magnitude of corrections, do not occur in the scalar
channel,\cite{cdrqcII} wherein the full kernel contains additional
attraction.\cite{pmpipi} Indeed, it follows that an interaction model
employed in rainbow-ladder truncation that simultaneously provides a good
description of scalar and pseudoscalar mesons must contain spurious degrees
of freedom.  Quantitative studies of the effect of the higher-order terms
have begun.\cite{pieterpiK,xbox1,xbox2,xbox3}
 
The placement of the rainbow-ladder truncation as the first term in a procedure 
that can systematically be improved explains clearly why this truncation has 
been successful, the boundaries of its success, why it has failed outside these 
boundaries, and why straightening out the failures will not undermine the 
successes. 
 
\subsection{Selected model-independent results} 
\addtocounter{subsection}{1} 
\noindent 
In the hadron spectrum the pion is identified as both a Goldstone mode, 
associated with DCSB, and a bound state composed of constituent $u$- and 
$d$-quarks, whose effective mass is $\sim 350\,$MeV.  Naturally, in quantum 
mechanics, one can fabricate a potential that yields a bound state whose mass 
is much less than the sum of the constituents' masses. However, that requires 
\textit{fine tuning} and, without additional fine tuning, such models predict 
properties for spin- and/or isospin-flip relatives of the pion which conflict
with experiment. A correct resolution of this apparent dichotomy is one of
the fundamental challenges to establishing QCD as the theory underlying
strong interaction physics, and the DSEs provide an ideal framework within
which to achieve that end, as we now explain following the proof of Ref.\
[\ref{Rmrt98}].
 
\subsubsection{A proof of Goldstone's theorem} 
\addtocounter{subsubsection}{1} 
\noindent 
Consider the BSE of Eq.\ (\ref{bsegen}) expressed for the isovector 
pseudoscalar channel: 
\begin{equation} 
\label{genbsepi} 
\left[\Gamma_\pi^j(k;P)\right]_{tu} =  \int^\Lambda_q 
\,[\chi_\pi^j(q;P)]_{sr} \,K^{rs}_{tu}(q,k;P)\,, 
\end{equation} 
with $\chi_\pi^j(q;P)=S(q_+) \Gamma_\pi^j(q;P) S(q_-)$ obvious from Eq.\ 
(\ref{chiM}) and $j$ labelling isospin, of which the solution has the general 
form 
\begin{eqnarray} 
\nonumber 
\Gamma_\pi^j(k;P) & = &  \tau^j \gamma_5 \left[ i E_\pi(k;P) + 
\gamma\cdot P F_\pi(k;P) \rule{0mm}{5mm}\right. \\ 
& & \left. \rule{0mm}{5mm}+ \gamma\cdot k \,k \cdot P\, G_\pi(k;P) + 
\sigma_{\mu\nu}\,k_\mu P_\nu \,H_\pi(k;P) \right]. \label{genpibsa} 
\end{eqnarray} 
It is again apparent that the dressed-quark propagator, the solution of Eq.\ 
(\ref{gendse}), is an important part of the BSE's kernel. 
 
In studying the pion it is crucial to understand chiral symmetry, and its 
explicit and dynamical breaking.  These features are expressed in the 
axial-vector Ward-Takahashi identity, Eq.\ (\ref{avwti}), which involves the 
axial-vector vertex: 
\begin{equation} 
\label{genave} 
\left[\Gamma_{5\mu}^j(k;P)\right]_{tu} = 
Z_2 \, \left[\gamma_5\gamma_\mu \frac{\tau^j}{2}\right]_{tu} \,+ 
\int^\Lambda_q \, [\chi_{5\mu}^j(q;P)]_{sr} \,K^{rs}_{tu}(q,k;P)\,, 
\end{equation} 
that has the 
general form 
\begin{eqnarray} 
\nonumber\Gamma_{5 \mu}^j(k;P) & = & 
\frac{\tau^j}{2} \gamma_5 
\left[ \gamma_\mu F_R(k;P) + \gamma\cdot k k_\mu G_R(k;P) 
- \sigma_{\mu\nu} \,k_\nu\, H_R(k;P) 
\right]\\ 
& & + 
 \tilde\Gamma_{5\mu}^{j}(k;P) 
+ \frac{P_\mu}{P^2 + m_\phi^2} \phi^j(k;P)\,, 
\label{genavv} 
\end{eqnarray} 
where $F_R$, $G_R$, $H_R$ and $\tilde\Gamma_{5\mu}^{i}$ are regular as 
$P^2\to -m_\phi^2$, $P_\mu \tilde\Gamma_{5\mu}^{i}(k;P) \sim {\rm O }(P^2)$ 
and $\phi^j(k;P)$ has the structure depicted in (\ref{genpibsa}).  This form 
admits the possibility of at least one pole term in the axial-vector vertex 
but does not require it. 
 
Substituting (\ref{genavv}) into (\ref{genave}) and equating putative pole 
terms, it is clear that, if present, $\phi^j(k;P)$ satisfies Eq.\ 
(\ref{genbsepi}).  Since this is an eigenvalue problem that only admits a 
$\Gamma_\pi^j \neq 0$ solution for $P^2= -m_\pi^2$, it follows that 
$\phi^j(k;P)$ is nonzero only for $P^2= -m_\pi^2$ and the pole mass is 
$m_\phi^2 = m_\pi^2$.  Hence, if $K$ supports such a bound state, the 
axial-vector vertex contains a pion-pole contribution whose residue, $r_A$, 
is not fixed by these arguments; i.e., Eq.\ (\ref{genavv}) becomes 
\begin{eqnarray} 
\nonumber 
\Gamma_{5 \mu}^j(k;P) & = & 
\frac{\tau^j}{2} \gamma_5 
\left[ \gamma_\mu F_R(k;P) + \gamma\cdot k k_\mu G_R(k;P) 
- \sigma_{\mu\nu} \,k_\nu\, H_R(k;P) \right]\\ 
& & + 
 \tilde\Gamma_{5\mu}^{i}(k;P) 
+ \frac{r_A P_\mu}{P^2 + m_\pi^2} \Gamma_\pi^j(k;P)\,. 
\label{truavv} 
\end{eqnarray} 
 
Consider now the chiral limit axial-vector Ward-Takahashi identity, Eq.\ 
(\ref{avwti}).  If one assumes $m_\pi^2=0$ in Eq.\ (\ref{truavv}), 
substitutes it into the l.h.s.\ of Eq.\ (\ref{avwti}) along with Eq.\ 
(\ref{sinvp}) on the right, and equates terms of order $(P_\nu)^0$ and 
$P_\nu$, one obtains the chiral-limit relations\cite{mrt98} 
\begin{eqnarray} 
\label{bwti} 
r_A E_\pi(k;0)  &= &  B(k^2)\,, \\ 
 F_R(k;0) +  2 \, r_A F_\pi(k;0)                 & = & A(k^2)\,, 
 \label{fwti}\\ 
G_R(k;0) +  2 \,r_A G_\pi(k;0)    & = & 2 A^\prime(k^2)\,,\\ 
\label{gwti} 
H_R(k;0) +  2 \,r_A H_\pi(k;0)    & = & 0\,. 
\end{eqnarray} 
We know $B(k^2) \equiv 0$ in the chiral limit [Eq.\ (\ref{Bpert0})] and that a 
$B(k^2)\neq 0$ solution of Eq.\ (\ref{gendse}) in the chiral limit signals 
DCSB.  Indeed, in this case 
\begin{equation} 
\label{Mchiral} 
M(p^2) \stackrel{{\rm large}-p^2}{=}\, 
\frac{2\pi^2\gamma_m}{3}\,\frac{\left(-\,\langle \bar q q \rangle^0\right)} 
           {p^2 
        \left(\sfrac{1}{2}\ln\left[p^2/\Lambda_{\rm QCD}^2\right] 
        \right)^{1-\gamma_m}}\,, 
\end{equation} 
where $\langle \bar q q \rangle^0$ is the renormalisation-point-independent
vacuum quark condensate.\cite{bankscasher} Furthermore, there is at least one
nonperturbative DSE truncation scheme that preserves the axial-vector
Ward-Takahashi identity, order by order.  Hence
Eqs. (\ref{bwti})-(\ref{gwti}) are exact quark-level Goldberger-Treiman
relations, which state that when chiral symmetry is dynamically
broken:\cite{mrt98}\\[1ex]
\centerline{\parbox{33em}{\flushleft 
\begin{enumerate} 
\item the homogeneous isovector pseudoscalar BSE has a massless, $P^2=0$, 
solution; \vspace*{-1ex} 
\item the Bethe-Salpeter amplitude for the massless bound state has a term 
proportional to $\gamma_5$ alone, with $E_\pi(k;0)$ completely determined by 
the scalar part of the quark self energy, in addition to other pseudoscalar 
Dirac structures, $F_\pi$, $G_\pi$ and $H_\pi$, that are nonzero; 
\vspace*{-1ex} 
\item and the axial-vector vertex is dominated by the pion pole for 
$P^2\simeq 0$. \vspace*{1ex} 
\end{enumerate}}} 
The converse is also true.  Hence DCSB is a sufficient and necessary condition 
for the appearance of a massless pseudoscalar bound state (of what can be 
very-massive constituents) that dominates the axial-vector vertex for $P^2\sim 
0$. 
 
\subsubsection{A mass formula} 
\addtocounter{subsubsection}{1} 
\noindent 
When chiral symmetry is explicitly broken the axial-vector Ward-Takahashi 
identity becomes: 
\begin{equation} 
\label{avwtim} 
P_\mu \Gamma_{5\mu}^j(k;P)  = S^{-1}(k_+) i \gamma_5\frac{\tau^j}{2} 
+  i \gamma_5\frac{\tau^j}{2} S^{-1}(k_-) 
- 2i\,m(\zeta) \,\Gamma_5^j(k;P) , 
\end{equation} 
where the pseudoscalar vertex is given by 
\begin{eqnarray} 
\label{genpve} 
\left[\Gamma_{5}^j(k;P)\right]_{tu} & = & 
Z_4\,\left[\gamma_5 \frac{\tau^j}{2}\right]_{tu} \,+ 
\int^\Lambda_q \, 
\left[ \chi_5^j(q;P)\right]_{sr} 
K^{rs}_{tu}(q,k;P)\,. 
\end{eqnarray} 
As argued in connection with Eq.\ (\ref{genave}), the solution of Eq.\ 
(\ref{genpve}) has the form 
\begin{eqnarray} 
 \nonumber 
i \Gamma_{5 }^j(k;P) & = & 
\frac{\tau^j}{2} \gamma_5 
\left[ i E_R^P(k;P) + \gamma\cdot P \, F_R^P 
+ \gamma\cdot k \,k\cdot P\, G_R^P(k;P) 
\right. \\ 
& & 
\left.+ \, \sigma_{\mu\nu}\,k_\mu P_\nu \,H_R^P(k;P) \right] 
+ \frac{ r_P }{P^2 + m_\pi^2} \Gamma_\pi^j(k;P)\,,\label{genpvv} 
\end{eqnarray} 
where $E_R^P$, $F_R^P$, $G_R^P$ and $H_R^P$ are regular as $P^2\to -m_\pi^2$; 
i.e., the isovector pseudoscalar vertex also receives a contribution from the 
pion pole.  In this case equating pole terms in the Ward-Takahashi identity, 
Eq.\ (\ref{avwtim}), entails\cite{mrt98} 
\begin{equation} 
\label{gmora} 
r_A \,m_\pi^2 = 2\,m(\zeta) \,r_P(\zeta)\,. 
\end{equation} 
This is another exact relation in QCD.  Now it is important to determine the 
residues $r_A$ and $r_P$. 
 
A consideration of the renormalised axial-vector vacuum polarisation 
shows:\cite{mrt98} 
\begin{eqnarray} 
\label{fpiexact} r_A \,\delta^{ij} \,  P_\mu = f_\pi \,\delta^{ij} \,  P_\mu 
&=& Z_2\,{\rm tr} \int^\Lambda_q \sfrac{1}{2} \tau^i \gamma_5\gamma_\mu S(q_+) 
\Gamma_\pi^j(q;P) S(q_-)\,; 
\end{eqnarray} 
i.e., the residue of the pion pole in the axial-vector vertex is the pion decay 
constant.  The factor of $Z_2$ on the r.h.s.\ in Eq.\ (\ref{fpiexact}) is 
crucial: it ensures the result is gauge invariant, and cutoff and 
renormalisation-point independent. Equation (\ref{fpiexact}) is the exact 
expression in quantum field theory for the pseudovector projection of the 
pion's wave function on the origin in configuration space. 
 
A close inspection of Eq.\ (\ref{genpve}), following its re-expression in 
terms of the renormalised, fully-amputated quark-antiquark scattering 
amplitude: $M = K + K (SS) K + \ldots$,  yields\cite{mrt98} 
\begin{equation} 
\label{cpres} i \delta^{ij} \, r_P = Z_4\,{\rm tr} \int^\Lambda_q \sfrac{1}{2} 
\tau^i \gamma_5 S(q_+) \Gamma_\pi^j(q;P) S(q_-)\,, 
\end{equation} 
wherein the dependence of $Z_4$ on the gauge parameter, the regularisation 
mass-scale and the renormalisation point is exactly that required to ensure: 1) 
$r_P$ is finite in the limit $\Lambda\to \infty$; 2) $r_P$ is gauge-parameter 
independent; and 3) the renormalisation point dependence of $r_P$ is just such 
as to guarantee the r.h.s.\ of Eq.\ (\ref{gmora}) is renormalisation point 
\textit{independent}.  Equation (\ref{cpres}) expresses the pseudoscalar 
projection of the pion's wave function on the origin in configuration space. 
 
Let us focus for a moment on the chiral limit behaviour of Eq.\ (\ref{cpres}) 
whereat, using Eqs.\ (\ref{genpibsa}), (\ref{bwti})-(\ref{gwti}), one readily 
finds 
\begin{equation} 
- \langle \bar q q \rangle_\zeta^0 = f_\pi r_P^0(\zeta) 
=  Z_4(\zeta,\Lambda)\, N_c\, {\rm tr}_{\rm D} \int^\Lambda_q  S_{\hat m =0}(q) 
\,. \label{qbq0} 
\end{equation} 
Equation (\ref{qbq0}) is unique as the expression for the chiral limit 
\textit{vacuum quark condensate}.  It is $\zeta$-dependent but independent of 
the gauge parameter and the regularisation mass-scale, and Eq.\ (\ref{qbq0}) 
thus proves that the chiral-limit residue of the pion pole in the pseudoscalar 
vertex is$\,$ $ (-\langle \bar q q \rangle_\zeta^0) /f_\pi$.  Now Eqs.\ 
(\ref{gmora}), (\ref{qbq0}) yield 
\begin{equation} 
\label{gmor} 
(f_\pi^0)^2 \, m_\pi^2 = - 2 \, m(\zeta)\, \langle \bar q q \rangle_\zeta^0 + 
{\rm O}(\hat m^2)\,, 
\end{equation} 
where $f_\pi^0$ is the chiral limit value from Eq.\ (\ref{fpiexact}).  Hence 
what is commonly known as the Gell-Mann--Oakes--Renner relation is a 
\textit{corollary} of Eq.\ (\ref{gmora}). 
 
One can now understand the results in Table \ref{masses}: a massless bound 
state of massive constituents is a necessary consequence of DCSB and will 
emerge in any few-body approach to QCD that employs a systematic truncation 
scheme which preserves the Ward-Takahashi identities. 
 
We stress that Eqs.\ (\ref{gmora})-(\ref{cpres}) are valid for any values of 
the current-quark masses and the generalisation to $N_f$ quark flavours 
is\cite{mr97,marisAdelaide,mishaSVY} 
\begin{equation} 
f_H^2 \, m_H^2 = - \, \langle \bar q q \rangle_\zeta {\cal M}_H^\zeta, 
\end{equation} 
${\cal M}^H_\zeta = m^\zeta_{q_1} + m^\zeta_{q_2}$ is the sum of the 
current-quark masses of the meson's constituents; 
\begin{equation} 
\label{fH} 
f_H \, P_\mu = Z_2 {\rm tr} \int_q^\Lambda \! \sfrac{1}{2} (T^H)^T \gamma_5 
\gamma_\mu {\cal S}(q_+)\, \Gamma^H(q;P)\, {\cal S}(q_-)\,, 
\end{equation} 
with ${\cal S}= {\rm diag}(S_u,S_d,S_s,\ldots)$, $T^H$ a flavour matrix 
specifying the meson's quark content, e.g., $T^{\pi^+}=\sfrac{1}{2} 
(\lambda^1+i\lambda^2)$, $\{\lambda^i\}$ are $N_f$-flavour generalisations of 
the Gell-Mann matrices, and 
\begin{equation} 
\label{qbqH} 
\langle \bar q q \rangle^H_\zeta  =  i f_H\, Z_4\, {\rm tr}\int_q^\Lambda \! 
\sfrac{1}{2}(T^H)^T \gamma_5 {\cal S}(q_+) \,\Gamma^H(q;P)\, {\cal 
S}(q_-) \,. 
\end{equation} 
Owing to its chiral limit behaviour, $\langle \bar q q \rangle^H_\zeta$ has 
been called an in-hadron condensate. 
 
In the heavy-quark limit, Eq.\ (\ref{fH}) yields the model-independent 
result\cite{marisAdelaide,mishaSVY} 
\begin{equation} 
\label{fHheavy} 
f_H \propto \frac{1}{\sqrt{M_H}}\,; 
\end{equation} 
i.e., it reproduces a well-known consequence of heavy-quark 
symmetry.\cite{neubert94} A similar analysis of Eq.\ (\ref{qbqH}) gives a new 
result 
\begin{equation} 
\label{qbqHheavy} 
- \langle \bar q q \rangle^H_\zeta = \mbox{constant} + 
  O\left(\frac{1}{m_H}\right) \mbox{~for~} \frac{1}{m_H} \sim 0\,. 
\end{equation} 
Combining Eqs.\ (\ref{fHheavy}),  (\ref{qbqHheavy}), one 
finds\cite{marisAdelaide,mishaSVY} 
\begin{equation} 
m_H \propto \hat m_f \;\; \mbox{for} \;\; \frac{1}{\hat m_f} \sim 0\,, 
\end{equation} 
where $ \hat m_f$ is the renormalisation-group-invariant current-quark mass of 
the flavour-nonsinglet pseudoscalar meson's heaviest constituent.  This is the 
result one would have anticipated from constituent-quark models but here we 
have reviewed a direct proof in QCD.  Equation (\ref{gmora}) is a single 
formula that unifies aspects of light- and heavy-quark physics and, as we shall 
illustrate, can be used to gain an insightful understanding of modern lattice 
simulations.

%% file: sect3.tex
 
\vspace*{1pt}\textlineskip  
\section{Foundation for a Description of Mesons}    
\addtocounter{section}{1} 
\setcounter{equation}{0} 
\setcounter{figure}{0} 
\setcounter{table}{0} 
\vspace*{-0.5pt} 
\noindent 
The \label{sect3label} renormalisation-group-improved rainbow-ladder
truncation has long been employed to study light mesons and [Sec.\ 2] it can
be a quantitatively reliable tool for vector and flavour nonsinglet
pseudoscalar mesons.  In connection with Eqs.\ (\ref{ladder}),
(\ref{rainbowdse}) we argued that the truncation preserves the ultraviolet
behaviour of the quark-antiquark scattering kernel in QCD but requires an
assumption about that kernel in the infrared; viz., on the domain $Q^2 \lsim
1\,$GeV$^2$, which corresponds to length-scales $\gsim 0.2\,$fm.  The
calculation of this behaviour is a primary challenge in contemporary hadron
physics and there is
progress.\cite{hawes,cdrvienna,tonyfinalgluon,blochmrgluon,fischer,jonivar,raya,langfeld}
However, at present the efficacious approach is to model the kernel in the
infrared, which enables quantitative comparisons with experiments that can be
used to inform theoretical analyses.
 
The most extensively applied model is specified by 
using\cite{mr97,pmspectra2} 
\begin{equation} 
\frac{\alpha(Q^2)}{Q^2} = \frac{4\pi^2}{\omega^6} D\, Q^2 {\rm 
e}^{-Q^2/\omega^2} + \, \frac{ 8\pi^2\, \gamma_m } { \ln\left[\tau + \left(1 + 
Q^2/\Lambda_{\rm QCD}^2\right)^2\right]} \, {\cal F}(Q^2) \,, \label{gk2} 
\end{equation} 
in Eqs.\ (\ref{ladder}), (\ref{rainbowdse}).  Here, ${\cal F}(Q^2)= [1 - 
\exp(-Q^2/[4 m_t^2])]/Q^2$, $m_t$ $=$ $0.5\,$GeV; $\tau={\rm e}^2-1$; $\gamma_m 
= 12/25$; and\cite{pdgold} $\Lambda_{\rm QCD} =
\Lambda^{(4)}_{\overline{\rm MS}}=0.234\,$GeV.\footnote{NB.\ Eq.\ 
(\protect\ref{gk2}) gives $\alpha(m_Z^2)= 0.126$. Comparison with a modern 
value:\protect\cite{pdg} $0.117 \pm 0.002$, means that a smaller $\Lambda_{\rm 
QCD}$ is acceptable in the model, if one wants to avoid overestimating the 
coupling in the ultraviolet, but not a larger value.} \ The true parameters in 
Eq.\ (\ref{gk2}) are $D$ and $\omega$, which together determine the integrated 
infrared strength of the rainbow-ladder kernel; i.e., the so-called interaction 
tension,\cite{cdrvienna} $\sigma^\Delta$. However, we emphasise that they are 
not independent:\cite{pmspectra2} in fitting to a selection of observables, a 
change in one is compensated by altering the other; e.g., on the domain 
$\omega\in[0.3,0.5]\,$GeV, the fitted observables are approximately constant 
along the trajectory\cite{raya} 
\begin{equation} 
\label{omegaD} 
\omega \,D = (0.72\,{\rm GeV})^3. 
\end{equation} 
This correlation: a reduction in $D$ compensating an increase in $\omega$, acts 
to keep a fixed value of the interaction tension.  Equation (\ref{gk2}) is thus 
a one-parameter model. 
 
\subsection{Rainbow gap equation} 
\addtocounter{subsubsection}{1} 
\noindent 
Inserting Eq.\ (\ref{gk2}) into Eq.\ (\ref{rainbowdse}) provides a model for 
QCD's gap equation and in applications to hadron physics one is naturally 
interested in the nonperturbative DCSB solution.  A familiar property of gap 
equations is that they only support such a solution if the interaction tension 
exceeds some critical value.  In the present case that value is\cite{cdrvienna} 
$\sigma_c^\Delta \sim 2.5\,$GeV/fm.  This amount of infrared strength is 
sufficient to generate a nonzero vacuum quark condensate \textit{but only 
just}.  An acceptable description of hadrons requires\cite{mr97} $\sigma^\Delta 
\sim 25\,$GeV/fm and that is obtained with\cite{pmspectra2} 
\begin{equation} 
\label{valD} 
D= (0.96\,{\rm GeV})^2 \,. 
\end{equation} 
 
This value of the model's infrared mass-scale parameter and the two 
current-quark masses 
\begin{equation} 
\label{mqs} 
\begin{array}{cc} 
m_u(1\,{\rm GeV})=5.5\, {\rm MeV} \,,\;&  m_s(1\,{\rm GeV})=125\, {\rm MeV}\,, 
\end{array} 
\end{equation} 
defined using the one-loop expression 
\begin{equation} 
\label{Zmone} 
\frac{m(\zeta)}{m(\zeta^\prime)} = Z_m(\zeta^\prime,\zeta) \approx \left( 
\frac{\ln[\zeta^\prime/\Lambda_{\rm 
QCD}]}{\ln[\zeta/\Lambda_{\rm QCD}]}\right)^\gamma 
\end{equation} 
to evolve $m_u(19\,{\rm GeV})=3.7\,$MeV, $m_s(19\,{\rm GeV})=85\,$MeV, were 
obtained in Ref.\ [\ref{Rpmspectra2}] by requiring a least-squares fit to the 
$\pi$- and $K$-meson observables listed in Table \ref{RGIpiK}.  The procedure 
was straightforward: the rainbow gap equation [Eqs.\ (\ref{renormS}), 
(\ref{rainbowdse}), (\ref{gk2})] was solved with a given parameter set and the 
output used to complete the kernels in the homogeneous ladder BSEs for the 
$\pi$- and $K$-mesons [Eqs.\ (\ref{ladder}), (\ref{rainbowdse}), 
(\ref{genbsepi}), (\ref{genpibsa}) with $\tau^j$ for the $\pi$ channel and 
$\tau^j \to T^{K^+} = \sfrac{1}{2}(\lambda^4+i\lambda^5)$ for the $K$].  These 
BSEs were solved to obtain the $\pi$- and $K$-meson masses, and the 
Bethe-Salpeter amplitudes.  Combining this information gives the leptonic decay 
constants via Eq.\ (\ref{fH}).  This was repeated as necessary to arrive at the 
results in Table \ref{RGIpiK}, which were judged satisfactory. The model gives 
a vacuum quark condensate 
\begin{equation} 
\label{qbq0val} 
-\langle \bar q q \rangle^0_{1\,{\rm GeV}}= (0.242\,{\rm GeV})^3\,, 
\end{equation} 
calculated from Eq.\ (\protect\ref{qbq0}) and evolved using the one-loop 
expression in Eq.\ (\ref{Zmone}). 
 
\begin{table}[t] 
\tcaption{\label{RGIpiK} Comparison of experimental values with results for 
$\pi$ and $K$ observables calculated using the renormalisation-group-improved 
rainbow-ladder interaction specified by Eq.\ (\protect\ref{gk2}), quoted in 
MeV.  The model's sole parameter and the current-quark masses were varied to 
obtain these results.  The best fit parameter values are given in Eqs.\ 
(\protect\ref{valD}), (\protect\ref{mqs}). (Adapted from Ref.\ 
[\protect\ref{Rpmspectra2}].)} \centerline{\smalllineskip 
\begin{tabular}{l | c c c c} 
         & $m_\pi$ & $m_K$ & $f_\pi$ & $f_K$ \\\hline 
Calc.\protect\cite{pmspectra2} & 138 & 497 & 93 & 109 \rule{0ex}{2.7ex}\\ 
Expt.\protect\cite{pdg} & 138 & 496 & 92 & 113\\\hline 
\end{tabular}} 
\end{table} 
 
With the model's single parameter fixed, and the dressed-quark propagator 
obtained, it is straightforward to compose and solve the homogeneous BSE for 
vector mesons and calculate properties analogous to those in Table 
\ref{RGIpiK}.  The predictions\cite{pmspectra2} of the model are reproduced in 
Table \ref{RGIvector}.  The expression in QCD for a vector meson's electroweak 
decay constant is\cite{mishaSVY} 
\begin{equation} 
\label{fV} f_H^V \, M_H^V = \sfrac{1}{3} Z_2 \, {\rm tr} \int_q^\Lambda \! 
(T^H)^T \gamma_\mu {\cal S}(q_+)\, \Gamma_\mu^H(q;P)\, {\cal S}(q_-)\,, 
\end{equation} 
where $M_H^V$ is the meson's mass.  This quantity characterises decays such 
as $\rho\to e^+ e^-$, $\tau \to K^\ast \nu_\tau$, and in the heavy-quark 
limit\cite{mishaSVY} 
\begin{equation} 
f_H^V \propto \frac{1}{\surd M_H^V}\,, \;\; f_H^V \approx f_H\,, 
\end{equation} 
reproducing additional familiar consequences of heavy-quark symmetry. 
 
Given the discussion in Sec.\ 2.2, the phenomenological success of the 
rainbow-ladder kernel, evident in the results in Tables \ref{RGIpiK} and 
\ref{RGIvector}, is unsurprising and, indeed, was to be expected. 
 
\begin{table}[t] 
\tcaption{\label{RGIvector} Experiment cf.\ predictions of a rainbow-ladder 
kernel for simple vector meson observables.  No parameters were varied to
obtain the results.  The root-mean-square error over predicted quantities is
$<8$\%.  NB.\ A charged particle normalisation is used for $f_H^V$ in Eq.\
(\protect\ref{fV}), which differs from that in Eq.\ (\protect\ref{fH}) by a
multiplicative factor of $\surd 2$.  (Adapted from Ref.\
[\protect\ref{Rpmspectra2}].)} \centerline{\smalllineskip
\begin{tabular}{l | c c c c c c } 
         & $m_\rho$ & $m_{K^\ast}$ & $m_{\phi}$ 
         & $f_\rho$ & $f_{K^\ast}$ & $f_\phi$  \\\hline 
Calc.\protect\cite{pmspectra2} & 742 & 936 & 1072 & 207 & 241 & 259 
\rule{0ex}{2.7ex}\\ 
Expt.\protect\cite{pdg}        & 771 & 892 & 1019 & 217 & 227 & 228 \\\hline 
Rel.-Error   & 0.04 & -0.05 & -0.05 & 0.05 & -0.06 & -0.14 \\\hline 
\end{tabular}} 
\end{table} 
 
\subsection{Comparison with lattice simulations} 
\begin{figure}[t] 
\centerline{\includegraphics[height=20em]{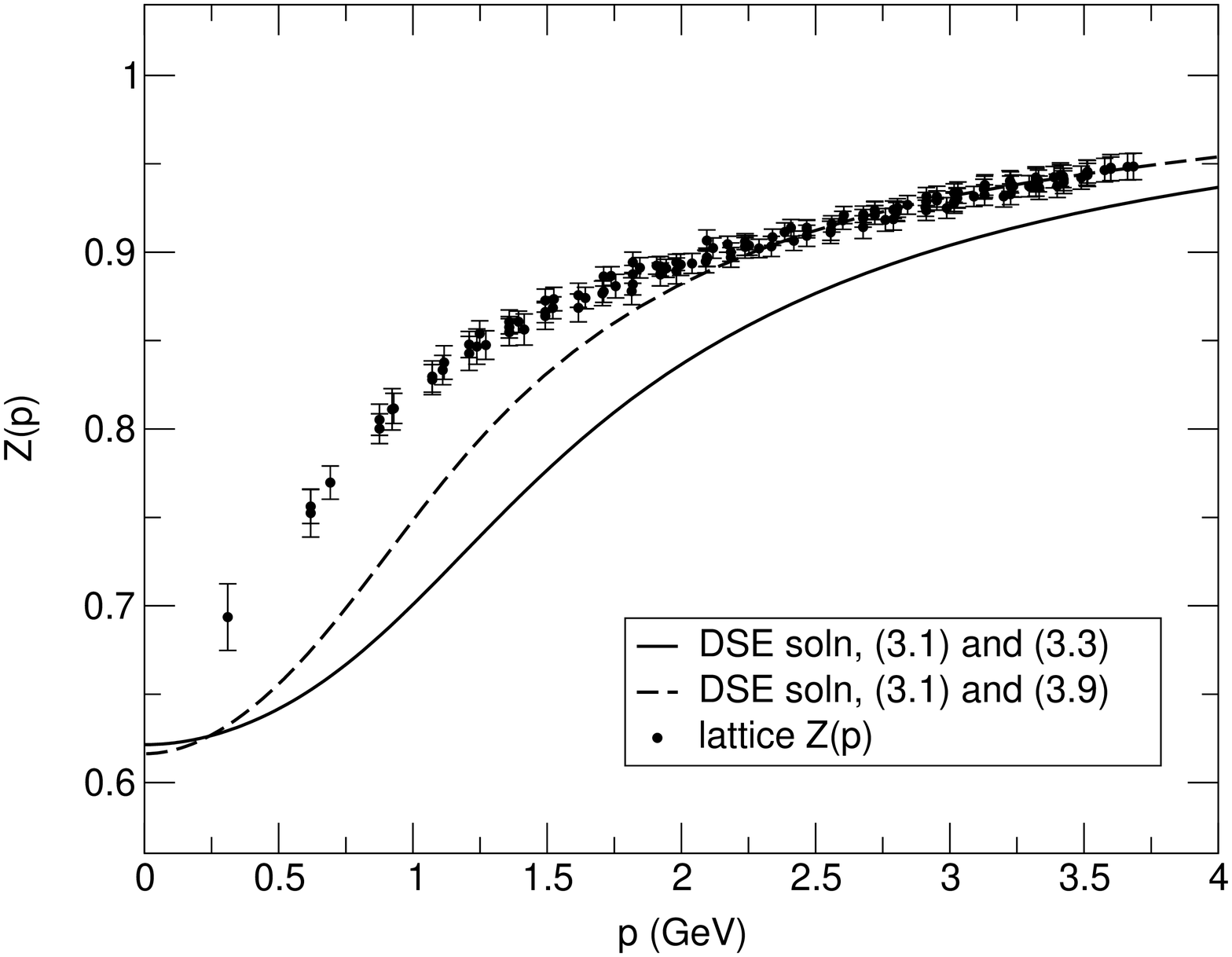}} 
 
 
%
\fcaption{\label{figZ} Wave function renormalisation.  Solid line: solution of 
the gap equation using Eqs.\ (\protect\ref{gk2}), (\protect\ref{valD}); data: 
lattice simulations,\protect\cite{latticequark} obtained with $m=0.036/a\sim 
60\,$MeV; dashed-line: gap equation solution using Eqs.\ (\protect\ref{gk2}), 
(\protect\ref{DomegaL}). The DSE study used a renormalisation point 
$\zeta=19\,$GeV and a current-quark mass $0.6\,m_s^{1\,{\rm GeV}}$ [Eq.\ 
(\protect\ref{mqs})] so as to enable a direct comparison with the lattice data. 
(Adapted from Ref.\ [\protect\ref{Rraya}].)} 
\end{figure}

\begin{figure}[t] 
\centerline{\includegraphics[height=20em]{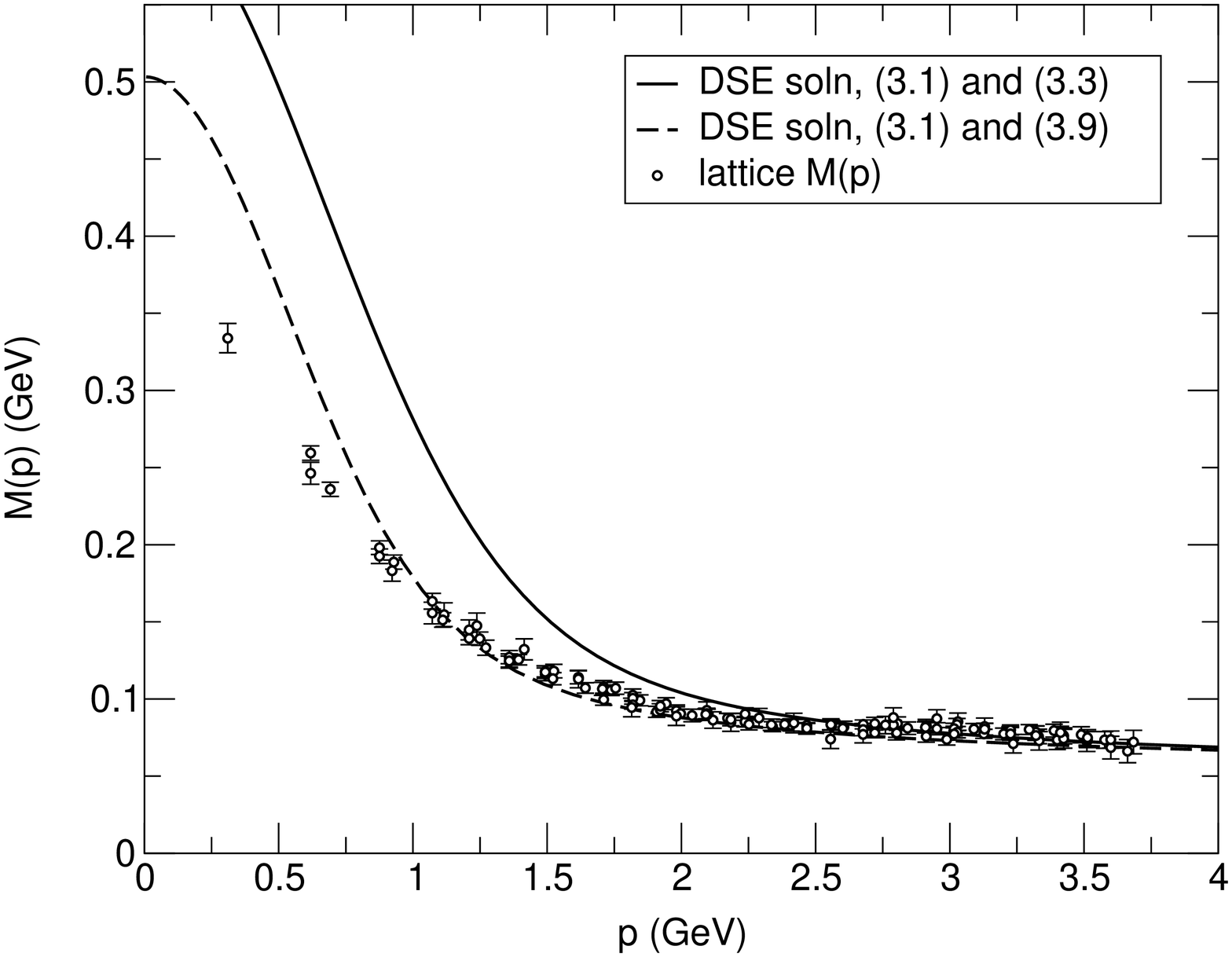}} 
 
 
\fcaption{\label{figM} Mass function.  Solid line: solution of the gap equation 
using Eqs.\ (\protect\ref{gk2}), (\protect\ref{valD}); data: lattice 
simulations,\protect\cite{latticequark} obtained with $m=0.036/a\sim 60\,$MeV; 
dashed-line: gap equation solution using Eqs.\ (\protect\ref{gk2}), 
(\protect\ref{DomegaL}). The DSE study used a renormalisation point 
$\zeta=19\,$GeV and a current-quark mass $0.6\,m_s^{1\,{\rm GeV}}$ [Eq.\ 
(\protect\ref{mqs})].  (Adapted from Ref.\ [\protect\ref{Rraya}].)} 
\end{figure}

\addtocounter{subsubsection}{1} \noindent %
Since the solution of the gap equation has long been of interest in grappling
with DCSB in QCD, in Figs.\ \ref{figZ}, \ref{figM} we depict the scalar
functions characterising the renormalised dres\-sed-quark propagator: the
wave function renormalisation, $Z(p^2)$, and mass function, $M(p^2)$,
obtained by solving Eq.\ (\ref{rainbowdse}) using Eq.\ (\ref{gk2}).  The
infrared suppression of $Z(p^2)$ and enhancement of $M(p^2)$ are longstanding
predictions of DSE studies,\cite{cdragw} which could have been anticipated
from Ref.\ [\ref{Rbjw}].  This prediction has recently been confirmed in
numerical simulations of quenched lattice-QCD, as is evident in the figures.
 
It is not yet possible to reliably determine the behaviour of lattice Schwinger 
functions for current-quark masses that are a realistic approximation to those 
of the $u$- and $d$-quarks.  Therefore a lattice estimate of $m_\pi$, $f_\pi$, 
$\langle \bar q q\rangle^0$ is absent.  To obtain such an estimate, Ref.\ 
[\ref{Rraya}] used the rainbow kernel described herein and varied $(D,\omega)$ 
in order to reproduce the lattice data.  A best fit was obtained with 
\begin{equation} 
\label{DomegaL} 
\begin{array}{cc} 
D=(0.74\,{\rm GeV})^2\,,\; & \omega=0.3\,{\rm GeV}\,, 
\end{array} 
\end{equation} 
at a current-quark mass of $0.6\,m_s^{1\,{\rm GeV}}\!\approx 14\,m_u$ [Eq.\ 
(\protect\ref{mqs})] chosen to coincide with that employed in the lattice 
simulation.  Constructing and solving the homogeneous BSE for a pion-like bound 
state composed of quarks with this current-mass yields 
\begin{equation} 
m_{\pi}^{m_q \sim 14 m_u} = 0.48\,{\rm GeV},\; f_{\pi}^{m_q \sim 14 m_u} = 
0.094\,{\rm GeV}\,. 
\end{equation} 
The parameters in Eq.~(\ref{DomegaL}) give chiral limit results:\cite{raya} 
\begin{equation} 
\label{latticechiral} 
f_\pi^0= 0.068\,{\rm GeV}\,,\; 
-\langle\bar q q\rangle^0_{1 \,{\rm GeV}} = (0.19\,{\rm GeV})^3\,, 
\end{equation} 
whereas Eqs.\ (\ref{omegaD}), (\ref{valD}) give $f_\pi^0=0.088\,$GeV, which
agrees with the estimate of chiral perturbation theory, and the vacuum quark
condensate in Eq.\ (\ref{qbq0val}).  Following this preliminary study it is
important to reanalyse the lattice data using different models for the
infrared behaviour of the scattering kernel, $K$, so that the quenching error
can be estimated.
 
\begin{figure}[t] 
\centerline{\includegraphics[height=20em]{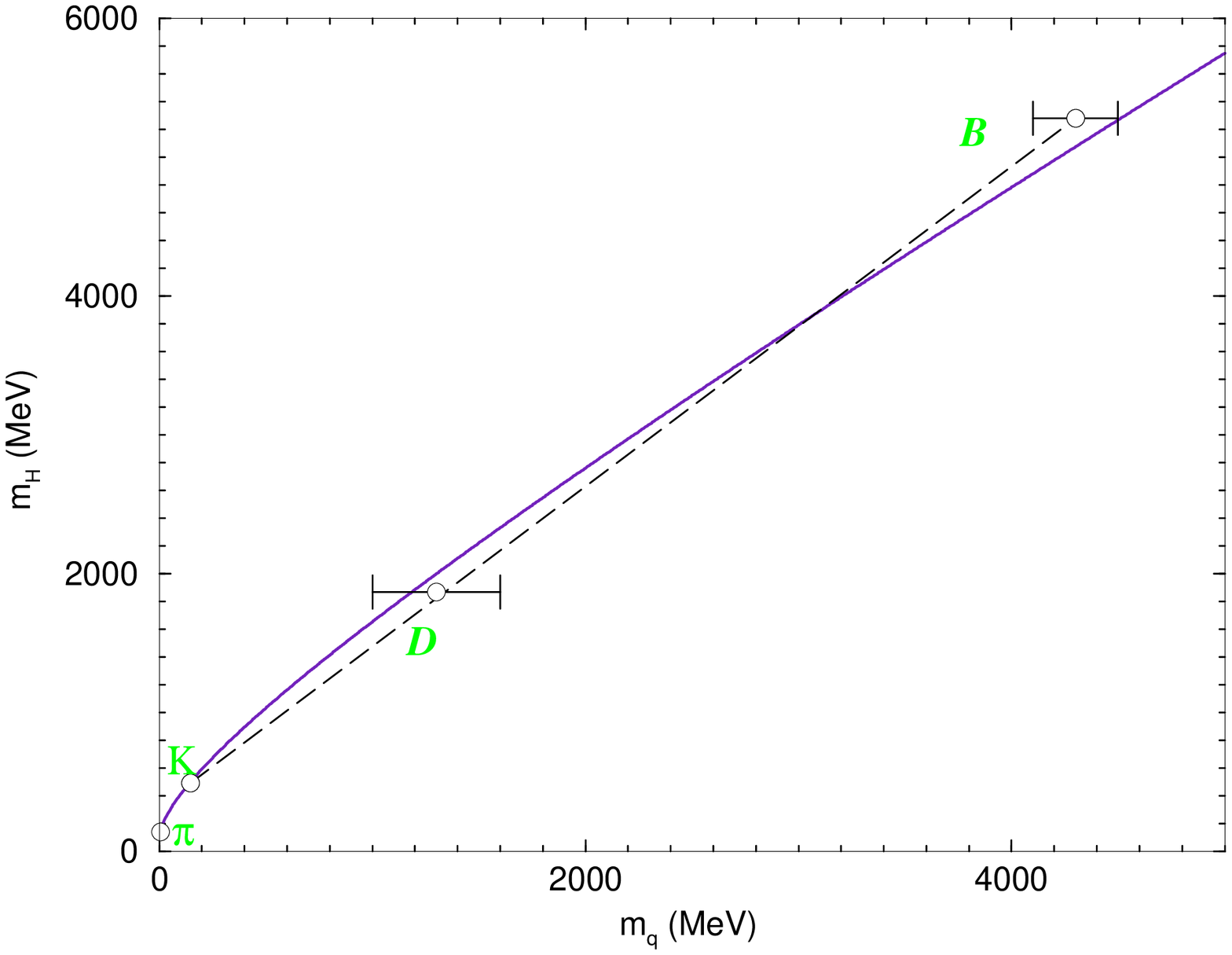}} 
 
\fcaption{\label{figmH} Solid line: pseudoscalar $u\bar q$ meson's mass as a 
function of $m_q^{\zeta}$, $\zeta=19\,{\rm GeV}$, with a fixed value of
$m_u^{\zeta}$ corresponding to $m_u^{1\,{\rm GeV}}=5.5\,$MeV,
Eq.~(\protect\ref{mqs}).  The experimental data points are from
Ref.~[\protect\ref{Rpdg}] as are the errors assigned to the associated
heavy-quark masses.  The dashed curve is a straight line drawn through the
$K$, $D$, $B$ masses.  (Adapted from Ref.\ [\protect\ref{Rcdrlc01}].  See
also Ref.\ [\protect\ref{RTandyErice}].)}
\end{figure}

The unification of light and heavy pseudoscalar meson masses via the mass
formula in Eq.\ (\ref{gmora}) has also been quantitatively explored using the
rainbow-ladder kernel.  That is illustrated in Fig.~\ref{figmH} wherein the
calculated mass of a $u \bar q$ pseudoscalar meson is plotted as a function
of $m_q(\zeta)$, with $m_u(\zeta)$ fixed via Eq.\ (\ref{mqs}).  The
calculations are depicted in the figure by the solid curve, which is, in
MeV,\cite{pmqciv}
\begin{equation} 
\label{pietermH} m_H = 83 + 500 \sqrt{\cal X} + 310\,{\cal X},\; 
{\cal X}= m_q^{\zeta}/\Lambda_{\rm QCD}. 
\end{equation} 
The curvature appears slight in the figure but that is misleading: the
nonlinear term in Eq.\ (\ref{pietermH}) accounts for almost all of $m_\pi$
(the Gell-Mann--Oakes-Renner relation is nearly exact for the pion) and
$80\,$\% of $m_K$.  NB.\ The dashed line in Fig.\ \ref{figmH} fits the $K$,
$D$, $B$ subset of the data exactly.  It is drawn to illustrate how easily
one can be misled.  Without careful calculation one might infer from this
apparent agreement that the large-$m_q$ limit of Eq.\ (\ref{gmora}) is
already manifest at the $s$-quark mass whereas, in reality, the linear term
only becomes dominant for $m_q \gsim 1\,$GeV, providing $50\,$\% of $m_D$ and
$67\,$\% of $m_B$.  The model predicts $m_c^{1\,{\rm GeV}}=1.1\,$GeV,
$m_b^{1\,{\rm GeV}}=4.2\,$GeV, values typical of Poincar\'e covariant
treatments.\cite{mishaSVY}
 
Equation (\ref{pietermH}) can be used as a basis for extrapolating the results 
of lattice simulations to realistic values of the light current-quark masses. 
It is a true expression of essential consequences of DCSB and the importance of 
incorporating such constraints in the analysis of lattice data is beginning to 
be appreciated.\cite{thomaslattice} The rainbow-ladder kernel has also been 
employed in an analysis of a trajectory of fictitious pseudoscalar mesons, all 
composed of equally massive constituents.\cite{pmqciv} (The only physical state 
on this trajectory is the pion.)  The DSE study predicts\cite{cdrqciv} 
\begin{equation} 
\frac{m_{H_{m=2 m_s}}}{m_{H_{m=m_s}}} = 2.2\,, 
\end{equation} 
in agreement with a result of recent quenched lattice
simulations.\cite{michaels} It provides an understanding of this result,
showing that the persistent dominance by the term nonlinear in the
current-quark mass owes itself to a large value of the in-meson condensates
for light-quark mesons; e.g.,\cite{mr97} $\langle \bar q q\rangle^{s\bar
s}_{1\,{\rm GeV}}= (-0.32\,{\rm GeV})^3$, and thereby confirms the
large-magnitude condensate version of chiral perturbation theory. (NB.\ This
last observation is also supported by Eq.\ (\ref{latticechiral}) and the
associated discussion.)  References [\ref{Rmqciv},\ref{RTandyErice}] provide
vector meson trajectories too.
 
\subsection{Pion's valence-quark distribution function} 
\addtocounter{subsubsection}{1} \noindent %
The momentum-dependent dressing of quark and gluon propagators is a fact.  It 
is certainly the keystone of DCSB, materially influences hadron observables and 
quite likely plays a central role in confinement.  This was anticipated in the 
Global Colour Model.\cite{reggcm,petergcm,gunnergcm} As we shall subsequently 
illustrate, the rainbow-ladder kernel is unique today in providing a direct 
description and unification of a wide range of meson phenomena in terms of a 
single parameter that characterises the long-range behaviour of the 
quark-antiquark interaction.  Establishing this is a reward for substantial 
effort.  However, it is a recent development.  Historically, algebraic 
parametrisations of the dressed-quark propagator and Bethe-Salpeter amplitudes, 
based on the known behaviour of numerical solutions, were used to expedite a 
comparison between theory and experiment.  Successes with that 
approach\cite{mishaSVY,conradsep,jacquesscalar,gcmpipi,cdrpion,racdr,thomson,mikeharry,Kl3,hechtmckellar,mikefred,peter98,mikepipi,a1b1} 
provided the impetus for refining the DSE method and its direct numerical 
applications. 
 
The utility of an algebraic form for the dressed-quark propagator and
Bethe-Salpeter amplitudes is self-evident: calculating the simplest elastic
form factor requires the repeated evaluation of a multidimensional integral
whose integrand is a complex-valued function, and a functional of the
propagator and amplitudes.  The expedient remains of use, notably in
connection with baryons, and can be illustrated by reviewing a recent
calculation of the pion's valence-quark distribution function.\cite{uvpi}
 
The cross section for deep inelastic lepton-hadron scattering can be 
interpreted in terms of the momentum-fraction probability distributions of 
quarks and gluons in the hadronic target, and since the pion is a two-body 
bound state with only $u$- and $d$-valence-quarks it is the least complicated 
system for which these distribution functions can be calculated.  However, in 
the absence of pion targets, their measurement is not straightforward and they 
have primarily been inferred from Drell-Yan measurements in pion-nucleus 
collisions.\cite{DYexp2} 
 
The distribution functions provide a measure of the pion's quark-gluon
substructure but they cannot be calculated perturbatively.  Fortunately, DSEs
furnish a sound theoretical description of pion properties, providing a
model-independent explanation of its essentially dichotomous nature as both a
Goldstone boson and a low-mass bound state of massive constituents, and also
supply an efficacious phenomenological tool. Hence they are ideal for
exploring the nature of the pion's parton distribution functions.  The
valence-quark distribution function, $u_v^\pi(x)$, is of particular interest
because its pointwise behaviour is affected by aspects of confinement
dynamics; e.g., by the mechanisms responsible for the finite extent and
essentially nonpointlike nature of the pion.  (NB.\ $u^{\pi^+} = \bar
d^{\pi^+}$ in the ${\cal G}$-parity symmetric limit of QCD.)
 
\subsubsection{Handbag contributions} 
\begin{figure}[t] 
\centerline{\includegraphics[height=17em]{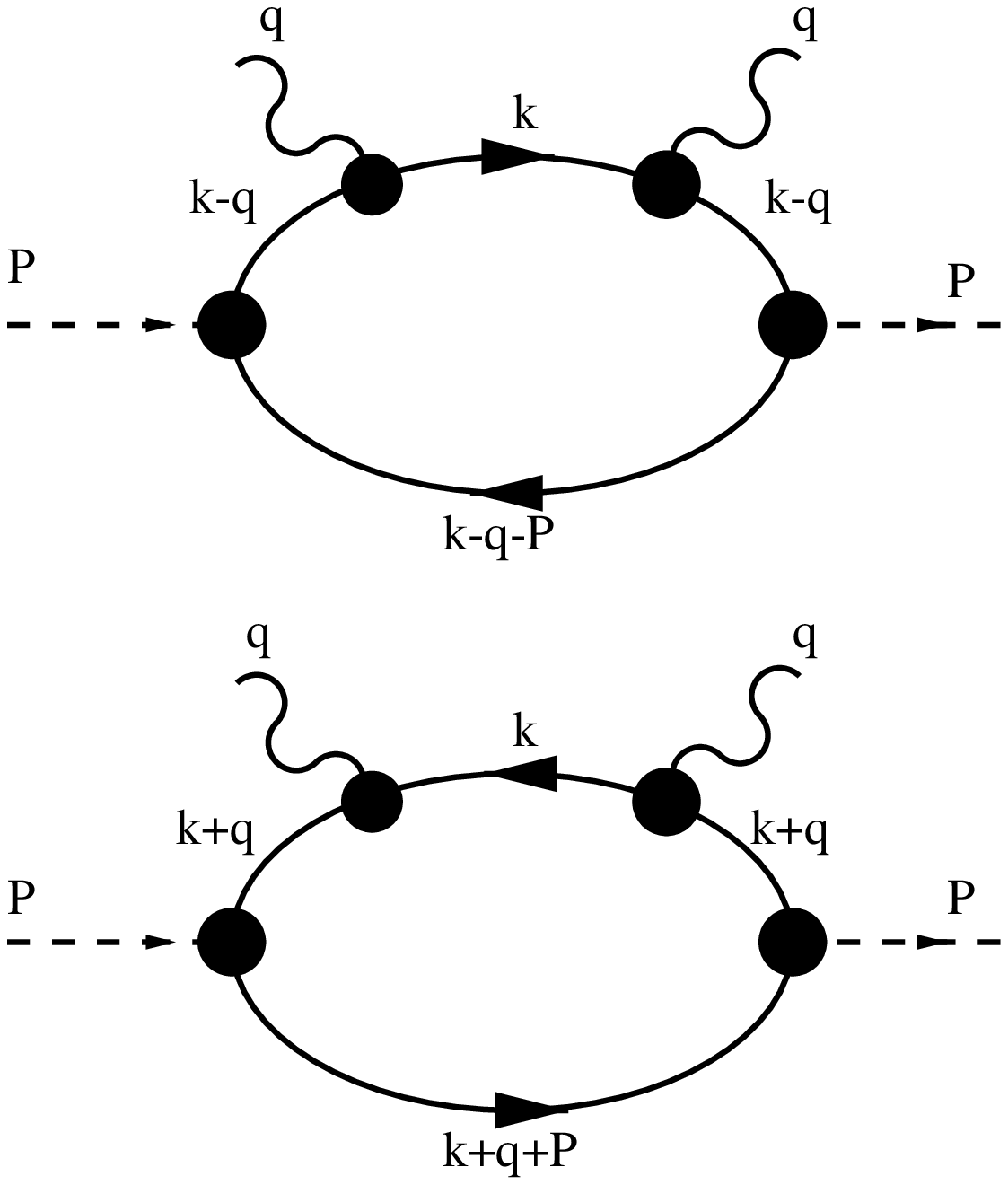}}\vspace*{1ex} 
 
\fcaption{\label{fighandbag} ``Handbag'' contributions to the virtual 
photon-pion forward Compton scattering amplitude.  $\pi$, dashed-line; 
$\gamma$, wavy-line; ${\cal S}$, internal solid-line, dressed-quark 
propagator, Eq.\ (\protect\ref{qprop}).  The filled circles represent the 
pion's Bethe-Salpeter amplitude, $\Gamma_\pi$ in Eqs.\ 
(\protect\ref{genpibsa}), (\protect\ref{Epimodel})-(\protect\ref{bsaG}), and 
the dressed-quark-photon vertex, $\Gamma_\mu$ in Eq.\ (\protect\ref{bcvtx}), 
depending on which external line they begin/end.  (Adapted from Ref.\ 
[\protect\ref{Ruvpi}].)} 
\end{figure}

\addtocounter{subsubsection}{1} 
\noindent 
Reference [\ref{Ruvpi}] focuses on a calculation of the ``handbag diagrams''
illustrated in Fig.\ \ref{fighandbag}, which are the only impulse
approximation contributions to virtual photon-pion forward Compton scattering
that survive in the deep inelastic Bjorken limit:
\begin{equation} 
\label{BJlim} 
q^2\to\infty\,,\; \; P\cdot q \to -\infty \;\;\mbox{but}\;\; x:= 
-\frac{q^2}{2 P\cdot q}\;\; \mbox{fixed.} 
\end{equation} 
The upper diagram represents the renormalised matrix element 
\begin{eqnarray} 
\nonumber \lefteqn{T^+_{\mu\nu}(q,P) 
= {\rm tr}\int^\Lambda_k \tau_-\Gamma_\pi(k_\Gamma;-P)\,}\\ 
&& \times \, S(k_t)\, ieQ\Gamma_\nu(k_t,k) 
\,S(k)\,ieQ\Gamma_\mu(k,k_t)\,S(k_t)\, \tau_+\Gamma_\pi(k_\Gamma;P)\,S(k_s)\,, 
\label{Tmunu} 
\end{eqnarray} 
where: $\tau_{\pm} = \sfrac{1}{2}(\tau_1 \pm i \tau_2)$; ${\cal S}(\ell)=
{\rm diag}[S_u(\ell),S_d(\ell)]$, with $S_u = S_d = S$, assuming isospin
symmetry; and $k_\Gamma= k-q-P/2$, $k_t=k-q$, $k_s=k-q-P$.  A new element in
Eq.\ (\ref{Tmunu}) is $\Gamma_\mu(\ell_1,\ell_2)$, the dressed-quark-photon
vertex, with $Q= {\rm diag}(2/3,-1/3)$ the quark-charge matrix.  It can be
obtained by solving the inhomogeneous vector BSE;\cite{pieterpion} i.e., Eq.\
(\ref{avbse}) with $\gamma_5\gamma_\mu \to \gamma_\mu$, or modelled, based on
symmetry considerations, as we describe below.  The matrix element
represented by the lower diagram is the crossing partner of Eq.\
(\ref{Tmunu}) and is obvious by analogy.
 
The hadronic tensor relevant to inclusive deep inelastic lepton-pion 
scattering can be obtained from the forward Compton process via the optical 
theorem: 
\begin{equation} 
\label{Wmunu} 
W_{\mu\nu}(q;P)= W^+_{\mu\nu}(q;P)+ W^-_{\mu\nu}(q;P)= \frac{1}{2\pi} {\bf\sf 
Im}\left[ T^+_{\mu\nu}(q;P) + T^-_{\mu\nu}(q;P) 
\right]. 
\end{equation} 
In the Bjorken limit one finds\cite{uvpi} 
\begin{equation} 
\label{WF1F2} 
W^+_{\mu\nu}(q;P) = F_1^+(x) \, t_{\mu\nu} + F_2^+(x) \, \frac{q_\mu^t 
q_\nu^t}{2 x} \,, 
\end{equation} 
$t_{\mu\nu} =\delta_{\mu\nu} - q_\mu q_\nu/q^2$, $q_\mu^t = q_\mu + 2 x 
P_\mu$, and 
\begin{equation} 
\label{CGrel} 
F_2^+(x)= 2 x F_1^+(x)\,,\; F_{1,2}^+(x) \to 0\;{\rm as}\;x\to 1\,. 
\end{equation} 
Combining these results with their analogues for $W^-_{\mu\nu}$, one recovers
Bjorken scaling of the deep inelastic cross section, namely, the cross
section depends only on $x$, and not on $P \cdot q$ and $q^2$ separately.
Furthermore, the derivation shows that, in the Bjorken limit, $x$ is truly
the fraction of the pion's momentum carried by the struck quark, and one may
therefore write
\begin{equation} 
F_2^{e\pi}(x) = F_2^+(x) + F_2^-(x) 
= \frac{4}{9} [ x u(x) + x \bar u(x) ] 
+ \frac{1}{9} [ x d(x) + x \bar d(x) ] + \ldots, 
\end{equation} 
where $u(x)$, $\bar u(x)$, etc., are the quark and antiquark distribution 
functions, and the ellipsis denotes contributions from the $s-$ and 
$c$-quarks.  (Heavier quarks are assumed not to contribute at all.) 
 
As we now explain, the calculation in Ref.\ [\ref{Ruvpi}] produces the
valence-quark distributions:
\begin{equation} 
q_v^\pi(x) := q^\pi(x) - \bar q^\pi(x) \,. 
\end{equation} 
It is plain from Eqs.\ (\ref{Tmunu}), (\ref{Wmunu}) that the hadronic tensor
depends on the dressed-quark propagator, pion Bethe-Salpeter amplitude and
dressed-quark-photon vertex, and although sea-quarks are implicitly contained
in these elements, the ``handbag'' impulse approximation diagrams in Fig.\
\ref{fighandbag} only admit a coupling of the photon to the propagator of the
dressed-quark constituent.  The quark's internal structure is not resolved.
The calculation therefore yields the valence-quark distribution at a scale
$q_0$ that is characteristic of the resolution and $\ell_0=1/q_0$ is a
length-scale that typifies the size of the valence quark.  As with all
calculations of this type hitherto, $q_0$ is an {\it a priori} undetermined
parameter, although one anticipates $0.3\lsim q_0\lsim 1.0\,$GeV ($0.7\gsim
\ell_0\gsim 0.2\,$fm), with the lower bound set by the constituent-quark mass
and the upper by the onset of the perturbative domain. A sea-quark
distribution is generated via the evolution equations when the valence
distribution is evolved to that $q^2$-scale appropriate to a given
experiment.  An explicit sea-quark distribution at the scale $q_0$ can arise
from non-impulse diagrams; e.g., contributions that one might identify with
photon couplings to intermediate-state quark-meson-loops that can appear as a
dressing of the quark propagator:
\begin{equation} 
\pi^+ = u\,\bar d \to (u \bar s s) \, \bar d = (K^+ s)\, 
\bar d\to u\,\bar d = \pi^+\,, 
\end{equation} 
with deep inelastic scattering in this instance taking place on the kaon: 
$\gamma K^+ s \to K^+ s\,\gamma$, etc.  Such intermediate states arise as 
vertex corrections in the quark-DSE and were neglected merely to simplify the 
first calculation of $u_v^\pi(x)$, the improvement of which is a modern 
challenge. With these observations in mind, 
\begin{equation} 
\label{F2uvx} 
F_2^+(x;q_0)  =  \sfrac{4}{9}\,x\, u_v^\pi(x;q_0) \,,\; 
F_2^-(x;q_0)  =  \sfrac{1}{9}\,x\, \bar d_v^\pi(x;q_0)\,. 
\end{equation} 
The calculations are required to yield 
\begin{equation} 
\label{uvnorm} 
\int_0^1\,dx\,u_v^\pi(x;q_0) = 1 = \int_0^1\,dx\,\bar d_v^\pi(x;q_0)\,; 
\end{equation} 
viz., to ensure that the $\pi^+$ contains one, and only one, $u$-valence-quark 
and one $\bar d$-valence-quark. 
 
\subsubsection{Algebraic parametrisations} 
\addtocounter{subsubsection}{1} 
\noindent 
To complete the calculation one must specify the elements in the integrand of 
Eq.\ (\ref{Tmunu}).  This is where the algebraic parametrisations appear. The 
dressed-light-quark propagator is 
\begin{eqnarray} 
\label{qprop} 
S(p) & = & -i\gamma\cdot p\, \sigma_V(p^2) + \sigma_S(p^2) 
 =  \label{defS} 
\left[i \gamma\cdot p \, A(p^2) + B(p^2)\right]^{-1}\,,\\ 
\label{ssm} 
\bar\sigma_S(x) & =&  2\,\bar m \,{\cal F}(2 (x+\bar m^2)) 
+ {\cal F}(b_1 x) \,{\cal F}(b_3 x) \, 
\left[b_0 + b_2 {\cal F}(\varepsilon x)\right]\,,\\ 
\label{svm} 
\bar\sigma_V(x) & = & \frac{1}{x+\bar m^2}\, \left[ 1 - {\cal F}(2 (x+\bar 
m^2))\right]\,, 
\end{eqnarray} 
with ${\cal F}(y) = (1-{\rm e}^{-y})/y$, $x=p^2/\lambda^2$, $\bar m$ = 
$m/\lambda$, $\bar\sigma_S(x) = \lambda\,\sigma_S(p^2)$ and $\bar\sigma_V(x) 
= \lambda^2\,\sigma_V(p^2)$.  The mass-scale, $\lambda=0.566\,$GeV, and 
dimensionless parameter values:\footnote{ 
$\varepsilon=10^{-4}$ in Eq.\ (\ref{ssm}) acts only to decouple the large- and 
intermediate-$p^2$ domains.  The study used Landau gauge because it is a 
fixed point of the QCD renormalisation group and $Z_2\approx 1$, even 
nonperturbatively.\protect\cite{mr97}} 
\begin{equation} 
\label{tableA} 
\begin{array}{ccccc} 
   \bar m& b_0 & b_1 & b_2 & b_3 \\\hline 
   0.00897 & 0.131 & 2.90 & 0.603 & 0.185 
\end{array}\;, 
\end{equation} 
were fixed in a least-squares fit to light-meson observables,\cite{thomson} 
and the dimensionless $u$-current-quark mass corresponds to 
\begin{equation} 
m_u({1\,{\rm GeV}}) = 5.1\,{\rm MeV}. 
\end{equation} 
The pointwise form of the dressed-quark wave function renormalisation and mass 
function obtained with this simple parametrisation is qualitatively identical 
to that of the numerical DSE solutions depicted in Figs.\ \ref{figZ}, 
\ref{figM}, and hence is consistent with lattice-QCD simulations.  However, it 
was proposed long before both, and provided the first clear evidence that 
nonperturbative momentum-dependent dressing of parton propagators is 
fundamentally important in QCD and provides a key to understanding hadron 
properties.  The parametrisation represents the propagator as an entire 
function\cite{entire,entireCJB} and thereby exhibits confinement through the 
violation of reflection positivity,\footnote{This sufficient condition for 
confinement is discussed at length in Sec.\ 6.2 of Ref.\ 
[\protect\ref{Rcdragw}], Ref.\ [\protect\ref{Rgastao}], Sec.\ 2.2 of Ref.\ 
[\protect\ref{Rrevbasti}] and Sec.\ 2.4 of Ref.\ [\protect\ref{Rrevreinhard}].} 
\ which means, loosely speaking, that the quark fragments before it can reach a 
detector;\cite{stingl} and manifests DCSB with 
\begin{equation} 
-\langle \bar qq \rangle_{1\,{\rm GeV}^2}  = 
\lambda^3\,\frac{3}{4\pi^2}\, 
\frac{b_0}{b_1\,b_3}\,\ln\frac{1}{\Lambda_{\rm QCD}^2} 
= (0.221\,{\rm GeV})^3\,, 
\end{equation} 
which is calculated directly from Eqs.\ (\ref{qbq0}), (\ref{Zmone}) after 
noting that Eqs.~(\ref{ssm}), (\ref{svm}) yield Eq.~(\ref{Mchiral}) with 
$\gamma_m = 1$. 
 
The general form of the pion's Bethe-Salpeter amplitude is given in Eq.\ 
(\ref{genpibsa}).  The behaviour of the invariant functions therein is largely 
constrained by the axial-vector Ward-Takahashi identity, as we explained in 
connection with Eqs.\ (\ref{bwti})-(\ref{gwti}).  These relations and numerical 
studies\cite{mr97} support the parametrisation\cite{Maris:1998hc} 
\begin{equation} 
\label{Epimodel} 
E_\pi(k;P) = \frac{1}{N_\pi}\,B_\pi(k^2)\,, 
\end{equation} 
where $B_\pi$ is obtained from Eqs.~(\ref{qprop})-(\ref{svm}), evaluated 
using $\bar m=0$ and 
\begin{equation} 
b_0\to b_0^\pi=0.19 
\end{equation} 
with the other parameters unchanged, and: 
\begin{eqnarray} 
\label{bsaF} 
F_\pi(k;P) & = & E_\pi(k;P)/(110 f_\pi)\,; \\ 
\label{bsaG} 
G_\pi(k;P) & = & 2 F_\pi(k;P)/[k^2+M_{\rm UV}^2]\,, 
\end{eqnarray} 
$M_{\rm UV}=10\,\Lambda_{\rm QCD}$; and $H_\pi(k;Q)\equiv 0$.  The amplitude is 
canonically normalised consistent with the impulse approximation: 
\begin{eqnarray} 
\nonumber \lefteqn{ P_\mu = 
\int\!\frac{d^4 k}{(2\pi)^4}\left\{\rule{0mm}{5mm} 
{\rm tr}_{\rm CD} \left[ 
{\Gamma}_\pi(k;-P) \frac{\partial S(k_+)}{\!\!\!\!\!\!\partial P_\mu} 
{\Gamma}_\pi(k;P) S(k_-) \right] 
\right. 
} \\ 
& & \left. 
\;\;\;\;\;\;\;\;\;\;\;\;\;\;\;\;\;\;\; 
 + {\rm tr}_{\rm CD} \left[ 
{\Gamma}_\pi(k;-P) S(k_+) {\Gamma}_\pi(k;P) 
        \frac{\partial S(k_-)}{\!\!\!\!\!\!\partial P_\mu}\right] 
\rule{0mm}{5mm}\right\}.\label{pinorm} 
\end{eqnarray} 
This fixes $N_\pi$.  The decay constant is subsequently obtained from Eq.\ 
(\ref{fpiexact}). 
 
The manner whereby an Abelian gauge boson couples to a dressed-fermion has
been studied extensively and a range of qualitative constraints have been
elu\-ci\-da\-ted.\cite{ayse97} This research supports an {\it
Ansatz}:\cite{bc80}
\begin{equation} 
i\Gamma_\mu(\ell_1,\ell_2)  = 
i\Sigma_A(\ell_1^2,\ell_2^2)\,\gamma_\mu 
+ 
(\ell_1+\ell_2)_\mu\,\left[\sfrac{1}{2}i\gamma\cdot (\ell_1+\ell_2) \, 
\Delta_A(\ell_1^2,\ell_2^2) + \Delta_B(\ell_1^2,\ell_2^2)\right]\,, 
\label{bcvtx} 
\end{equation} 
wherein 
\begin{eqnarray} 
\Sigma_F(\ell_1^2,\ell_2^2)  =  \sfrac{1}{2}\,[F(\ell_1^2)+F(\ell_2^2)]\,,\; 
&& \; 
\Delta_F(\ell_1^2,\ell_2^2)  = 
\frac{F(\ell_1^2)-F(\ell_2^2)}{\ell_1^2-\ell_2^2}\,, 
\end{eqnarray} 
with $F= A, B$ the scalar functions in Eq.\ (\ref{qprop}), which preserves 
many of the constraints, in particular, the vector Ward-Takahashi identity. 
Furthermore, the \textit{Ansatz} is expressed solely in terms of the 
dressed-quark propagator. 
 
Using the elements just described, one obtains the following calculated 
values for an illustrative range of pion observables (adapted from Ref.\ 
[\ref{RMaris:1998hc}]): 
\begin{equation} 
\begin{array}{l | ccc} 
 & m_\pi \, ({\rm GeV}) & f_\pi \, ({\rm GeV}) & r_\pi \, ({\rm fm}) \\\hline 
{\rm Calc.} & 0.139 & 0.090 & 0.56 \rule{0ex}{2.7ex}\\ 
{\rm Expt.}^{\protect\ref{Rpdg},\protect\ref{RAmendolia:1986wj}} 
            & 0.138 & 0.092 & 0.66 
\end{array} 
\end{equation} 
with the pion's charge radius, $r_\pi$, calculated in impulse 
approximation.\cite{cdrpion} 
 
\subsubsection{Calculated distribution function} 
\begin{figure}[t] 
\centerline{\includegraphics[height=17em]{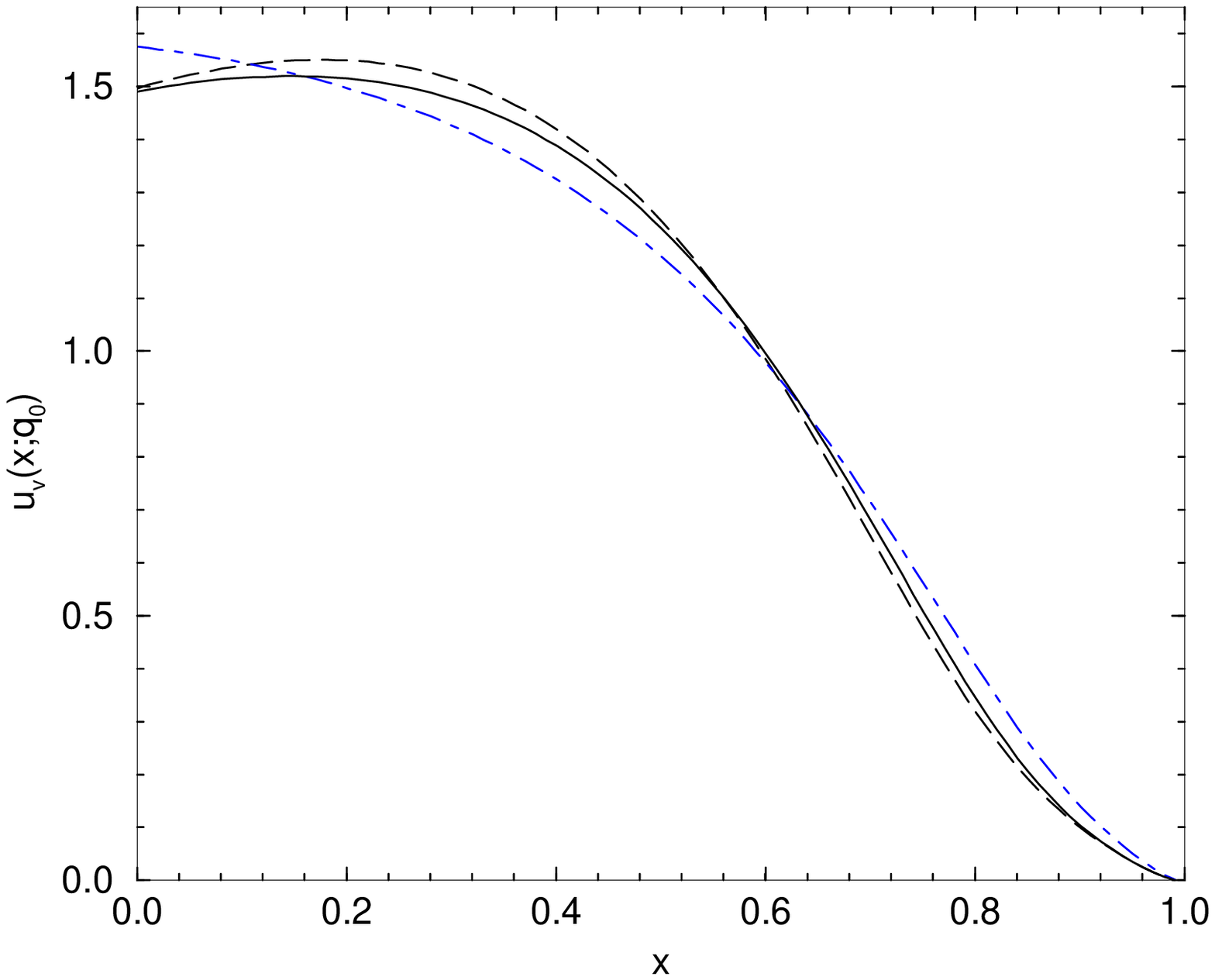}} 
\fcaption{\label{uvxpic} Solid line: $u_v(x;q_0)$ calculated using the DSE 
model described in Sec.\ 3.3.  The resolving scale $q_0=0.54\,{\rm 
GeV}=1/(0.37\,{\rm fm})$ is fixed as described in connection with Eq.\ 
(\protect\ref{q0scale}).  Dashed line: $u_v(x;q_0)$ calculated in the absence 
of the pseudovector components of the pion's Bethe-Salpeter amplitude; i.e., 
$F=0=G$ in Eq.\ (\protect\ref{genpibsa}) instead of Eqs.\ (\protect\ref{bsaF}) 
and (\protect\ref{bsaG}).  Dot-dashed line: distribution calculated with 
$m_\pi\to 0.1\,m_\pi$, $\check M=0.36\,$GeV.  (Adapted from Ref.\ 
[\protect\ref{Ruvpi}].)} 
\end{figure}

\addtocounter{subsubsection}{1} 
\noindent 
In the Bjorken limit, Eq.\ (\ref{Tmunu}) reduces to a one-dimensional integral 
that depends parametrically on the valence-quark mass, $\check M$, and using 
the parametrisations described above that equation yields the valence-quark 
distribution function via Eqs.\ (\ref{Wmunu}), (\ref{WF1F2}), (\ref{F2uvx}). 
The value 
\begin{equation} 
\check M = 0.30\,{\rm GeV} 
\end{equation} 
is fixed by normalisation, Eq.\ (\ref{uvnorm}), and gives the distribution 
function depicted in Fig.\ \ref{uvxpic}.  It vanishes at $x=1$, in accordance 
with the kinematic constraint expressed in Eq.\ (\ref{CGrel}), and corresponds 
to a finite value of $F_1(x=0)$, which signals the absence of sea-quark 
contributions.  Unsurprisingly, since the pion is a light bound state of heavy 
constituents, the shape of the distribution is characteristic of a strongly 
bound system: cf.\ for a weakly bound system\cite{piller} $u_v(x) \approx 
\delta(x-\sfrac{1}{2})$. 
 
\begin{figure}[t] 
\centerline{\includegraphics[height=20em]{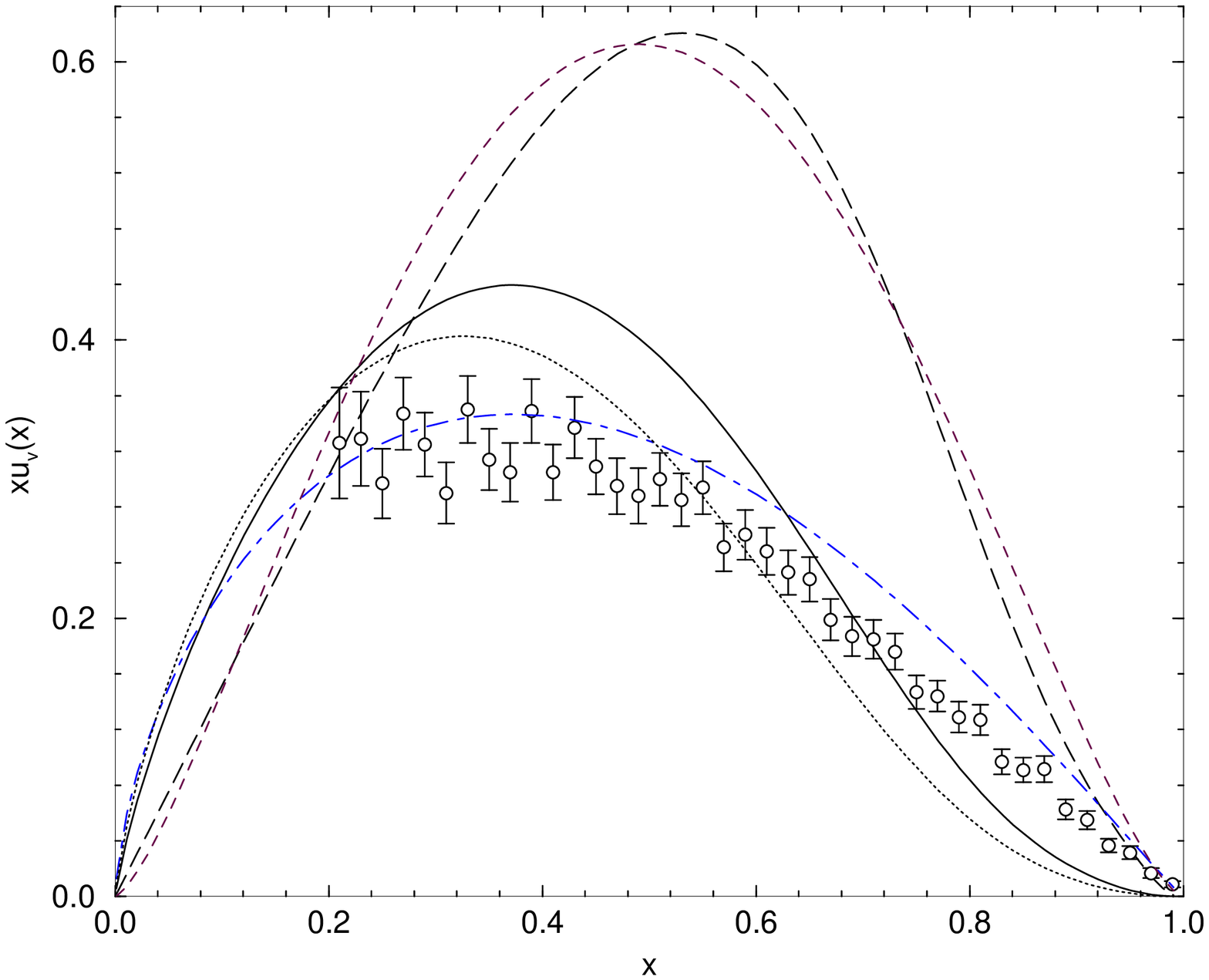}} 
\fcaption{\label{xuvxpic} Dashed line:\protect\cite{uvpi} $x u_v(x;q_0)$; 
short-dashed line: fit of Eqs.\ (\protect\ref{fitxuvx}) and 
(\protect\ref{fitparams}).  Solid line: the evolved distribution, $x 
u_v(x;q=2\,{\rm GeV})$; dotted line: $x u_v(x;q=4.05\,{\rm GeV})$, evolved 
with a $4$-flavour value of $\Lambda_{\rm QCD}= 0.204\,$GeV; and dot-dashed 
line: the phenomenological fit of Ref.\ [\protect\ref{Rsutton}].  The 
Drell-Yan data are from Ref.\ [\protect\ref{RDYexp2}].  (Adapted from Ref.\ 
[\protect\ref{Ruvpi}].)} 
\end{figure}

The momentum-fraction carried by the valence-quarks at this resolving scale is 
\begin{equation} 
\langle x_q\rangle^\pi=\int_0^1\!dx\,x\,[u_v(x;q_0)+\bar d_v(x;q_0)] = 0.71\,, 
\end{equation} 
with the remainder, $\langle x_g\rangle^\pi=0.29$, carried by the gluons that 
bind the pion bound state, which are invisible to the electromagnetic 
probe.\footnote{NB.\ The parametrised pionic parton distributions in Ref.\ 
[\protect\ref{RGRVpi}] yield a gluon momentum-fraction of $\langle 
x_g\rangle^\pi=0.29$ at $q_0= 0.51\,$GeV.} \ As with all calculations of parton 
distributions hitherto, in Ref.\ [\ref{Ruvpi}] the resolving scale 
\begin{equation} 
\label{q0scale} 
q_0 = 0.54\,{\rm GeV} = 1/(0.37\,{\rm fm}) 
\end{equation} 
was chosen so that when using the nonsinglet evolution equations (see, e.g., 
Ref.\ [\ref{Rpdg}]) to evolve the distribution in Fig.\ \ref{uvxpic} up to 
$q=2\,$GeV, agreement was obtained between the first and second moments of the 
calculated distribution and those determined from a phenomenological fit to 
data, in this case the fit of Ref.\ [\ref{Rsutton}], viz. 
\begin{equation} 
\begin{array}{l|lll} 
        &  \langle x_q\rangle_{2\,{\rm GeV}}^\pi 
        &  \langle x_q^2 \rangle_{2\,{\rm GeV}}^\pi 
        &  \langle x_q^3 \rangle_{2\,{\rm GeV}}^\pi  \\\hline 
{\rm Calc.}^{\protect\ref{Ruvpi}}   &  0.24       & 0.098 & 0.049 
\rule{0ex}{2.7ex}    \\ 
{\rm Fit}^{\protect\ref{Rsutton}}& 0.24\pm 0.01    & 0.10\pm 0.01 & 
                0.058 \pm 0.004      \\ 
{\rm Latt.}^{\protect\ref{Rlattice}} 
                & 0.27 \pm 0.01                 & 0.11 \pm 0.03 & 
                0.048 \pm 0.020 
\end{array}\label{moments} 
\end{equation} 
The original and evolved distributions are depicted in Fig.\ \ref{xuvxpic}. 
 
A fit to $x \,u_v(x;q)$, acceptable for the estimation of moments, is obtained 
with 
\begin{equation} 
\label{fitxuvx} x\,u_v^{\rm mom}(x;q) = {\cal A}_{\alpha,\beta}\, x^\alpha \, 
(1-x)^\beta ,\; 
{\cal 
A}_{\alpha,\beta}=\Gamma(1+\alpha+\beta)/[\Gamma(\alpha)\,\Gamma(1+\beta)], 
\end{equation} 
whereupon the moments of $u_v^{\rm mom}(x;q)$ are given by 
\begin{equation} 
\langle x^n \rangle = \prod_{i=1}^n\,\frac{ i+ \alpha 
-1}{i+\alpha+\beta}\,. 
\end{equation} 
The Drell-Yan data\cite{DYexp2} are described by Eq.\ (\ref{fitxuvx}) with $q=2 
\,$GeV values 
\begin{equation} 
\label{abDY} 
\alpha^{\rm DY} = 0.57 \pm 0.03\,,\; \beta^{\rm DY} = 1.27 \pm 0.04\,, 
\end{equation} 
while a global fit to Drell-Yan and prompt photon data yields the consistent 
result\cite{sutton} 
\begin{equation} 
\label{abglobal} 
\alpha^{\rm fit} = 0.64 \pm 0.03\,,\; \beta^{\rm fit} = 1.15 \pm 0.02\,. 
\end{equation} 
 
The calculation of Ref.\ [\ref{Ruvpi}] is described by the values 
\begin{equation} 
\begin{array}{c|lll} 
q \,({\rm GeV}) & 0.57 & 2.0   & 4.05 \\\hline 
\alpha                  & 1.34 & 0.92 &  0.84    \\ 
\beta                   & 1.31 & 1.80  & 1.98 
\end{array}\\ 
\label{fitparams} 
\end{equation} 
A material feature of this DSE result is the value of $\beta \simeq 2$ because 
although perturbative QCD cannot be used to obtain the pointwise dependence of 
the distribution function it does predict\cite{uvxpQCD} the power-law 
dependence at $x\simeq 1$: 
\begin{equation} 
\label{pQCD} 
\mbox{pQCD:}\;\;\; u_v^\pi(x) \stackrel{x\sim 1}{\propto} (1-x)^2\,, 
\end{equation} 
in agreement with the DSE result (corrections are logarithmic).  However, as
is apparent in Fig.\ \ref{xuvxpic} and emphasised by Eqs.\ (\ref{abDY}),
(\ref{abglobal}), this prediction \textit{disagrees} with extant experimental
data.\footnote{This is in spite of the fact that the first four moments
agree: Eq.\ (\protect\ref{moments}).  Plainly, the low moments contain little
information about the distribution on the valence-quark
domain. Lattice-simulations are currently limited to the low moments.} \ That
is very disturbing because a verification of the experimental result would
present a profound threat to QCD, even challenging the assumed
vector-exchange nature of the force underlying the strong interaction.
 
\begin{figure}[t] 
\centerline{\includegraphics[height=21em]{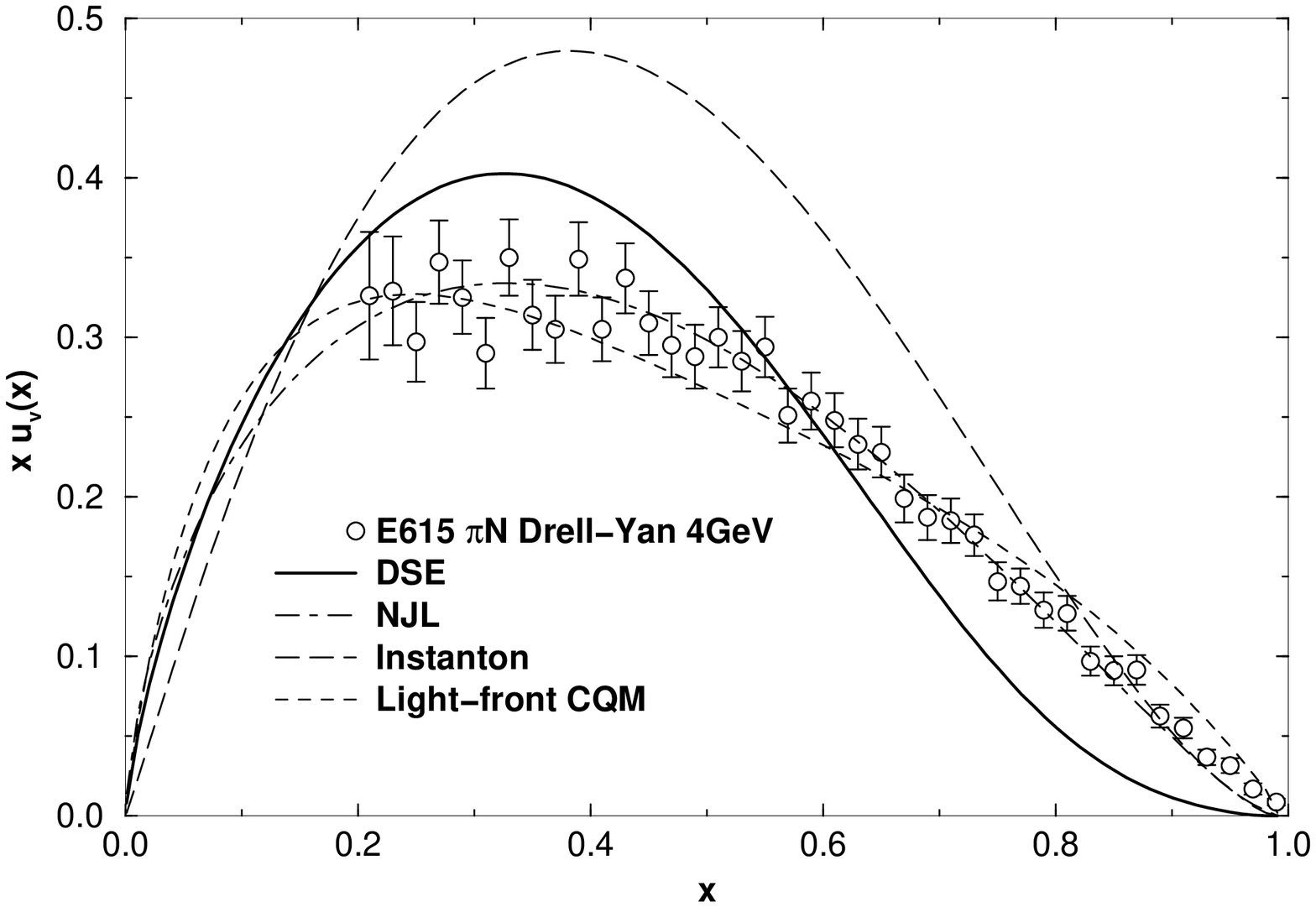}} 
\fcaption{\label{xuvxpicall} Fuller comparison of theory with experiment. 
Solid line: DSE result;\protect\cite{uvpi} dot-dashed line: NJL
model;\protect\cite{njluvx} long-dashed line: instanton
model;\protect\cite{dorokhov} short-dashed line: light-front constituent
quark model.\protect\cite{miller} All calculations have been evolved to
$q=4.0\,$GeV using a $4$-flavour value of $\Lambda_{\rm QCD}= 0.204\,$GeV.
The Drell-Yan data are from Ref.\ [\protect\ref{RDYexp2}].  (Adapted from
Ref.\ [\protect\ref{Rkrishni}].)}
\end{figure} 
In Fig.\ \ref{xuvxpicall} we illustrate that the only extant calculation which 
agrees with the distribution inferred from $\pi N$ Drell-Yan data is that 
performed with a particular regularisation of the Nambu--Jona-Lasinio 
model.\cite{njluvx} That calculation yields a distribution 
\begin{equation} 
u_v^{\rm pt}(x;q_0^{\rm NJL}=0.35\,{\rm GeV}) = \theta(x) \, \theta(1-x) 
\end{equation} 
which corresponds to the valence-quark carrying each and every fraction of
the pion's momentum with equal probability.  The result is an artefact tied
to this model's representation of the pion bound state by a momentum-{\it
independent} Bethe-Salpeter amplitude;\cite{uvpi,cudell} i.e., representing
the pion as a point-particle, which is a necessary consequence of the model's
momentum-independent interaction.  It is clearly the hardest distribution
that is physically possible.
 
\begin{figure}[t] 
\centerline{\includegraphics[height=9em]{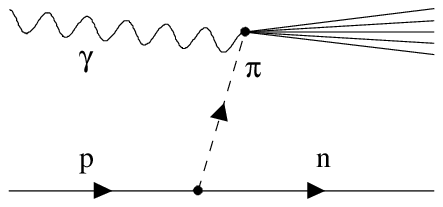}} 
\fcaption{\label{figsullivan} Deep inelastic scattering from the proton's 
$\pi$-cloud (Sullivan process) could provide a means of measuring $u_v^\pi(x)$ 
at JLab. (Adapted from Ref.\ [\protect\ref{Rkrishni}].)} 
\end{figure}

This observation underscores the serious nature of the discrepancy described 
above in connection with Eq.\ (\ref{pQCD}), and in highlighting that 
disagreement the DSE study\cite{uvpi} has catalysed interest in $u^\pi_v(x)$ 
and proposals for its remeasurement.  One proposal that could use existing 
facilities would employ the process depicted in Fig.\ \ref{figsullivan} at 
JLab.\cite{krishni}  This process could also be used efficaciously at a future 
electron-proton collider to accurately probe $u_v^\pi(x)$ on the valence-quark 
domain.\cite{royEIC} 
 
\subsubsection{Summary} 
\addtocounter{subsubsection}{1} 
\noindent 
This application illustrates the power of using algebraic parametrisations of 
key DSE elements: one can proceed rapidly to an insightful analysis of hadronic 
phenomena and arrive at robust conclusions.  Now, however, this type of 
analysis is supported, improved and in some cases superseded by a direct 
application of the one-parameter rainbow-ladder model specified by Eq.\ 
(\ref{gk2}).   

%% file: sect4.tex
 
\vspace*{1pt}\textlineskip  
\section{{\bf\sl Ab Initio} Calculation of Meson Properties}
\addtocounter{section}{1} 
\setcounter{equation}{0} 
\setcounter{figure}{0} 
\setcounter{table}{0} 
\vspace*{-0.5pt} 
\noindent 
In this section we review the systematic application of the 
\label{sect3pmlabel} renormalisation-group-improved rainbow-ladder kernel 
defined by Eq.\ (\ref{gk2}) to a wide range of meson observables.  We
emphasise at the outset that each result is a prediction, in the sense that
the model's mass-scale is fixed, Eq.\ (\ref{valD}), and every element in each
calculation is completely determined by, and calculated from, that kernel.
 
\subsection{Elastic electromagnetic form factors} 
\addtocounter{subsection}{1} 
\noindent 
We first consider processes that involve only three external particles, at 
least one of which is a hadron.  The simplest of these are the pion and kaon 
electromagnetic form factors.  The matrix element describing the coupling of 
a photon with momentum $q$ to a pseudoscalar meson composed of quark $a$ with 
electric charge $e_a$ and antiquark $\bar{b}$, electric charge $e_{\bar b}$, 
can be written as the sum of two terms 
\begin{equation} 
\Lambda^{a\bar{b}}_\nu(P,q) = 
                e_a \, \Lambda^{a\bar{b}a}_\nu(P,q) + e_{\bar b} \, 
                \Lambda^{a\bar{b}\bar{b}}_\nu(P,q) \,, 
\label{mesonff} 
\end{equation} 
which, respectively, describe the coupling of the photon to the quark and the 
antiquark.  In the Breit frame, where the incoming meson has total momentum 
$P-q/2$, 
\begin{equation} 
\label{2PF} 
\Lambda^{a\bar{b}}_\nu(P,q) = 2\, P_\nu \, F(q^2) \,, 
\end{equation} 
where $F(q^2)$ is the meson's elastic electromagnetic form factor.  ($r^2 = -6 
F^\prime(0)$ is the square of the meson's charge radius.)  Each term on the 
r.h.s.\ of Eq.\ (\ref{mesonff}) has this property and hence one can also write 
\begin{eqnarray} 
\Lambda^{a\bar{b}}_\nu(P,q) &=& 2 \, P_\nu \left[ e_a \, F^a(q^2) + e_{\bar b} 
\, F^{\bar b}(q^2) \right], \label{FaFb} 
\end{eqnarray} 
and thereby explicate the contribution of each quark to the total elastic form 
factor. 
 
The matrix element describing the scattering process is most easily calculated 
using the renormalised impulse approximation,\cite{cdrpion} in which, e.g., 
\begin{equation} 
\Lambda^{a\bar{b}\bar{b}}_\nu(P,q) = 
        N_c \, {\rm tr}_{\rm D}\int^\Lambda_k \!S^a(\ell) \, 
        \Gamma^{a\bar{b}}(r_+;P_-) \, S^b(\ell_+) i 
        \Gamma^{b}_\nu(\ell_+,\ell_-)\, S^b(\ell_-) \, 
        \Gamma^{b\bar{a}}(r_-; - P_+)  , 
\label{triangle} 
\end{equation} 
where: $\ell = k+P/2$, $\ell_\pm = k-P/2 \pm q/2$, $r_\pm = k\pm q/4$, $P_\pm = 
P\pm q/2$; and, similar to what we saw with Eq.\ (\ref{Tmunu}), $S^{a,b}$ is a 
dressed-quark propagator, $\Gamma^{a\bar{b}}$ is the meson's Bethe-Salpeter 
amplitude, and $\Gamma_\mu^{a,b}$ is a dressed-quark-photon vertex. 
 
In this section the dressed-quark propagators are obtained by solving the 
mass-dependent rainbow gap equations, Eq.\ (\ref{rainbowdse}), and the 
Bethe-Salpeter amplitudes come from Eqs.\ (\ref{ladder}), (\ref{genbsepi}). 
Furthermore, the dressed-quark-photon vertex is calculated by solving a 
renormalised inhomogeneous BSE, viz.\ 
\begin{equation} 
\Gamma^a_\mu(k_+,k_-) = Z_2\, \gamma_\mu + 
    \int^\Lambda_\ell K(k,\ell;q)\, S^a(\ell_+) \, 
    \Gamma^a_\mu(\ell_+,\ell_-) \, S^a(\ell_-) \,, 
\label{verBSE} 
\end{equation} 
wherein $k_\pm = k \pm q/2$, $\ell_\pm= \ell\pm q/2$, with the kernel again 
given by Eq.\ (\ref{ladder}).  At this point it is important to remember that 
because the rainbow-ladder truncation is the first term in a systematic DSE 
truncation scheme [Sec.\ 2.2] the vertex thus obtained satisfies the 
Ward-Takahashi identity 
\begin{equation} 
i\,q_\mu \,\Gamma^a_{\mu}(k_+,k_-) = S_a^{-1}(k_+) - S_a^{-1}(k_-) \,. 
\label{wtid} 
\end{equation} 
(NB.\ This is only true if a Poincar\'e covariant regularisation scheme is 
employed.) 
 
\subsubsection{Current conservation} 
\addtocounter{subsubsection}{1} 
\noindent 
It is electromagnetic current conservation that reduces to one the number of 
form factors on the r.h.s.\ of Eq.\ (\ref{2PF}) and it also requires $F^a(0) = 
1 = F^{\bar b}(0)$.  As with the realisation of Goldstone's theorem, these 
consequences of symmetry are exhibited by the electromagnetic matrix elements 
if, and only if, there is an intimate relation between each of the elements in 
the calculation.  The impulse approximation, calculated with propagators and 
vertices obtained using the rainbow-ladder truncation, preserves these 
constraints without fine-tuning;\cite{cdrpion} i.e., independent of the 
detailed form of the interaction, Eq.\ (\ref{gk2}). 
 
\begin{figure}[t] 
\centerline{\includegraphics[width=29em]{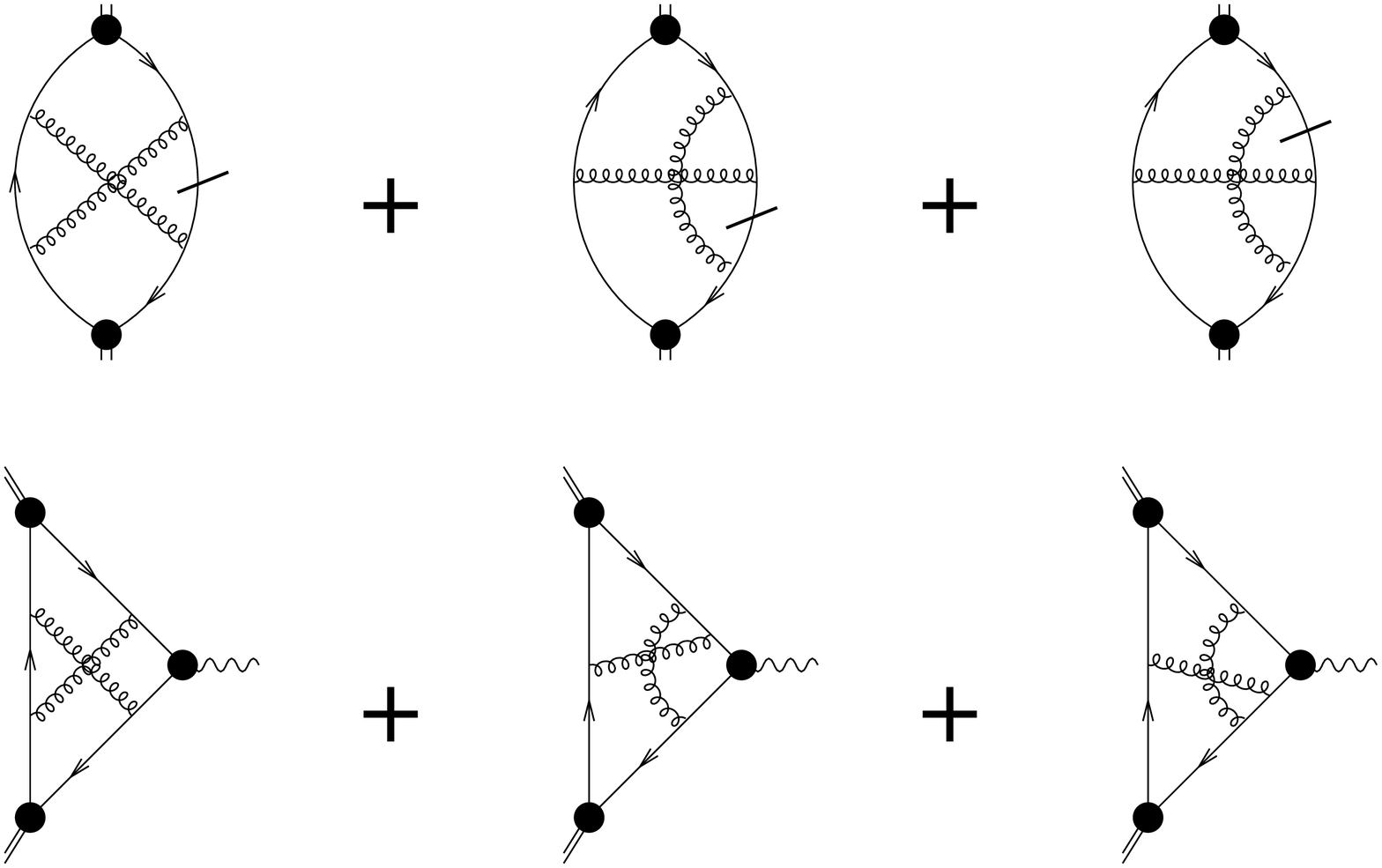}} 
\fcaption{\label{fig:beyond} Top row: The three additional contributions to the 
normalisation of the Bethe-Salpeter amplitude, Eq.\ (\protect\ref{pinorm}), 
generated when the second term in Eq.\ (\protect\ref{vtxexpand}) is included in 
the gap equation's kernel.  (Without loss of generality, for this illustration 
the integration variable is chosen such that the total momentum, $P$, flows 
only through the antiquark line, upon which the slash represents $\partial 
S/\partial P_\mu$.) Bottom row: The corrections to the impulse approximation, 
Eq.\ (\protect\ref{triangle}), necessary to maintain current conservation. 
(Adapted from Ref.\ [\ref{RpieterpiK}].)} 
\end{figure}

Suppose now that one adds the second term of Eq.\ (\ref{vtxexpand}) to the 
kernel of the gap equation.  Following the systematic procedure reviewed in 
Sec.\ 2.2, that term generates three additional contributions to the kernel of 
the BSE and also introduces a dependence on the bound state's total momentum. 
That in turn gives rise to a modification of the canonical normalisation 
condition for the Bethe-Salpeter amplitude; viz.,\cite{llewellyn} the integrals 
represented by the three diagrams in the top row of Fig.\ \ref{fig:beyond} must 
be added to the r.h.s.\ of Eq.\ (\ref{pinorm}). 
The dressed-quark-photon vertex calculated from Eq.\ (\ref{verBSE}) with the
modified kernel automatically satisfies the Ward-Takahashi identity, Eq.\
(\ref{wtid}), and current conservation is guaranteed without fine-tuning
provided one augments the impulse approximation by the three diagrams
depicted in the bottom row of Fig.\ \ref{fig:beyond}.\cite{pieterpiK} That
these diagrams are necessary and sufficient is obvious once one realises
that, according to the differential Ward identity: $i\Gamma_{\mu}(k,k)=
\partial S(k)/\partial k_\mu$, the derivative of a quark line is equivalent
to the insertion of a zero momentum photon. The generalisation of this
procedure is straightforward and it is therefore apparent that for any given
gap equation kernel it is systematically possible to construct the
current-conserving matrix element that describes the coupling of the photon
to a composite meson.
 
\subsubsection{Pion and kaon electromagnetic form factors} 
\addtocounter{subsubsection}{1} 
\noindent 
To be concrete we review a calculation of the $\pi$ and $K$ form 
factors:\cite{pieterpiK} 
\begin{eqnarray} 
\label{fpi} 
 F_{\pi}(q^2) &=& \frac{2}{3}F_\pi^{u}(q^2) 
        + \frac{1}{3}F_\pi^{\bar{d}}(q^2)  \,,\\ 
\label{fKplus} 
 F_{K^+}(q^2) &=& \frac{2}{3}F_{K^+}^{u}(q^2) 
        + \frac{1}{3}F_{K^+}^{\bar s}(q^2)            \,,\\ 
\label{fK0} 
 F_{K^0}(Q^2) &=& -\frac{1}{3}F_{K^0}^{d}(q^2) 
        + \frac{1}{3}F_{K^0}^{\bar s}(q^2)            \,. 
\end{eqnarray} 
So far as the strong interaction is concerned $u$ and $d$ quarks are identical 
in the isospin symmetric limit and hence there are only three independent form 
factors, $F_\pi^{u}(q^2)=F_\pi^{\bar{d}}(q^2)$, 
$F_{K^+}^{u}(q^2)=F_{K^0}^{d}(q^2)$, and $F_{K^+}^{\bar s}(q^2)=F_{K^0}^{\bar 
s}(q^2)$, which in impulse approximation are given by Eqs.\ (\ref{FaFb}), 
(\ref{triangle}). 
 
These form factors were calculated in Ref.\ [\ref{RpieterpiK}] as an
\textit{ab initio} and parameter-free application of the model defined by
Eqs.\ (\ref{gk2}), (\ref{valD}).  Therein the dressed-quark propagators were
calculated and used in constructing the kernel of the BSE; the BSE was solved
to determine the mesons' bound state amplitudes \textit{completely}, i.e.,
all the amplitudes in Eq.\ (\ref{genpibsa}) were calculated and shown to play
an important role; and then the elements were combined to yield a manifestly
Poincar\'e covariant prediction of meson electromagnetic form factors.

\begin{figure}[t] 
\centerline{\includegraphics[width=4in]{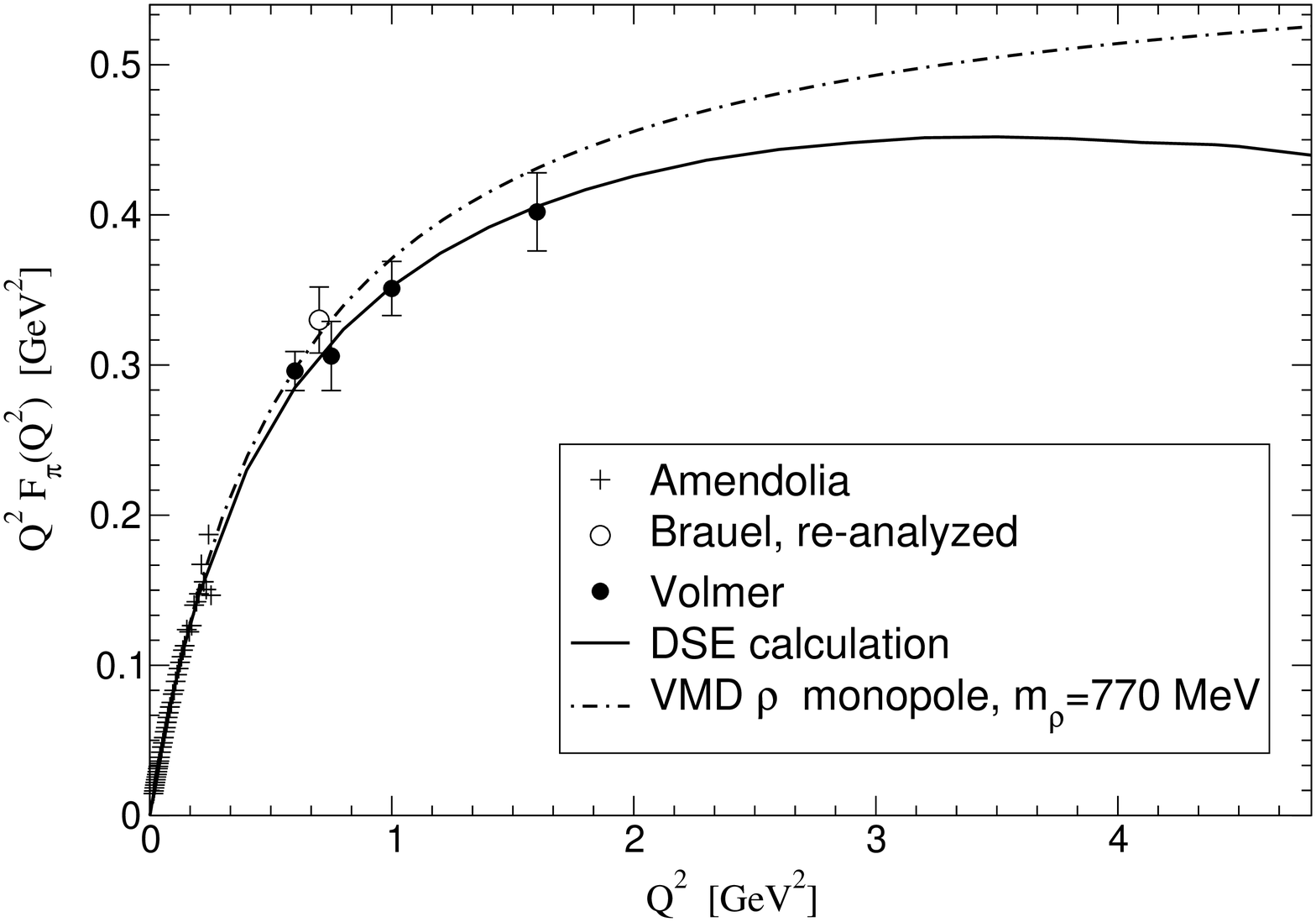}} 
\fcaption{\label{fig:fpi} Impulse approximation DSE prediction for $q^2 
F_\pi(q^2)$ obtained in an \textit{ab initio}, parameter-free application of 
the renormalisation-group-improved rainbow-ladder truncation, Eqs.\ 
(\protect\ref{gk2}), (\protect\ref{valD}). The data are from Refs.\ 
[\protect\ref{RAmendolia:1986wj},\protect\ref{RBrauel:1979zk},\protect\ref{RVolmer:2000ek}] 
(Adapted from Ref.\ [\protect\ref{RMaris:2001am}].)} 
\end{figure}

The result for the pion form factor is depicted in Fig.\
\ref{fig:fpi},\footnote{The nature and meaning of vector dominance is discussed 
in Sec.\ 2.3.1 of Ref.\ [\protect\ref{Rcdrpion}], Sec.\ 2.3 of Ref.\
[\protect\ref{Rrevbasti}] and Sec.\ 4.3 below: the low-$q^2$ behaviour of the
pion form factor is necessarily dominated by the lowest mass resonance in the
$J^{PC}=1^{--}$ channel.  Any realistic calculation will predict that and
also a deviation from dominance by the $\rho$-meson pole alone as
spacelike-$q^2$ increases.} \ \ wherein it is compared with the most recent
experimental data.\cite{Volmer:2000ek} The pion's calculated charge radius is
compared with experiment in Table \ref{tab:radii}.  The excellent agreement
is misleading because $\pi$-$\pi$ final-state interactions add $\lsim 15$\%
to the impulse approximation result for $r_\pi$.\cite{alkoferpiloop} (NB.\
This rescattering contribution to the form factor vanishes with increasing
$q^2$.)  However, that only makes the result more plausible by leading to a
disagreement with experiment commensurate with that one would anticipate
based on the discussion in Sec.\ 2.2.
 
It is evident in Table \ref{tab:radii} that the calculated charged kaon form
factor, Eq.\ (\ref{fKplus}), agrees with available
data,\cite{Amendolia:1986ui} which unfortunately is not a stringent test of
theory because it covers only a small low-$q^2$ domain and has large
errors. The $q^2$-dependence of the form factor again deviates from a simple
monopole on $q^2\gsim 2\,$GeV$^2$.
 
The neutral kaon form factor, Eq.\ (\ref{fK0}), as the difference between two
terms, reacts most to a model's details.  Of course, $F_{K^0}(q^2=0)=0$ as a
direct result of current conservation and Ref.\ [\ref{RpieterpiK}] is
guaranteed to reproduce that.  However, the evolution away from $q^2=0$ is
sensitive to differences between the dressed $d$- and $s$-quark mass
functions.  (This is obvious because if they were the same then
$F_{K^0}\equiv 0$, just as $F_{\pi^0}\equiv 0$.)  The calculation yields
$r_{K^0}^2<0$, Table \ref{tab:radii}, which is readily understood: the
heavier and positively charged $\bar s$-quark in the $K^0$ is more often to
be found at the meson's core so that the lighter $d$-quark provides an excess
of negative charge at the surface.  The result is therefore qualitatively
reliable.  However, as the squared-charge radius is small,\cite{Molzon:py}
$K^0$-$\bar K^0$ final state interactions, neglected in Ref.\
[\ref{RpieterpiK}], can have more of an impact on this observable. These
effects disappear with increasing $q^2$ but data are difficult to obtain for
$q^2\neq 0$.
 
It is a model independent prediction\cite{Maris:1998hc} of the DSE framework
reviewed herein that the elastic electromagnetic meson form factors display
\begin{equation} 
\label{q2Fq2} q^2 F(q^2) = {\rm constant},\; q^2 \gg \Lambda_{\rm QCD}^2, 
\end{equation} 
with calculable $\ln^d q^2/\Lambda_{\rm QCD}^2$ corrections, where $d$ is an
anomalous di\-men\-sion.\footnote{Corrections to the rainbow-ladder-impulse
approximation, such as those depicted in Fig.\ \protect\ref{fig:fpi},
contribute to the anomalous dimension but do not modify the power law
dependence.} \ %
This agrees with earlier perturbative QCD
analyses.\cite{Farrar:1979aw,Lepage:1980fj} However, to obtain this result in 
covariant gauges it is crucial to retain the pseudovector components of the
Bethe-Salpeter amplitude in Eq.\ (\ref{genpibsa}): $F_\pi$, $G_\pi$.  (NB.\
The quark-level Goldberger-Treiman relations, Eqs. (\ref{fwti}),
(\ref{gwti}), prove them to be nonzero.)  Without these
amplitudes,\cite{cdrpion} $q^2 F(q^2)
\propto 1/q^2$. Similar statements are true of the role played by nonleading 
components in the Bethe-Salpeter amplitudes of vector mesons.  The
calculation of Ref.\ [\ref{RMaris:1998hc}], which uses the propagator and
vertex parametrisations described in Sec.\ 3.3, suggests that the
perturbative behaviour of Eq.\ (\ref{q2Fq2}) is unambiguously evident for
$q^2\gsim 15\,$GeV$^2$.  Owing to challenges in the numerical analysis, the
\textit{ab initio} calculations of Ref.\ [\ref{RpieterpiK}] cannot yet make a
prediction for the onset of the perturbative domain but progress in remedying
that is being made.\cite{pichowskypoles}
 
\begin{table}[t] 
\tcaption{\label{tab:radii} Comparison of 
calculated\protect\cite{pieterpion,pieterpiK} squared-charge-radii (in ${\rm 
fm}^2$) with data.\protect\cite{Amendolia:1986wj,Amendolia:1986ui,Molzon:py} 
The $\gamma^*\pi^0\gamma$ interaction radius is also 
included.\protect\cite{Maris:2002mz,Behrend:1990sr} (Adapted from Ref.\ 
[\ref{RpieterpiK}].)\smallskip} 
\centerline{\smalllineskip 
\begin{tabular}{l|lll|l} 
    & $r^2_\pi$     & $r^2_{K^+}$   & $r^2_{K^0}$ 
    & $r^2_{\pi\gamma\gamma}$           \\[1ex] \hline 
Calc.   &  0.45 &  0.38    & $-0.086$ & 0.41                  \\ 
Expt.   & $0.44\pm0.01$ & $0.34\pm0.05$ & $-0.054\pm0.026$ 
    & $0.42\pm0.04$                 \\ \hline 
Rel.-Error 
    &  0.023    &  0.118    &  0.59     & 0.024 \\ \hline 
\end{tabular}} 
\end{table} 
 
\subsection{$K_{l3}$ decays} 
\addtocounter{subsection}{1} 
\noindent 
Related to the elastic form factors are the semileptonic transition form 
factors describing $K^+\to \pi^0 \ell \nu_\ell$ [$K_{\ell 3}^+$], $K^0\to \pi^- 
\ell \nu_\ell$ [$K_{\ell 3}^0$] and $\pi^+\to \pi^0 e \nu_e$ [$\pi_{e 3}$]. 
They proceed via the flavour-changing vector piece of the $V-A$ electroweak
interaction, in particular $j_\mu^{su}$ and $j_\mu^{du}$. The axial-vector
component does not contribute because the two mesons involved have the same
parity. Neither $j_\mu^{su}$ nor $j_\mu^{du}$ is conserved and the symmetry
breaking term measures the dressed-quark mass difference [see Eq.\
(\ref{Gsuwti})].  These processes can therefore be employed to probe flavour
symmetry violation,\cite{Kl3,Isgur:hu} and this more effectively than the
neutral kaon form factor.
 
\begin{figure}[t] 
\centerline{\includegraphics[width=28em]{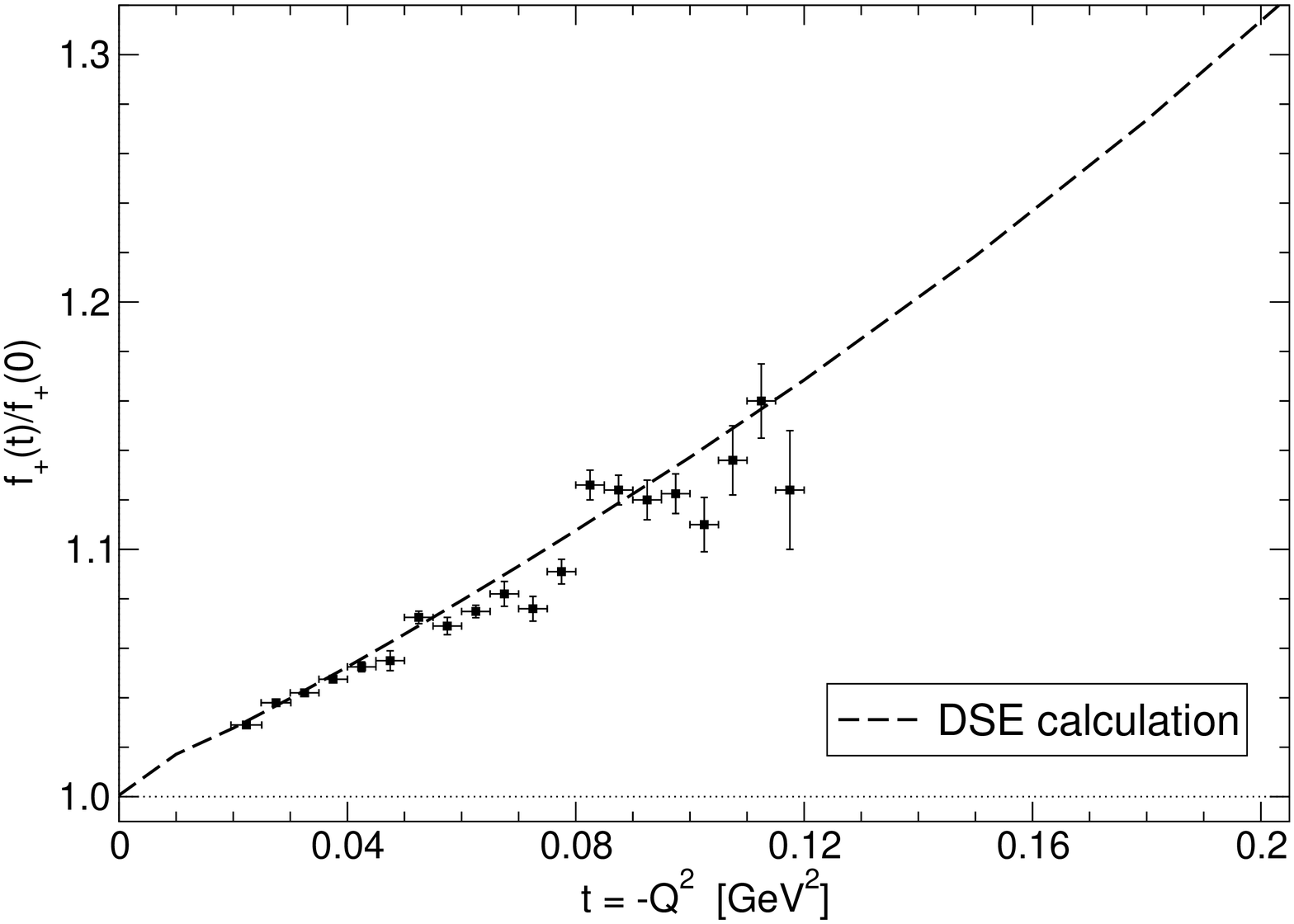} } 
\fcaption{\label{fig:ewff} 
DSE results for the $K_{l3}$ form factor $f_+(t=-q^2)$.  Experimental data
from CPLEAR.\protect\cite{Apostolakis:1999gs} (Adapted from
Ref.~[\protect\ref{RJi:2001pj}].)}
\end{figure}

The matrix element for the $K_{\ell 3}^0$ transition is characterised by two 
form factors: 
\begin{equation} 
\Lambda_{\mu}^{dsu}(p_K,q,-p_\pi)= \langle\pi^-(p_\pi)|\bar{s}\gamma_\mu 
u|K^0(p_K)\rangle = 2 \, P_\mu \,f_+^{K^0}(q^2) + q_\mu\, f^{K^0}_-(q^2)\,, 
\label{eq:fkl3} 
\end{equation} 
where $p_\pi= P-q/2$, $p_K= (P+q/2)$ and $q$ is the $W$-boson's momentum,
with $t=-q^2$.  In the isospin symmetric case considered in Refs.\
[\ref{RKl3},\ref{RJi:2001pj}], $m_u=m_d$ and $j_\mu^{du}$ is conserved so
that $f_+^{\pi}(t)=-F_\pi(t)$, $f_-^{\pi}\equiv 0$.  In addition,
$f_\pm^{K^0}\equiv f_\pm^{K^+}$ and hence all new information is contained in
the two form factors of Eq.\ (\ref{eq:fkl3}).  Their calculation in impulse
approximation involves only one element not already used in Sec.\ 4.1.2,
namely, the dressed-$suW$-vertex, which replaces the
dressed-quark-photon vertex.  That piece which contributes to $K_{\ell 3}$
decays is obtained from
\begin{equation} 
 \Gamma^{s u}_\mu(k_+,k_-) = Z_2 \, \gamma_\mu +
\int^\Lambda_\ell K(k,\ell;q) \,S_{s}(\ell_+) \, 
\Gamma^{su}_\mu(\ell_+,\ell_-) \, S_{u}(\ell_-)\, , 
\end{equation} 
where again, for a consistent truncation, the renormalisation-group-improved
ladder kernel must be used.  The vertex thus calculated satisfies
\begin{equation} 
\label{Gsuwti} 
i q_\mu \, \Gamma^{s u}_\mu(k_+,k_-) = S_{s}^{-1}(k_+) - S_{u}^{-1}(k_-) - 
(m_s - m_u)\, \Gamma^{su}_{\mbox{\footnotesize\boldmath $1$}}(k_+,k_-)\,, 
\end{equation} 
where $\Gamma^{su}_{\mbox{\footnotesize\boldmath $1$}}$ is a 
flavour-dependent scalar vertex analogous to the pseudoscalar vertex in Eq.\ 
(\ref{genpve}).  The parameter-free prediction\cite{Ji:2001pj} for $f_+^{K}$ 
is depicted in Fig.\ \ref{fig:ewff}. 
 
On the physical domain for $K_{\ell 3}$ decays: $m_\ell^2 < t < (m_K-m_\pi)^2 
= 0.13\;{\rm GeV}^2$, the calculated form factors\cite{Ji:2001pj} are well 
approximated by a straight line: 
\begin{equation} 
f(t) = f(0) \left[ 1 + \frac{\lambda}{m_\pi^2} \, t \right]\,, \; \lambda := 
- m_\pi^2 f'(m_\ell^2)/f(0) \,, 
\end{equation} 
supporting an assumption common in the analysis of experimental
data.\cite{pdg} 

The last term in Eq.\ (\ref{Gsuwti}) indicates the manner in which $K_{\ell
3}$ decays are sensitive to the current-quark-mass difference and its
enhancement through nonperturbative effects.\cite{Kl3} Experimentally this is
best explored via the scalar form factor
\begin{equation} 
 f_0(q^2) = f_+(q^2) - \frac{q^2}{m_K^2 - m_\pi^2} f_-(q^2) \,, 
\end{equation} 
which measures the divergence $ q_\mu \Lambda_{\mu}^{dsu}(p_K,q,-p_\pi)$.
Current algebra predicts the value of $f_0^K$ at the Callan-Treiman
point,\cite{CT66} $t=m_K^2-m_\pi^2=: \Delta$, $f_0^K(\Delta)= -
f_K/f_\pi=-1.23$, while a systematic analysis of corrections
yields:\cite{GL85} $f_0^K(\Delta)= -1.18$.  The Callan-Treiman point is not
experimentally accessible but the robust nature of its derivation makes the
value of $f_0(\Delta)$ a tight constraint on any theoretical framework.
 
In Table \ref{tab:weak} we display a representative sample of the results 
calculated in Ref.\ [\ref{RJi:2001pj}].  We note that $|f_+(0)|\approx 1$ and 
that is consistent with the Ademollo-Gatto theorem,\cite{AG64} which states 
that flavour symmetry breaking effects are suppressed at $t=0$.  Furthermore, 
it was observed in Ref.\ [\ref{RIsgur:hu}] and confirmed in Ref.\ 
[\ref{RKl3}], that $f_-(0)$ is a measure of the $s$:$u$ constituent-quark 
mass ratio.  Applying this notion to Ref.\ [\ref{RJi:2001pj}], in which the 
calculated value of that ratio is $\sim 1.25$, one estimates $f_-(0)\approx 
-0.15$, from Fig.\ 1 of Ref.\ [\ref{RIsgur:hu}].  The calculated value is 
actually $f_-(0)=-0.10$.  These observations again emphasise the power of a 
systematic symmetry preserving DSE truncation. 
 
\begin{table}[t] 
\tcaption{\label{tab:weak} 
Calculated\protect\cite{Ji:2001pj} $K_{l3}$ observables, compared with
experimental data\protect\cite{pdg} and chiral perturbation
theory.\protect\cite{GL85} The partial widths are calculated in the usual
way.  (Adapted from Ref.~[\protect\ref{RJi:2001pj}].)\smallskip}
\begin{center} 
\begin{tabular}{l|l|c|l} 
        & {\rm DSE} 
        & {\rm Expt.} 
        & {\rm ChPT}  \\[1ex]\hline 
$-f_+(0)$ & 0.96 &    & 0.98 \\ 
$-f_0(0)$ & 0.10 &    & 0.16 \\ 
$-f_0(\Delta)$ & 1.18 &      & 1.18 \\ 
$\lambda_+^{K^0}$ & 0.027 & 0.0300 $\pm$ 0.0020 & 0.031 \\ 
$\lambda_+^{K^+}$ & 0.027 & 0.0282 $\pm$ 0.0027 & 0.031 \\ 
$\lambda_0^{K^0}$ & 0.018 & 0.030 $\pm$ 0.005 & 0.017 \\ 
$\lambda_0^{K^+}$ & 0.018 & 0.013 $\pm$ 0.005 & 0.017 \\ 
$\Gamma(K^0_{e3})_{\times 10^{-8}\,{\rm eV}}$  & 0.49 & 0.494 $\pm$ 0.005 &      \\ 
$\Gamma(K^0_{\mu3})_{\times 10^{-8}\,{\rm eV}}$  & 0.32 & 0.346 $\pm$ 0.004 &      \\ 
$\Gamma(K^+_{e3})_{\times 10^{-8}\,{\rm eV}}$  & 0.24 & 0.259 $\pm$ 0.003 & 
\\ 
$\Gamma(K^+_{\mu3})_{\times 10^{-8}\,{\rm eV}}$ & 0.16 & 0.174 $\pm$ 0.003 &     \\ \hline 
\end{tabular} 
\end{center} 
\end{table} 
 
\subsection{Form factors in the timelike region} 
\addtocounter{subsection}{1} 
\noindent 
The electromagnetic pion form factor, $F_\pi(Q^2)$, exhibits a peak at $q^2=
-t = -m_\rho^2$ that is associated with $e^+ e^- \to
\rho \to
\pi^+\pi^-$.  This is a general feature of the electroweak form factors of 
pseudoscalar mesons, all of which can be expressed as 
\begin{equation} 
\label{rhopole} 
F_P(t) = g_{VPP} \, \frac{1}{m_V^2 - i m_V \Gamma_V - t} \, \frac{m_V^2}{g_V} 
\end{equation} 
in the neighbourhood of the relevant flavour channel's lowest-mass vector 
meson resonance, assuming it is narrow.  In Eq.\ (\ref{rhopole}), $g_{VPP}$ 
is the coupling constant modulating the $V \to P \bar P$ decay, $\Gamma_V$ is 
the total width of the vector meson and $m_V^2/g_V$ characterises the 
strength of the vector-meson--photon coupling. 
 
This behaviour arises naturally in applications of the rainbow-ladder kernel.
In the case of the pion and kaon form factors the poles appear as a
straightforward consequence of solving the inhomogeneous BSE,
Eq.~(\ref{verBSE}); a result we have already illustrated in connection with
the axial-vector vertex, Eq.\ (\ref{genavv}).  In the neighbourhood of the
pole the solutions of Eq.\ (\ref{verBSE}) assume the form\cite{pieterpion}
\begin{equation} 
 \Gamma^a_\mu(p_+,p_-) \approx f_V m_V 
 \frac{1}{m_V^2-t}\, \Gamma_\mu^{a\bar{a}}(p;q)  \; , 
\label{comres} 
\end{equation} 
where $\Gamma_\mu^{a\bar{a}}$ is the vector meson's Bethe-Salpeter amplitude
and $f_V$ was introduced in Eq.\ (\ref{fV}).\footnote{$\,f_V$ and $g_V$ are
algebraically related; e.g., for the $\rho^0$, $f_\rho m_\rho =
\sqrt{2} m_\rho^2 /g_\rho$; and $f_\phi m_\phi = 3 \, m_\phi^2 /g_\phi$.} \ \
The absence of a width in Eq.\ (\ref{comres}) is a defect of the 
rainbow-ladder truncation, which can be remedied without overcounting by 
including diagrams in the kernel that represent the vector meson's coupling 
to the $P \bar P$ intermediate state.  Alternatively, if the width:mass ratio 
is small, that can be done via bound state perturbation 
theory.\cite{mikepipi,Hollenberg:nj,Leinweber:1993yw,Mitchell:1996dn} 
 
\begin{figure}[t] 
\centerline{\includegraphics[width=4in]{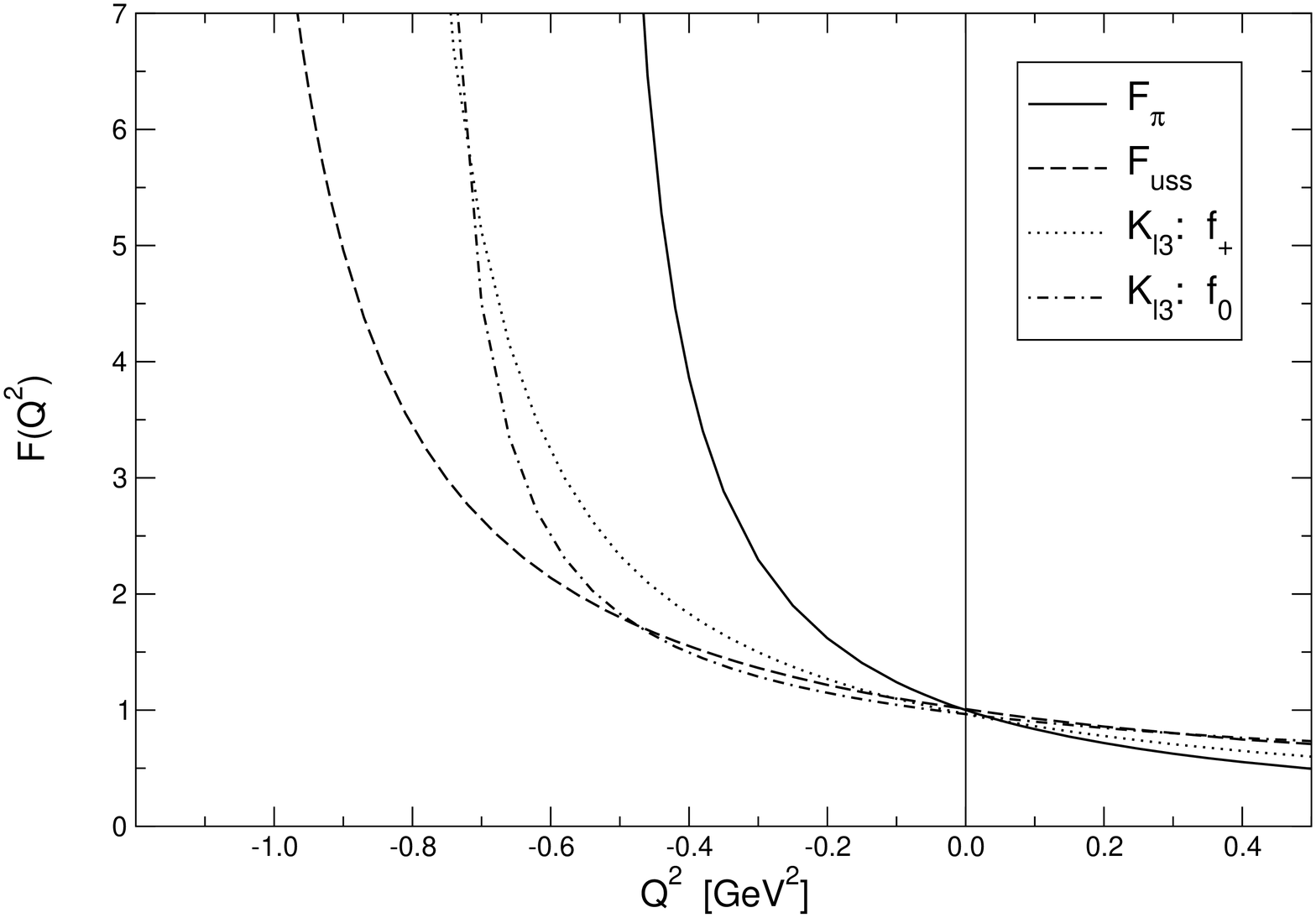}} 
\fcaption{\label{fig:timeff} 
Timelike ($t=-Q^2$) behaviour of the individual form factors describing 
electroweak $\pi$ and $K$ transitions. The vector meson poles are apparent. 
(Adapted from Ref.\ [\protect\ref{RMaris:2001rq}].)} 
\end{figure} 
 
The timelike behaviour of the form factors describing electroweak $\pi$ and
$K$ transitions is depicted in Fig. \ref{fig:timeff}.  It is apparent that
$F_\pi$ is singular at $t=(0.74\,{\rm GeV})^2$ and $F_{K^+}^{\bar s}$ at $t=
(1.05\,{\rm GeV})^2$; namely, at the masses of the $\rho$- and $\phi$-mesons
(see Table \ref{RGIvector}).  Since every other element has already been
calculated, the coupling constants $g_{\rho\pi\pi}$, $g_{\phi K K}$ can be
determined by fitting Eq.\ (\ref{rhopole}) in the neighbourhood of the
pole.\cite{Maris:2001rq} Alternatively, one can directly calculate the
coupling constants using impulse approximation three-hadron triangle
diagrams.  The results coincide\cite{Jarecke:2002xd} and are listed in Table
\ref{tab:strong}.  The agreement with experiment at the $9$\% level is good, 
given the rainbow-ladder truncation's established accuracy.
\begin{table}[b] 
\tcaption{\label{tab:strong} 
Calculated coupling constants for the two-pseudoscalar decays of vector 
mesons.  The rms rel.-error is $9$\%.  (Adapted from Ref.\ 
[\protect\ref{RJarecke:2002xd}].)\smallskip} 
\begin{center} 
\begin{tabular}{l|llll} 
Calc.   &  5.2      &  4.3          &  4.3      &  4.1  \\ 
Expt.   & $5.99\pm0.02$ & $4.48\pm0.04$ & $4.58\pm0.05$ 
                        & $4.59\pm0.05$ \\ \hline 
Rel.-Error 
    & 0.13      & 0.04      & 0.07      & 0.11  \\ \hline 
\end{tabular} 
\end{center} 
\end{table} 
 
Figure \ref{fig:timeff} also reveals poles in the $K_{\ell 3}$ electroweak
transition form factors.  The transverse form factor, $f_+(t)$, exhibits a
pole at $t= (0.94\,{\rm GeV})^2$; i.e., at the mass of the $K^\star$ meson
(see Table \ref{RGIvector}), as expected, and it is this singularity that
facilitates the successful description illustrated in Table
\ref{tab:weak}.\cite{Kl3}   The scalar form factor, $f_0(t)$, possesses a 
pole at $t= (0.89\,{\rm GeV})^2$.  As the discussion makes clear, this is the
model's prediction for the mass of the lightest $0^+_{u\bar s}$ meson.  The
model also yields a $0^+_{u\bar u+d\bar d}$ meson with mass
$0.67\,$GeV.\cite{pmdiquark} However, as observed in Sec.\ 2.2, the
rainbow-ladder truncation is known to be unreliable in the scalar channel
and, until a quantitative analysis of the corrections is complete, it is
impossible to associate these dressed-quark-antiquark bound states with any
mesons in the hadron spectrum or with the scalar states in model
studies.\cite{Oller:1998zr,penningtonscalar,Black:2000tk}
\pagebreak

\subsection{Vector meson transition form factors} 
\addtocounter{subsection}{1} 
\noindent 
Vector meson transition form factors, such as $\gamma^\ast \pi \to \rho$ and
$\gamma^\ast \pi \to \omega$, are important in the description of hadron
photo- and electro-production reactions, and their analysis can help in
developing a deeper understanding of the relation between QCD and efficacious
meson-exchange models of the $N$-$N$ interaction.\cite{Hechtphoto} A large
variety of these form factors have been calculated in impulse
approximation,\cite{Maris:2001am,Maris:2002mz} which for the $\rho \to
\pi\gamma$ transition is 
\begin{eqnarray} 
\Lambda^{\rho \pi \gamma}_{\mu\nu}(P;q) &=& 
      \frac{e\,N_c}{3} \, {\rm tr}_{\rm D}\!\! \int^\Lambda_k \!  S(q_2) \, 
        \Gamma^\pi(r_{+-};-P-q) \, S(q_1) \Gamma^\rho_\mu(r_{+0};P)\, S(k)
        \, i  
        \Gamma^\gamma_\nu(k,q_2) 
\nonumber\\ 
        &=& e\, \frac{g_{\rho \pi\gamma} }{m_{\rho}} 
        \, \epsilon_{\mu \nu \alpha \beta } \, P_{\alpha } q_{\beta } 
        \, F_{\rho\pi\gamma}(q^2) \,, 
\label{eq:rhopig} 
\end{eqnarray} 
where $P$ is the $\rho$-meson's momentum, $q$ is that of the photon and $q_1
= k + P$, $q_2 = k - q$, $r_{\alpha\beta}=k+\alpha P/2+\beta q/2$.  The
coupling $g_{\rho \pi\gamma}$ is defined such that
$F_{\rho\pi\gamma}(q^2=0)=1$ and then
\begin{equation} 
\Gamma_{\rho \to \pi \gamma} = 
        \frac{\alpha_{\rm em}}{24} \, g_{\rho \pi \gamma}^2 
        \; m_\rho \left(1 - \frac{m_\pi^2}{m_\rho^2} \right)^3\! , 
\label{rhoradwidth} 
\end{equation} 
where $\alpha_{\rm em}$ is QED's fine structure constant.  For $m_u = m_d$, 
$\Gamma_{\rho^\pm \to \pi^\pm \gamma}=\Gamma_{\rho^0 \to \pi^0 \gamma}$ and 
$\Gamma_{\omega \to \pi \gamma} = 9\, \Gamma_{\rho \to \pi 
\gamma}$.  The analysis of $K^\star \to K \gamma$ is also 
straightforward and in the limit of $SU(3)$-flavour
symmetry:\cite{Maris:2001am} $\Gamma_{K^{\star \pm} \to K^\pm \gamma}=
\Gamma_{\rho^\pm \to \pi^\pm \gamma}$ and $\Gamma_{K^{\star 0} \to K^0 \gamma}= 
4\, \Gamma_{K^{\star \pm} \to K^\pm \gamma}$, so one anticipates
$\Gamma_{K^{\star 0} \to K^0 \gamma} > \Gamma_{K^{\star \pm} \to K^\pm
\gamma}$ in reality.

\begin{table}[t] 
\tcaption{\label{tab:vecraddec} 
Calculated\protect\cite{Maris:2001am,Maris:2002mz} coupling constants for
radiative vector meson decays and associated widths compared with
data.\protect\cite{pdg} The calculations yield the ratios $g_{V P\gamma}/m_V$,
which are tabulated: the rms rel.-error is $12$\%.  Errors in the calculated
values of meson masses propagate into the values of the widths [see Eq.\
(\protect\ref{rhoradwidth})] (Adapted from
Ref.~[\protect\ref{RMaris:2001am}].) \smallskip}
\centerline{\smalllineskip 
\begin{tabular}{l|llll} 
$\frac{g_{V P\gamma}}{m_V}$ (GeV$^{-1}$) 
    & $\rho^\pm\to\pi^\pm\gamma$ 
    & $\omega\to\pi\gamma$ 
    & $K^{\star\pm}\to K^\pm\gamma$ 
    & $K^{\star 0}\to K^0\gamma$        \\ \hline 
Calc.   &  0.69     &  2.07     &  0.99     &  1.19         \\ 
Expt.   & $0.73\pm0.05$ & $2.35\pm0.07$     & $0.84\pm0.05$ & $1.27\pm0.07$ 
                                            \\ \hline 
Rel.-Error& 0.055 & 0.119 & 0.18 & 0.063   \\[1ex] \hline \hline 
\multicolumn{2}{l}{ $\Gamma_{V\to P\gamma}$ (keV) } 
                                        &    \\ \hline 
Calc.   &  53       &  479 
    &  90       &  130          \\ 
Expt.   & $67\pm7.5$    & $734\pm35$ 
    & $50.3\pm4.6$   & $117\pm10$       \\ \hline 
\end{tabular} } 
\end{table} 

\begin{figure}[t] 
\centerline{\includegraphics[width=2.4in]{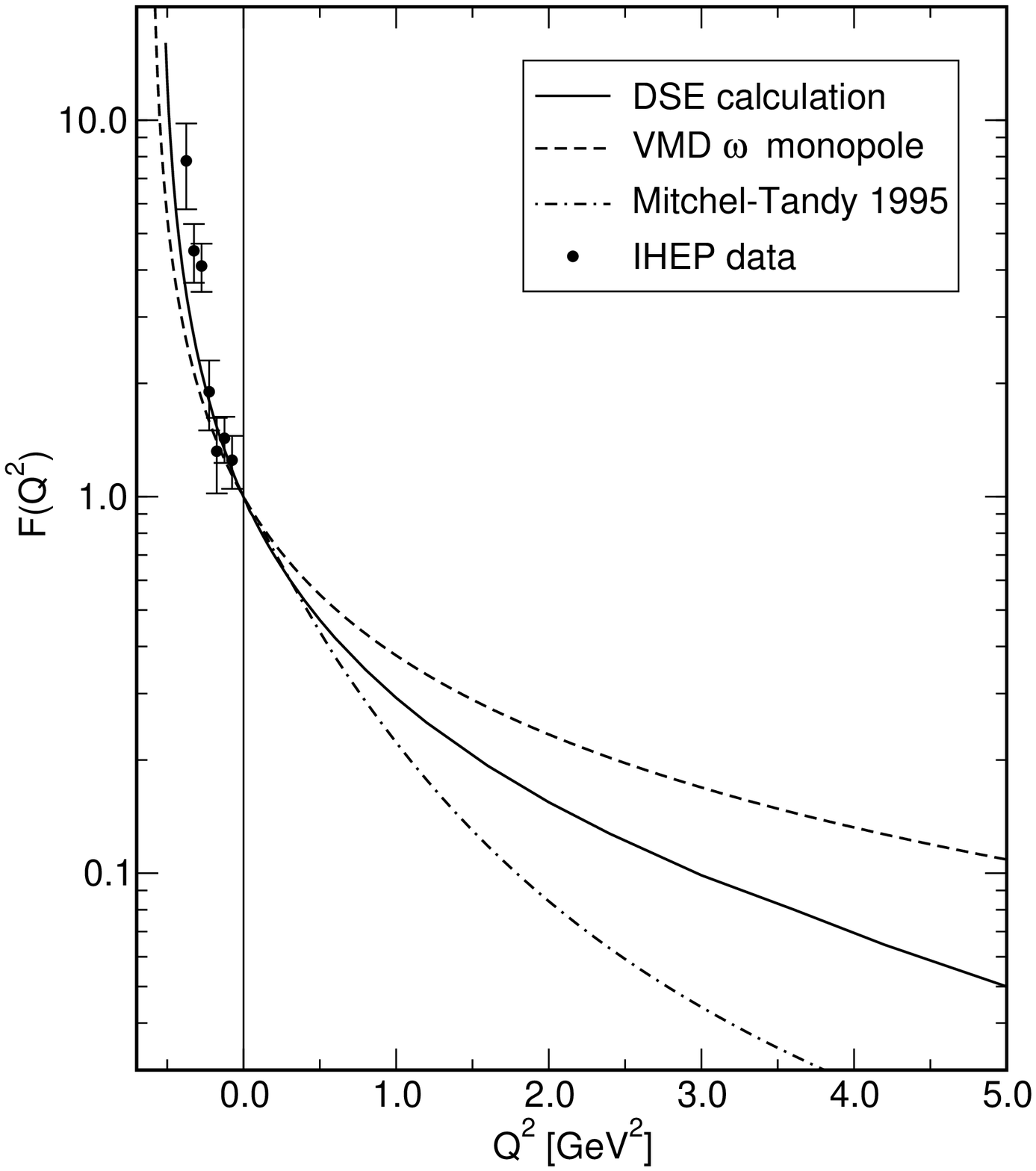} 
\includegraphics[width=2.4in]{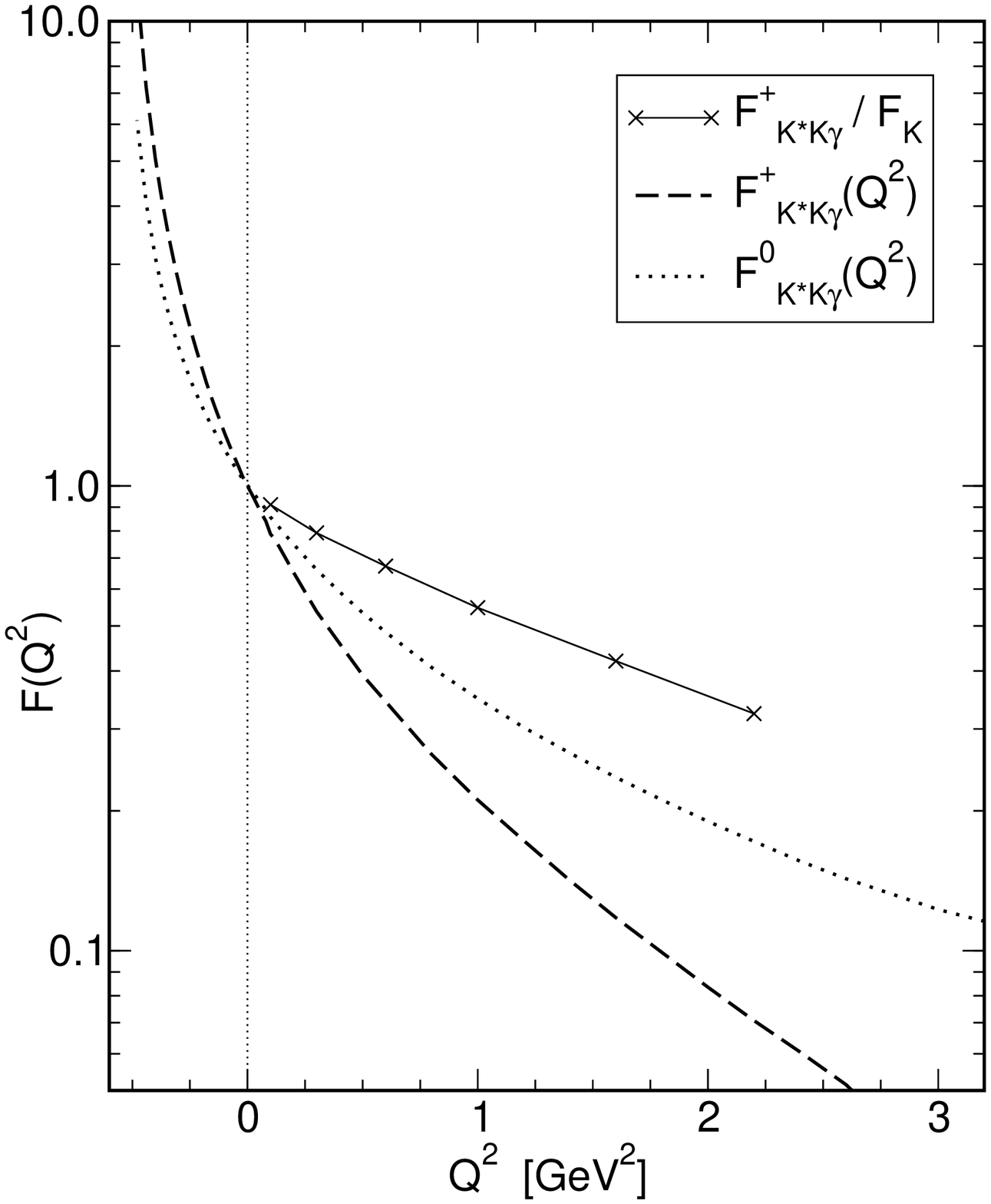} } 
\fcaption{\label{fig:transff} Left panel: DSE result for 
$F_{\omega\pi\gamma}(q^2)=F_{\rho\pi\gamma}(q^2)$ compared with experimental 
data.\protect\cite{Dzhelyadin:1980tj} The dot-dashed line is an earlier, 
rudimentary DSE calculation\protect\cite{kevinpeter} proceeding from 
parametrisations of the propagators and vertices after the fashion described 
in Sec.\ 3.3.  Right panel: DSE result for $F_{K^{\star}K\gamma}(q^2)$, 
charged (dashed line) and neutral (dotted line).  The solid line is the ratio 
$F_{K^{\star +}K^+\gamma}(q^2)/F_K(q^2)$.  (Adapted from 
Refs.~[\protect\ref{RMaris:2002mz},\protect\ref{RMaris:2002na}].)} 
\end{figure}

The calculated results\cite{Maris:2001am,Maris:2002mz} are summarised in
Table \ref{tab:vecraddec}.  The agreement is consistent with the established
accuracy of the rainbow-ladder truncation.  The transition form factors are
depicted in Fig.\ \ref{fig:transff} and, on the physical domain,
$F_{\rho\pi\gamma}(q^2)$ agrees with extant data\cite{Dzhelyadin:1980tj} to a
degree consistent with the absence of a width for the $\rho$-meson in
rainbow-ladder truncation.  An algebraic analysis of the behaviour of the
form factors at large spacelike-$q^2$ has not been completed but they do
appear to fall faster than $1/q^2$ (see the $\omega$-dominance curve in the
figure).  This is consistent with analyses in perturbative QCD, which
indicate that such processes are suppressed because they do not conserve
hadron helicity.\cite{hadronhelicity} The charged $K^\star K \gamma$ form
factor falls very rapidly because of cancellations between the contributing
$\gamma$-$u$- and $\gamma$-$s$-quark diagrams.  The analogous contributions
for $K^{\star 0}$ interfere constructively.
 
\subsection{\mbox{\boldmath $\pi^0 \to \gamma\gamma$}} 
\addtocounter{subsection}{1} 
\noindent 
The decay $\pi_0\to\gamma\gamma$ and associated $\gamma \gamma^\ast \to 
\pi^0$ transition form factor are closely related to the processes discussed 
in the last subsection and furthermore they have long been recognised to
possess a number of unique features that are especially important for testing
QCD.\cite{Lepage:1980fj}
 
In impulse approximation the transition form factor is obtained from the
vertex
\begin{eqnarray} 
\nonumber 
\lefteqn{\Lambda^{\pi\gamma\gamma}_{\mu\nu}(k_1;k_2) }\\ 
\nonumber 
& = & \alpha_{\rm em} \frac{4\pi \sqrt{2} }{3} \, N_c \,{\rm tr}_{\rm 
    D}\int^\Lambda_\ell \! S(\ell_1) \, \Gamma^\pi(\hat \ell; -P) \, S(\ell_2) 
    i\Gamma_\mu(\ell_2,\ell_{12})\, S(\ell_{12}) \, 
    i\Gamma_\nu(\ell_{12},\ell_1) \\ 
& = & 2 i \frac{\alpha_{\rm em} \,g_{\pi\gamma\gamma}}{\pi f_\pi} \epsilon_{\mu 
\nu \rho \sigma }\, k_{1\rho } k_{2\sigma} \, F_{\pi \gamma 
\gamma}(k_1;k_2)\,, 
\label{eq:pigg} 
\end{eqnarray} 
where $\ell_1 = \ell - k_1$, $\ell_2 = \ell + k_2$, $\hat \ell = 
(\ell_1+\ell_2)/2$, $\ell_{12}=\ell-k_1+k_2$, and the coupling constant 
$g_{\pi\gamma\gamma}$ is defined such that $F_{\pi \gamma 
\gamma}(k_1=0=k_2) =1$.  For $k_1^2=0=k_2^2$, the vertex describes $\pi^0$ 
decay and yields the width 
\begin{equation} 
 \Gamma_{\pi^0 \gamma \gamma} = 
\frac{m_\pi^3}{16 \pi} \left(\frac{\alpha_{\rm em}}{\pi 
f_\pi}\right)^2 \, g^2_{\pi\gamma\gamma}. 
\label{piwidth} 
\end{equation} 
It is a textbook example that $g_{\pi\gamma\gamma}\equiv 0$ and $\pi^0 \to 
\gamma \gamma$ is forbidden in the absence of the Abelian anomaly; i.e., if 
the axial-vector quark current $J_{5\mu}^3=\bar q \sfrac{1}{2}\lambda^3 
\gamma_5 \gamma_\mu q$ is conserved in the chiral limit.  However, this 
current is anomalous, as may be demonstrated in many ways, and one arrives
instead at the model-independent chiral limit result: $g_{\pi\gamma\gamma}
=\sfrac{1}{2}$, wherewith Eq.\ (\ref{piwidth}) predicts $\Gamma_{\pi^0
\gamma \gamma}=7.73\;{\rm eV}$ cf.\ the experimental value:\cite{pdg} $7.84
\pm 0.56$, which corresponds to $g^{\rm expt.}_{\pi\gamma\gamma}=0.504\pm
0.018$.
 
It is a key feature of the DSEs that, with a systematic and nonperturbative
truncation scheme, all consequences of the Wess-Zumino term and Abelian
anomaly are obtained exactly, without fine tuning.  A true representation of
DCSB and the preservation of Ward-Takahashi identities is crucial in
achieving this.\cite{cdrpion,racdr,Maris:1998hc,WZterm} Hence, one obtains
algebraically $g_{\pi\gamma\gamma}(m_\pi=0)=
\sfrac{1}{2}$ and\cite{Maris:2002mz}
\begin{equation}
g_{\pi\gamma\gamma}(m_\pi)= 0.502\,.
\end{equation}
 
\begin{figure}[t] 
\centerline{\includegraphics[width=28em]{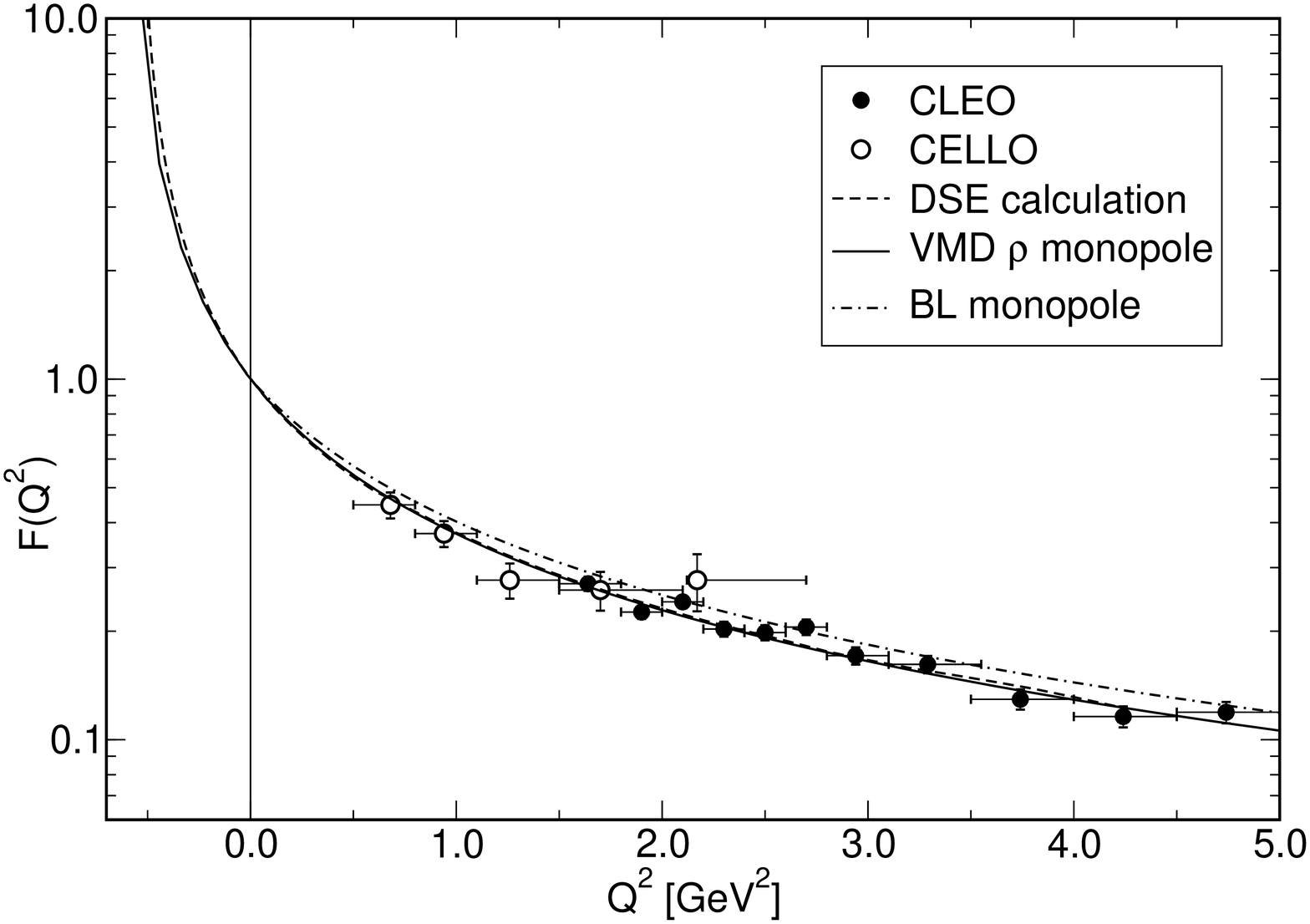} } 
\fcaption{\label{fig:pigg} DSE prediction,\protect\cite{Maris:2002mz} dashed 
line, for the $\gamma^* \pi\to \gamma$ transition form factor; solid line, a
monopole with mass scale $m_\rho^2 = 0.59\;{\rm GeV}^2$; and dot-dashed line,
monopole based on the asymptotic form:\protect\cite{Brodsky:1981rp}
$1/(1+Q^2/[8 \pi^2 f_\pi^2])$.  The data are from the
CELLO\protect\cite{Behrend:1990sr} and CLEO\protect\cite{Gronberg:1997fj}
collaborations (Adapted from Ref.[\protect\ref{RMaris:2002mz}].)}
\end{figure}

The $\gamma^* \pi \to \gamma$ transition form factor, which was measured by
the CELLO\cite{Behrend:1990sr} and CLEO\cite{Gronberg:1997fj} collaborations,
is defined as
\begin{equation} 
F_{\gamma^*\pi\gamma}(q^2) = F_{\pi\gamma\gamma}(k_1^2=q^2,k_2^2=0); 
\end{equation} 
i.e., as the form factor in Eq.\ (\ref{eq:pigg}) evaluated with one photon 
on-shell, and the impulse approximation result is plotted in Fig.\ 
\ref{fig:pigg}.  The calculated interaction radius 
$r_{\pi\gamma\gamma}^2 := -6 F_{\gamma^*\pi\gamma}^\prime(0) = 0.41~{\rm
fm}^2$ agrees with the experimental estimate:\cite{Behrend:1990sr} $0.42 \pm
0.04$.  It is evident that the constraints of DCSB are very tight.
 
The behaviour of the form factor at large spacelike-$q^2$ has been analysed
in perturbative QCD,\footnote{NB.\ The anomalous dimension is zero; i.e., at
leading order, in contrast to the elastic electromagnetic pion form factor,
$F_{\gamma^*\pi\gamma}$ does not depend on the strong running coupling,
$\alpha_s(q^2)$.} \ with the leading order
result:\cite{Lepage:1980fj,Brodsky:1981rp,radyushkin}
\begin{eqnarray} 
\label{piggpQCD} 
F_{\gamma^*\pi\gamma}^{\rm pQCD}(q^2) & = & J(1) \,\frac{4 \pi^2 f_\pi^2}{q^2} + 
{\rm O}\left(\frac{\alpha(q^2)}{q^2},\frac{1}{q^4}\right),\\ 
J(w) & = & \frac{4}{3} \int_0^1 dx \, \frac{\phi_\pi(x;\ln q)}{1 - w^2 (2 x 
-1)^2} , 
\end{eqnarray} 
where $w = (k_1^2-k_2^2)/(k_1^2+k_2^2)=1$ for $\gamma^* \pi \to \gamma$, and
$\phi_\pi(x;\ln q)$ is the pion's light-cone quark distribution amplitude,
$\int_0^1 \! dx\, \phi_\pi(x; \ln q) = 1$.  In this context accurate
measurements of the transition form factor can be interpreted as constraining
the $x$-dependence of $\phi_\pi(x; \ln q)$ at the experimental scale, $\ln
q$.  At very large $q^2$; viz., $\ln [q^2/\Lambda_{\rm QCD}^2]
\gg 1$, $\phi_\pi(x;\infty) = 6 \,x (1-x)$ and consequently $J(1)=2$. 
Existing data at $\ln [q/\Lambda_{\rm QCD}] \sim 2$
favour\cite{Gronberg:1997fj} $J(1)\approx 1.5$, a value which requires either
$\phi_\pi(x;\sim 2)$ to be far from asymptotic (e.g, $\phi_\pi(x;\sim 2) =
630 \, x^4 (1-x)^4 $ gives $J(w)=1.5$) or material corrections to the leading
order result.\cite{stanggcoll}
 
The expression in Eq.\ (\ref{eq:pigg}) has also been used to analyse the 
large spacelike-$q^2$ behaviour of the $\gamma^* \pi 
\to \gamma$ transition form factor, with the 
result\cite{peter98,klabucar,robertsfizika} 
\begin{equation} 
\label{piggasymp} 
F_{\gamma^*\pi\gamma}(q^2) = \frac{4}{3} \, \frac{4 \pi^2 f_\pi^2}{q^2}. 
\end{equation} 
This was obtained by assuming that for $k_2^2=0$ and $k_1^2=q^2\to 
\infty$, $2 k_1\cdot k_2 = - q^2$ because $P = k_1+k_2$ and the pion is 
on-shell, and
\begin{equation} 
\label{piggsymm} 
S(\ell_{12}) = \frac{1}{Z_2} \, \frac{1}{q^2}\, i \gamma\cdot (k_1-k_2)\,,\; 
\Gamma_\mu(\ell_2,\ell_{12}) = Z_1\, \gamma_\mu = 
\Gamma_\mu(\ell_{12},\ell_1)\,. 
\end{equation} 
The effect of Eq.\ (\ref{piggsymm}) is to treat the photons symmetrically in 
the integrand of Eq.\ (\ref{eq:pigg}).  In reality that might effect the 
situation $k_1^2=q^2=k_2^2$, which corresponds to $w=0$.  Since $J(w=0)=4/3$, 
this interpretation would reconcile Eqs.\ (\ref{piggpQCD}) and 
(\ref{piggasymp}).  In essence, this expresses the relevant observation in 
Ref.\ [\ref{RAnikin:1999cx}] and suggests that Eq.\ (\ref{eq:pigg}) should be 
reanalysed with a more conscious focus on the asymmetric $\gamma^*\pi\to 
\gamma$ transition. 
 
The calculations reported in Ref.\ [\ref{RMaris:2002mz}] are not affected by 
these considerations.  They simply provide the prediction of the 
rainbow-ladder DSE model for $F_{\gamma^*\pi\gamma}(q^2)$ on the domain 
covered by extant experiments, which do not extend into the truly asymptotic 
domain, $\ln [q^2/\Lambda_{\rm QCD}^2] \gg 1$.  The result is plotted in Fig.\ 
\ref{fig:pigg}, and on the scale of the figure is well approximated by a 
simple $\rho$-meson monopole.  It agrees with existing data, within errors, 
and interpreted in terms of Eq.\ (\ref{piggpQCD}), corresponds to $J(w=1;\ln 
q/\Lambda_{\rm QCD}\sim 2) \approx 1.7$. 
 
\subsection{Scattering processes} 
\addtocounter{subsection}{1} 
\noindent 
Consider now the more complex situation of four external lines.  We have
repeatedly emphasised that the DSEs provide for a resolution of the dichotomy
of the pion as both a Goldstone mode and a bound state of massive
dressed-quarks, and the role played in this by the systematic,
nonperturbative truncation scheme reviewed in Sec.\ 2.  As our exemplar we
therefore choose $\pi$-$\pi$ scattering.  It will immediately be appreciated
that this process has long been of interest and, indeed, that its particular
features and the effective Lagrangian that describe
them\cite{Weinberg:1966kf} are a keystone of chiral perturbation theory.
 
\begin{figure}[t] 
\centerline{\includegraphics[width=4in]{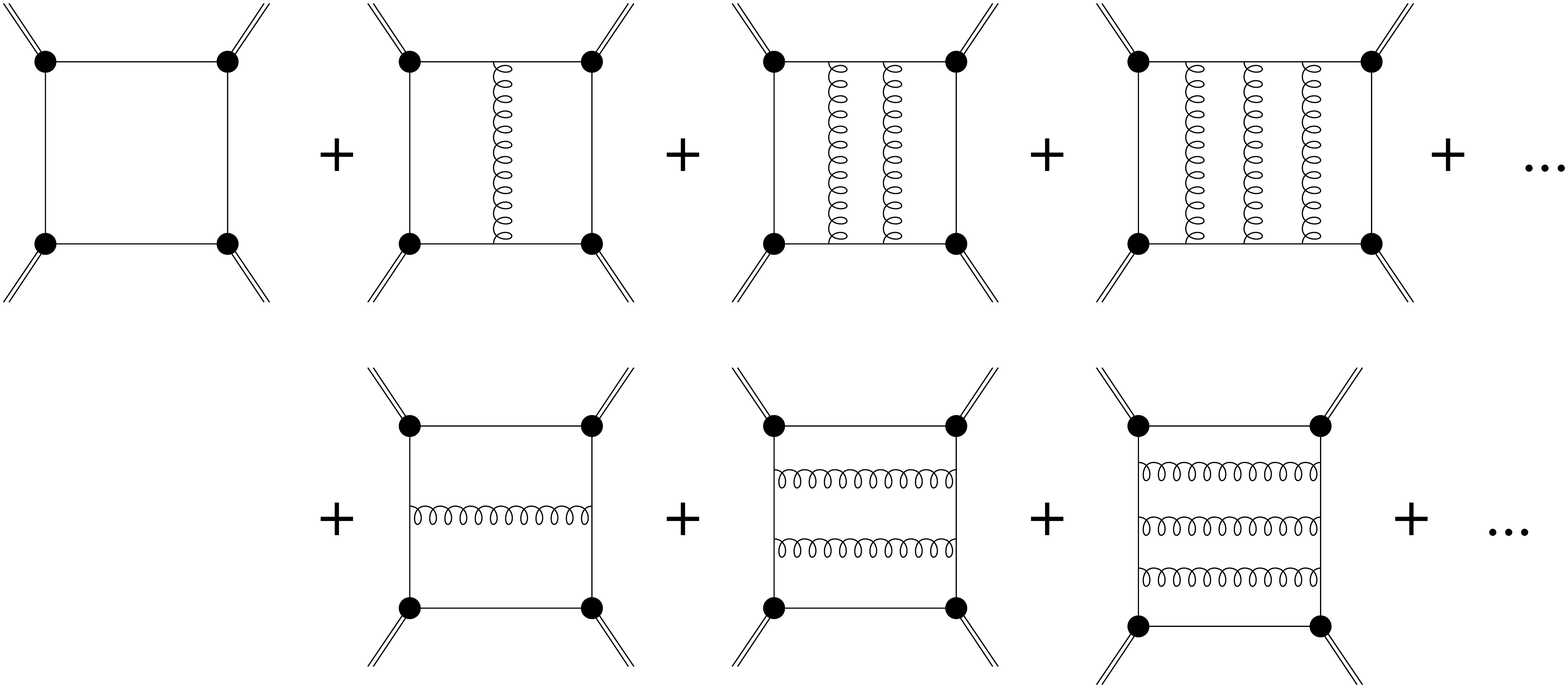} } 
\fcaption{\label{fig:pipiscat} Sum of terms required to describe 
$\pi$-$\pi$ scattering when the rainbow-ladder truncation is used to
calculate the dressed-quark propagator and pion Bethe-Salpeter amplitude.  In
every case, the interaction line is specified by Eq.\ (\protect\ref{gk2}).
The first diagram on the top line, $D_{\diamond}(s,t,u)$, is expressed in
Eq.\ (\protect\ref{eq:genericfour}); the remaining sum on the top line is
denoted by $D_{{\cal T}_s}(s,t,u)$; and the sum on the bottom line by $
D_{{\cal T}_t}(s,t,u)$. (Adapted from Ref.[\protect\ref{Rpmpipi}].) }
\end{figure}

Based on the analyses reviewed in the preceding subsections one would naively
suppose that the dominant contribution to $\pi$-$\pi$ scattering arises from
the first diagram in Fig.\ \ref{fig:pipiscat}, which represents the
renormalised expression
\begin{eqnarray} 
\nonumber \lefteqn{D_{\diamond}(s,t,u)=  {\rm tr} \int_\ell^\Lambda \! 
S(\ell_{--+}) \, \Gamma_\pi(\ell;-p_4) \, 
S(\ell_{++-}) }\\ 
&& \times \, \Gamma_\pi(\ell_{++0};-p_3) \, 
  S(\ell_{+++}) \, \Gamma_\pi(\ell_{+0+};p_2) \, 
  S(\ell_{+-+}) \, \Gamma_\pi(\ell_{0-+};p_1) \,, 
\label{eq:genericfour} 
\end{eqnarray} 
with $s=-(p_1+p_2)^2$, $t=-(p_1-p_3)^2$, $u=-(p_1-p_4)^2$, $p_1+p_2=p_3+p_4$ 
and $\ell_{\alpha \beta \gamma}= \ell + 
\sfrac{\alpha}{2} p_1+ \sfrac{\beta}{2} p_2+ \sfrac{\gamma}{2} p_3$. 
However, it follows immediately from Eqs.\ (\ref{genpibsa}), (\ref{bwti}), 
(\ref{fpiexact}) that in the chiral limit 
\begin{equation} 
D_{\diamond}(0,0,0) = {\rm tr}_{\rm isospin}[\tau^i \tau^j \tau^m \tau^n] 
\, \frac{4 N_c}{f_\pi^4} \, 
\int_\ell \frac{B^4(\ell^2)}{[\ell^2 A^2(\ell^2)+ B^2(\ell^2)]^2} 
\end{equation} 
and hence $D_{\diamond}(0,0,0) \neq 0$ when chiral symmetry is dynamically
broken.  It is clear now that this contribution alone violates the current
algebra results for low-energy $\pi$-$\pi$ scattering; viz., at threshold the
scattering amplitude must vanish as $m_\pi^2/f_\pi^2$.
 
The observation is not new.  In fact, it has long been known that in the
chiral limit $D_{\diamond}(0,0,0)$ is cancelled by contributions that may be
described as scalar-isoscalar two-pion correlations.  That mechanism is
readily realised; e.g., by analysing the auxiliary field effective action in
four-fermion interaction theories.\cite{gcmpipi,WZterm,schulzepipi} It can
also be achieved directly using a systematic truncation of the DSEs, by which
manner it is firmly placed in context with the material reviewed herein.
 
For the latter, if one employs a dressed-quark propagator and Bethe-Salpeter
amplitude obtained in the rainbow-ladder truncation, then the combination of
diagrams depicted in Fig.\ \ref{fig:pipiscat} is guaranteed to reproduce the
current-algebra results for near-threshold $\pi$-$\pi$
scattering.\cite{pmpipi,Bicudo:2001jq} To illustrate, we note that the sum in
the figure can be expressed
\begin{equation} 
D(s,t,u) = D_\diamond(s,t,u) + D_{{\cal T}_s}(s,t,u) + D_{{\cal T}_t}(s,t,u) 
\end{equation} 
and, in the chiral limit, it is readily established\footnote{To complete the 
proof one uses the axial-vector Ward-Takahashi identity, Eq.\ 
(\protect\ref{avwti}), the quark-level Goldberger-Treiman relation, Eq.\ 
(\ref{bwti}), and the Bethe-Salpeter equation for the fully-amputated 
quark-antiquark scattering matrix, Eq.\ (20) of Ref.\ 
[\protect\ref{Rmrt98}].} \ that if any one of the external pion momenta 
vanish then 
\begin{equation} 
D_{{\cal T}_s}(0,0,0) = -\sfrac{1}{2} D_\diamond(0,0,0) = D_{{\cal T}_t}(0,0,0) 
\end{equation} 
and thus $D(0,0,0) = 0$.  Extending this, one finds at leading order in
powers of mass and momenta\cite{Bicudo:2001jq}
\begin{equation} 
D(s,t,u) = \frac{1}{8 f_\pi^2} \left( s + t -u  \right)\,, 
\end{equation} 
from which follows the isospin-zero, -one and -two scattering amplitudes: 
\begin{equation}
\begin{array}{lll}
A_0(s,t,u) = & 3 \, D(s,t,u) + 3 \, D(s,u,t) - D(u,t,s) & \displaystyle =
\frac{2 s -  m_\pi^2}{2 f_\pi^2} \,, \\[2ex] 
A_1(s,t,u) = & 2\, D(s,t,u) - 2\, D(s,u,t) & \displaystyle =
\frac{ t - u}{2 f_\pi^2}\,,\\[2ex]
A_2(s,t,u) = & 2\, D (u,t,s) &
\displaystyle =   \frac{2 m_\pi^2 - s }{2
f_\pi^2}\,;  
\end{array}
\end{equation} 
viz., precisely the current algebra results for low-energy $\pi$-$\pi$
scattering, \textit{independent} of the detailed form of the interaction.

\begin{figure}[t] 
\centerline{\includegraphics[width=28em]{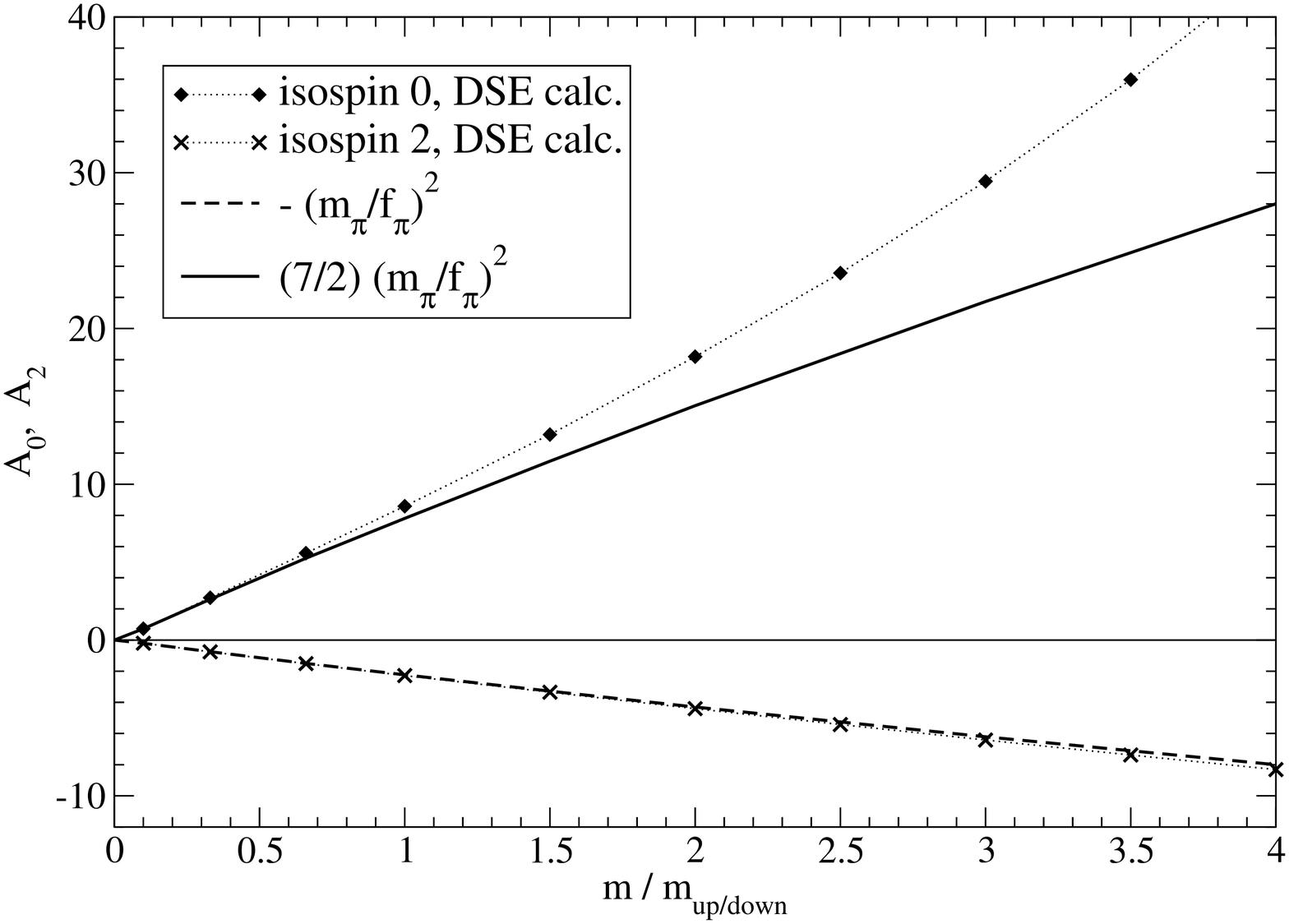}}
\fcaption{\label{fig:pipires} Current-quark-mass-dependence of the isospin-zero 
and -two scattering amplitudes at threshold: $A_0(4m_\pi^2,0,0)$, 
$A_2(4m_\pi^2,0,0)$, respectively, calculated by summing the contributions in 
Fig.\ \ref{fig:pipiscat} (Adapted from Ref.~[\protect\ref{RBicudo:2001jq}].)} 
\end{figure} 
 
In Fig.\ \ref{fig:pipires} we plot the mass-dependence of the isospin-zero
and -two scattering amplitudes at threshold.  At the physical value of the
light current-quark-mass, $95$\% of $D_\diamond(4m_\pi^2,0,0)$ is cancelled
by the remaining terms in Fig. \ref{fig:pipiscat}.  Furthermore, the direct
calculation yields $A_0= 8.6$, $A_2=-2.3$ cf.\ the current-algebra results:
$A_W^0= 7 m_\pi^2/(2 f_\pi^2)=7.8$, $A_W^2= - m_\pi^2/f_\pi^2= - 2.2$; i.e.,
the complete result exhibits a $10$\% increase over the leading order value
in the isospin-zero channel, which is an harbinger of the resonant behaviour
of $\pi$-$\pi$ scattering in this channel.
 
For a detailed comparison with experiment, as with meson charge
radii,\cite{alkoferpiloop} the DSE calculation should be augmented by the
inclusion of pion initial and final state interactions. Omitting them, the
results described herein are kindred to those obtained at tree-level in
chiral perturbation theory.  Bearing this in mind, in Table \ref{tab:pipi} we
compare the DSE predictions for $\pi$-$\pi$ scattering lengths with fits to
data using tree-level\cite{Donoghue:xa} and two-loop\cite{Colangelo:2001df}
chiral perturbation theory.  The comparison shows that final state
interactions only materially affect the scalar-isoscalar scattering length.
It is this channel that putatively exhibits a low-mass scalar meson.
 
\begin{table}[b] 
\tcaption{\label{tab:pipi} Rainbow-ladder DSE prediction for 
$\pi$-$\pi$ scattering lengths,\protect\cite{pmpipi} compared with those
fitted in analyses of data using chiral perturbation theory at
tree-level\protect\cite{Donoghue:xa} and at
second-order.\protect\cite{Colangelo:2001df}\smallskip}
\centerline{\smalllineskip 
\begin{tabular}{l|ccc} 
        & $a_0^0$ & $a_1^1$ & $a_0^2$       \\\hline 
DSE calc.   & 0.17 & 0.036 & -0.045\\ \hline 
Tree-level ChPT  & 0.15 &  0.036 & -0.045\\ 
2nd order        & 0.22 &  0.038 & -0.044\\ 
\hline 
\end{tabular} } 
\end{table} 
 
It is an opportune point to reemphasise a key aspect of the framework,
namely, because it works explicitly from a single enunciated kernel, final
state interactions of the type we have just described can be incorporated
cleanly and without overcounting.  For instance, in Sec.\ 4.3 we saw that the
dressed-quark-photon vertex exhibits a real-axis pole when $s=-q^2$ coincides
with a bound state mass and that a width was acquired only after including
meson rescattering effects.  In precisely the same manner, the sum of terms
in Fig.\ \ref{fig:pipiscat} produces the resonance poles one expects in
$\pi$-$\pi$ scattering; e.g., a $\rho$-meson pole appears in the isospin-one
scattering amplitude.\cite{pmpipi} The systematic nature of the truncation
means that one can subsequently add loops that represent initial and/or final
state interactions between the DSE-described rainbow-ladder pions and be
certain that these contributions are truly new and previously unaccounted
for, and just those terms needed to provide the strong width of the
$\rho$-meson.
 
It is important to note that in a Ward-Takahashi identity preserving
implementation of the rainbow-ladder truncation, contributions analogous to
the additional diagrams in Fig.\ \ref{fig:pipiscat}, which we have denoted
$D_{{\cal T}_s}(s,t,u)$, $D_{{\cal T}_t}(s,t,u)$, can only appear in the
calculation of amplitudes describing processes with \textit{four or more}
external particles.  In those involving only two or three external lines, the
addition of a single gluon rung connecting any two quarks can be absorbed
into the ladder sum that generated the vertex or bound state amplitude which
joins the two quark lines; i.e., it adds to what was already present, and is
therefore overcounting.  Hitherto, $\pi$-$\pi$ scattering is the only
amplitude with four external lines that has been fully explored and many
other processes are worthy of study or reexamination.  Of particular
interest, perhaps, are the anomalous $\gamma\pi\pi\pi$ amplitude, wherein the
interplay between the vector and axial-vector Ward-Takahashi identities is
known to be important,\cite{racdr} and, indeed, the Wess-Zumino
five-pseudoscalar interaction term itself.\cite{WZterm}
 
\subsection{Summary} 
\addtocounter{subsubsection}{1} 
\noindent 
We reiterate that every DSE result reviewed in this section is a prediction.
All follow from the strict implementation of the
renormalisation-group-improved rainbow-ladder truncation defined in Eqs.\
(\ref{ladder}), (\ref{rainbowdse}), with the model of the effective coupling
specified in Eq.\ (\ref{gk2}) whose single parameter takes the value in Eq.
(\ref{valD}), which was chosen to fit $f_\pi$, $m_\pi$, $f_K$, $m_K$. The
analyses demonstrate that the DSEs provide a mature approach that links
hadron physics experiments with elementary properties of QCD and allows for
their direct interpretation as critical data on the nature of quark
confinement and the quark/gluon wave functions of hadrons.  The success is
firm evidence that the quark-quark interaction possesses significant
integrated strength on the infrared domain $k^2\lsim 2\,$GeV$^2$, and
lattice-QCD simulations and DSE studies of QCD's gauge sector are now
searching for its origin.

%% file: sect5.tex
 
\vspace*{1pt}\textlineskip  
\section{On Baryons}        
\addtocounter{section}{1} 
\setcounter{equation}{0} 
\setcounter{figure}{0} 
\setcounter{table}{0} 
\vspace*{-0.5pt} 
\noindent 
Contemporary \label{sect4label} experimental facilities employ large momentum 
transfer reactions to probe the structure of hadrons and thereby attempt to 
elucidate the role played by quarks and gluons in building them.  Since the 
proton is a readily accessible target its properties have been studied most 
extensively and hence an understanding of a large fraction of the available 
data requires a Poincar\'e covariant theoretical description of the nucleon. 
 
The material we have reviewed thus far demonstrates that the properties of 
light mesons are well described by a Poincar\'e covariant rainbow-ladder 
truncation of QCD's DSEs.  An extension to baryons begins with a Poincar\'e 
covariant Faddeev equation.  That, too, requires an assumption about the 
interaction between quarks.  An analysis\cite{regbos} of the Global Colour 
Model\cite{reggcm,petergcm,gunnergcm} suggests that the nucleon can be viewed 
as a quark-diquark composite.  Pursuing that picture yields\cite{regfe} a 
Faddeev equation, in which two quarks are always correlated as a 
colour-antitriplet diquark quasiparticle (because ladder-like gluon exchange 
is attractive in the $\bar 3_c$ quark-quark scattering channel) and binding 
in the nucleon is effected by the iterated exchange of roles between the 
dormant and diquark-participant quarks. 
 
A first numerical study of the Faddeev equation for the nucleon was reported 
in Ref. [\ref{Rcjbfe}], and there have subsequently been numerous more 
extensive analyses; e.g., Refs. [\ref{Rbentz},\ref{Roettel}].  In particular, 
the formulation of Ref.~[\ref{Roettel}] employs confined quarks, and confined 
pointlike-scalar and -axial-vector diquark correlations, to obtain a spectrum 
of octet and decuplet baryons in which the rms-deviation between the 
calculated mass and experiment is only $2$\%.  The model also reproduces 
nucleon form factors over a large range of momentum transfer,\cite{oettel2} 
and its descriptive success in that application is typical of such Poincar\'e 
covariant treatments.\cite{cdrqciv,jacquesA,jacquesmyriad,nedm} 
 
However, these early successes were achieved without considering the role 
played by light pseudoscalar mesons.  In the context of spectroscopy, studies 
using the Cloudy Bag Model (CBM)\cite{tonyCBM} indicate that the 
dressed-nucleon's mass receives a negative contribution of as much as 
$300$-$400\,$MeV from pion self-energy corrections; 
i.e.,\cite{tonyANU,bruceCBM} $\delta M_+ = -300\,$ to $-400\,$MeV. Furthermore, 
a perturbative study, using the Faddeev equation, of the mass shift induced by 
pointlike $\pi$-exchange between quark and diquark constituents of the nucleon 
obtains\cite{ishii} $\delta M_+ = -150$ to $-300\,$MeV.  Such corrections much 
diminish the value of the $2$\% spectroscopic accuracy obtained using only 
quark and diquark degrees of freedom.  The size and qualitative impact of 
contributions from light-pseudoscalars to baryon masses may therefore provide 
material constraints on the development of a realistic quark-diquark picture. 
This has recently been explored in detail\cite{NpiN} and we now review the 
findings. 
 
\subsection{Diquarks and a Faddeev equation} 
\addtocounter{subsection}{1} \noindent 
The rainbow-ladder DSE truncation yields asymptotic diquark states in the 
strong interaction spectrum.\cite{conradsep,pmdiquark} Such states are not 
observed.  Their appearance is an artefact of this lowest-order truncation. 
Higher order terms in the quark-quark scattering kernel\footnote{NB.\ We saw 
in Sec.\ 2.2 that such contributions to the quark-\textit{antiquark} 
scattering kernel do not materially affect the properties of vector and 
flavour nonsinglet pseudoscalar mesons, a result which underlies the 
successes reviewed in Sec.\ 4.} \ (crossed-box and vertex corrections) act to 
ensure that QCD's quark-quark scattering matrix does not exhibit 
singularities that correspond to asymptotic diquark bound 
states.\cite{truncscheme,detmold} Nevertheless, studies with kernels that 
don't produce diquark bound states, do support a physical interpretation of 
the masses obtained using the rainbow-ladder truncation: $m_{qq}$ plays the 
role of a confined-quasiparticle mass in the sense that $l_{qq}=1/m_{qq}$ may 
be interpreted as a range over which the diquark can propagate inside a 
baryon.  These observations motivate the following \textit{Ansatz} for the 
quark-quark scattering matrix: 
\begin{equation} 
[M_{qq}(k,q;K)]_{rs}^{tu} = \sum_{J^P=0^+,1^+,\ldots} 
\Gamma^{J^P}(q;K)\, \Delta^{J^P}(K) \, 
\bar\Gamma^{J^P}(k;-K)\,, \label{AnsatzMqq} 
\end{equation} 
wherein $\Delta^{J^P}(K)$ act as diquark propagators and $\Gamma^{J^P}(q;K)$ 
are Bethe-Salpeter-like amplitudes describing the relative momentum 
correlation of the quarks constituting the diquark. 
 
\begin{table}[t] 
\tcaption{\label{tabledqmasses} Diquark pseudoparticle masses (in GeV) 
calculated in Ref.\ [\protect\ref{Rconradsep}].  The magnitudes and ordering 
are characteristic and model independent: see, e.g., Refs.\ 
[\ref{Rpmdiquark},\ref{Rgunner}] and recent lattice-QCD 
estimates.\protect\cite{latticediquark} (Adapted from Ref.\ 
[\protect\ref{RNpiN}].)\smallskip} 
\centerline{\smalllineskip 
\begin{tabular}{l | lllll| lll} 
$(qq)_{J}^P$ & $(ud)_{0}^{+}$ & $(us)_{0}^{+}$ & $(uu)_{1}^{+}$ & 
$(us)_{1}^{+}$ & $(ss)_{1}^{+}$ & $(uu)_{1}^{-}$ & $(us)_{1}^{-}$ & 
$(ss)_{1}^{-}$ 
\\\hline 
$m_{qq}$     & 0.74 & 0.88 & 0.95 & 1.05 & 1.13 & 1.47 & 1.53 & 1.64 
\end{tabular}} 
\end{table} 
 
The validity of Eq.\ (\ref{AnsatzMqq}) is key to the derivation of a 
quark-diquark Faddeev equation for baryons.  The simplification is amplified 
when the summation can be truncated after only a few terms.  That is indeed 
possible, as can be argued from Table \ref{tabledqmasses}, which lists 
calculated diquark pseudoparticle masses.\cite{conradsep} It is apparent that 
for octet and decuplet baryons it should be a good approximation to retain only 
the scalar and axial-vector diquark correlations because they alone have masses 
less than the bound states they will constitute.  (Naturally, spin-$3/2$ 
baryons cannot be described unless pseudovector correlations are retained.) 
Capitalising on this, one arrives at a remarkably simple matrix-integral 
equation. 
 
For example, one represents the bound state nucleon by a Faddeev amplitude: 
\begin{equation} 
\Psi = \Psi_1 + \Psi_2 + \Psi_3 \,, 
\end{equation} 
in which the subscript identifies the dormant quark and, e.g., $\Psi_{1,2}$ 
are obtained from $\Psi_3$ by a correlated cyclic permutation of all the 
quark labels.  According to the above assumption, the individual 
sub-amplitudes are written as just a sum of scalar and pseudovector diquark 
correlations: 
\begin{equation} 
\label{Psi} \Psi_3(p_i,\alpha_i,\tau_i) = \Psi_3^{0^+} + \Psi_3^{1^+}, 
\end{equation} 
with $(p_i,\alpha_i,\tau_i)$ the momentum, spin and isospin labels of the 
quarks constituting the nucleon, and $P=p_1+p_2+p_3$ the nucleon's total 
momentum.  The scalar diquark component in Eq.~(\ref{Psi}) is 
\begin{equation} 
\Psi_3^{0^+}(p_i,\alpha_i,\tau_i) 
= [\Gamma^{0^+}(\frac{1}{2}p_{[12]};K)]_{\alpha_1 
\alpha_2}^{\tau_1 \tau_2}\, \Delta^{0^+}(K) \,[{\cal S}(\ell;P) u(P)]_{\alpha_3}^{\tau_3}\,,\label{calS} 
\end{equation} 
where: the spinor satisfies $(i\gamma\cdot P + M)\, u(P) =0= \bar u(P)\,
(i\gamma\cdot P + M)$, with $M$ the mass obtained in solving the Faddeev
equation, and is also a spinor in isospin space, with $\varphi_+= {\rm
col}(1,0)$ for the proton and $\varphi_-= {\rm col}(0,1)$ for the neutron;
and $K= p_1+p_2=: p_{\{12\}}$, $p_{[12]}= p_1 - p_2$, $\ell := (-p_{\{12\}} +
2 p_3)/3$.  The pseudovector component is
\begin{equation} 
\Psi^{1^+}(p_i,\alpha_i,\tau_i)=   [{\tt 
t}^i\,\Gamma_\mu^{1^+}(\frac{1}{2}p_{[12]};K)]_{\alpha_1 
\alpha_2}^{\tau_1 \tau_2}\,\Delta_{\mu\nu}^{1^+}(K)\, 
[{\cal A}^{i}_\nu(\ell;P) u(P)]_{\alpha_3}^{\tau_3}\,, 
\label{calA} 
\end{equation} 
where the symmetric isospin-triplet matrices are: %
$ {\tt t}^+ = 
\frac{1}{\surd 2}(\tau^0+\tau^3) \,,\; 
{\tt t}^0 = \tau^1\,,\; 
{\tt t}^- = \frac{1}{\surd 2}(\tau^0-\tau^3)\,, 
$ 
with $(\tau^0)_{ij}=\delta_{ij}$ and $\tau^{1,3}$ the usual Pauli matrices. 
The colour antisymmetry of $\Psi_3$ is implicit in $\Gamma^{J^P}\!$. 
 
The Faddeev equation satisfied by $\Psi_3$ now reduces to a set of coupled 
equations for the matrix valued functions ${\cal S}$, ${\cal A}_\nu^i$ in 
Eqs.\ (\ref{calS}), (\ref{calA}): 
\begin{equation} 
\left[ \begin{array}{r} 
{\cal S}(k;P)\, u(P)\\ 
{\cal A}^i_\mu(k;P)\, u(P) 
\end{array}\right] 
= -4\,\int\frac{d^4\ell}{(2\pi)^4}\,{\cal M}(k,\ell;P) 
\left[ 
\begin{array}{r} 
{\cal S}(\ell;P)\, u(P)\\ 
{\cal A}^j_\nu(\ell;P)\, u(P) 
\end{array}\right], 
\label{FEone} 
\end{equation} 
wherein the kernel is 
\begin{equation} 
\label{calM} {\cal M}(k,\ell;P) = \left[\begin{array}{cc} 
{\cal M}_{00} & ({\cal M}_{01})^j_\nu \\ 
({\cal M}_{10})^i_\mu & ({\cal M}_{11})^{ij}_{\mu\nu}\rule{0mm}{3ex} 
\end{array} 
\right] 
\end{equation} 
with, e.g., 
\begin{equation} 
\label{calM00} 
{\cal M}_{00} = \Gamma^{0^+}(k_q-\ell_{qq}/2;\ell_{qq})\, 
S^{\rm T}(\ell_{qq}-k_q)\,\Gamma^{0^+}(\ell_q-k_{qq}/2;-k_{qq})\, 
S(\ell_q)\,\Delta^{0^+}(\ell_{qq}) \,, 
\end{equation} 
$\ell_q=\ell+P/3$, $k_q=k+P/3$, $\ell_{qq}=-\ell+ 2P/3$, $k_{qq}=-k+2P/3$, 
and $S$ is the propagator of the dormant quark constituent of the nucleon. 
The other entries are also expressed merely in terms of the 
Bethe-Salpeter-like amplitudes and propagators. 
 
It is implicit in Eqs.\ (\ref{FEone})-(\ref{calM00}) that $u(P)$ is a 
normalised average of $\varphi_\pm$ so that, e.g., the equation for the proton 
is obtained by projection on the left with $\varphi^\dagger_+$ and ${\cal 
M}_{01}$ generates an isospin coupling between $u(P)_{\varphi_+}$ on the 
l.h.s.\ of Eq.~(\ref{FEone}) and, on the r.h.s., 
\begin{equation} 
\surd 2\,{\cal A}^+_\nu\, u(P)_{\varphi^-} - {\cal A}^0_\nu 
\,u(P)_{\varphi_+}\,. 
\end{equation} 
This is merely the Clebsch-Gordon coupling of 
isospin-$1\oplus\,$isospin-$\frac{1}{2}$ to total isospin-$\frac{1}{2}$ and 
means that the scalar diquark amplitude in the proton, $(ud)_{0^+}\,u$, is 
coupled to itself {\it and} the linear combination: $\surd 2\, (uu)_{1^+}\, d 
- (ud)_{1^+} \, u$. 
 
The matrix-valued functions ${\cal S}$ and ${\cal A}_\mu^i$ are 
Bethe-Salpeter-like amplitudes that describe the momentum-space correlation 
between the quark and diquark in the nucleon and can, in general, assume 
quite complex forms.\cite{oettel} However, reduced forms can be used to good 
effect:\cite{NpiN} 
\begin{eqnarray} 
\label{calSE} 
{\cal S}(\ell;P) & =& f_1(\ell^2,P^2)\,I_{\rm D} + \frac{1}{M}\left(i\gamma\cdot 
\ell 
- \ell \cdot \hat P\, I_{\rm D}\right)\,f_2(\ell^2,P^2)\,, \\ 
\label{calAE} 
{\cal A}_\mu^i(\ell;P) & =& a_1^i(\ell^2,P^2) \, \gamma_5\gamma_\mu + 
a_2^i(\ell^2,P^2)\,\gamma_5 \gamma\cdot\hat\ell \,\hat\ell_\mu\, , 
\end{eqnarray} 
where $(I_{\rm D})_{rs}= \delta_{rs}$, $\hat P^2= - 1$, $\;\hat\ell^2=1$, 
and, assuming isospin symmetry, $a_j^1=a_j^2=a_j^3$, $j=1,2$.  In the 
nucleon's rest frame, $f_{1,2}$ in Eq.\ (\ref{calSE}) describe the upper, 
lower component of the bound-state nucleon's spinor.  It will readily be 
appreciated that a sizable value of $f_2/f_1$ corresponds to a significant 
amount of the nucleon's spin being stored as quark orbital angular momentum. 
In this connection the amplitude $a_2^i$ corresponds to a $D$-wave in the 
nucleon's wave function: in the rest frame, the quark and diquark spins are 
coupled to total-spin $\sfrac{3}{2}$, which must be combined with angular 
momentum $L=2$ to obtain a $J=\sfrac{1}{2}$ nucleon.  The physical content is 
frame invariant because the Faddeev amplitudes are Poincar\'e covariant. 
 
Equation (\ref{FEone}) can be solved once the kernel is specified and for 
that Ref.\ [\ref{RNpiN}] adopted the expedient, described in Sec.\ 3.3, of 
using algebraic parametrisations of the propagators and Bethe-Salpeter 
amplitudes.  The rationale is the same: DSE studies of baryons are not yet 
mature and hence there is merit in employing sound simplifications.  The 
dressed-quark propagator of Eqs.\ (\ref{qprop})-(\ref{tableA}) was used, 
without change.  The Bethe-Salpeter-like amplitudes describing the 
relative-momentum correlations of quarks within the diquark were represented 
by 
\begin{eqnarray} 
\label{Gamma0p} \Gamma^{0^+}(k;K) &=& \frac{1}{{\cal N}^{0^+}} \, 
H^a\,C i\gamma_5\, i\tau_2\, {\cal F}(k^2/\omega_{0^+}^2) \,, \\ 
\label{Gamma1p} {\tt t}^i \Gamma^{1^+}_\mu (k;K) &=& \frac{1}{{\cal N}^{1^+}}\, 
H^a\,i\gamma_\mu C\,{\tt t}^i\, {\cal F}(k^2/\omega_{1^+}^2)\,, 
\end{eqnarray} 
with ${\cal F}(y)$ given after Eq.\ (\ref{svm}), the colour matrix 
$[H^{c_3}]_{c_1 c_2}=\epsilon_{c_1 c_2 c_3}$ and ${\cal N}^{J^P}\!$, the 
canonical normalisation, fixed by an expression very much like Eq.\ 
(\ref{pinorm}); and the propagators for the confined diquarks by: 
\begin{eqnarray} 
\Delta^{0^+}(K) & = & \frac{1}{m_{0^+}^2}\,{\cal F}(K^2/\omega_{0^+}^2)\,,\\ 
\Delta^{1^+}_{\mu\nu}(K) & = & \left(\delta_{\mu\nu} + \frac{K_\mu 
K_\nu}{m_{1^+}^2}\right) \, 
\frac{1}{m_{1^+}^2}\, {\cal F}(K^2/\omega_{1^+}^2) . 
\end{eqnarray} 
These expressions define a four-parameter model: $m_{J^P}$ are diquark masses 
and $\omega_{J^P}$ are widths characterising the size of the diquark 
correlation inside baryons. 
 
\begin{table}[t] 
\tcaption{\label{tableMass} Calculated nucleon and $\Delta$ 
masses.\protect\cite{NpiN} The results in the first and third rows were 
obtained using scalar and pseudovector diquark correlations: 
$m_{1^+}=0.90\,$GeV in row 1, $m_{1^+}=0.94\,$GeV in row 3. 
($m_{0^+}=0.74\,$GeV, always.)  Pseudovector diquarks were omitted in the 
second and fourth rows.  $\omega_{f_{1,2}}$ are discussed after 
Eq.~(\protect\ref{lqq}), and \mbox{\sc r} in and after 
Eq.~(\protect\ref{scR}).  All dimensioned quantities are in GeV. (Adapted 
from Ref.\ [\protect\ref{RNpiN}].) \smallskip} 
\centerline{\smalllineskip 
\begin{tabular} 
{l|ccccccc} 
     & $\omega_{0^+}$ & $\omega_{1^+}$ & $M_N$ & $M_\Delta$  & 
     $\omega_{f_1}$ & $\omega_{f_2}$ & {\sc r}\\\hline 
$0^+$ \& $1^+$ & 0.64 & 1.19~ & 0.94 & 1.23~  &  0.49 & 0.44 & 0.25\\ 
$0^+$ & 0.64 & - & 1.59 & -  & 0.39 & 0.41 & 1.28\\\hline 
$0^+$ \& $1^+$ & 0.45 & 1.36~ & 1.14 & 1.33~ & 0.44 & 0.36& 0.54\\ 
$0^+$ & 0.45 & - & 1.44 & -  & 0.36 & 0.35 & 2.32\\ \hline 
\end{tabular}} 
\end{table} 
The nucleon's Faddeev equation is complete with these definitions.  An 
analogous equation for the $\Delta$ is readily obtained and does not require 
additional input.  The equations can be solved for the nucleon and $\Delta$ 
masses, and also yield their Faddeev amplitudes with which one can define an 
impulse approximation to $N$ and $\Delta$ elastic and transition form 
factors.\cite{jacquesA,mikepN} Regarding the four parameters, the scalar 
diquark mass can be taken from Table \ref{tabledqmasses}, as may the constraint 
$m_{1^+}/m_{0^+}\approx 1.3$.  That leaves the diquark width parameters, 
$\omega_{J^P}$. 
 
One goal of Ref.\ [\ref{RNpiN}] was to illustrate and emulate the success of 
Ref.\ [\ref{Roettel}] by showing there are intuitively reasonable values of the 
parameters for which one obtains the nucleon and $\Delta$ masses: $M_N = 
0.94\,{\rm GeV}$, $M_\Delta = 1.23\,{\rm GeV}$.  The results of that exercise 
are listed in Table \ref{tableMass}.  It is clear that the observed masses are 
easily obtained using solely the dressed-quark and -diquark degrees of freedom 
described above.  The first two rows of the table show that the additional 
quark exchange associated with the presence of pseudovector correlations 
provides considerable attraction.  Here it reduces the nucleon's mass by 41\% 
and, of course, the $\Delta$ would not be bound in this approach without the 
$1^+$ correlation.  Furthermore, in agreement with intuition, the nucleon and 
$\Delta$ masses increase with increasing $m_{J^P}$.  The diquark width 
parameters are also reasonable.  For example, with calculated results 
\begin{equation} 
\label{lqq} l_{0^+} := 1/\omega_{0^+} = 0.31\,{\rm fm} > l_{1^+} := 
1/\omega_{1^+} = 0.17\,{\rm fm}\,, 
\end{equation} 
these correlations lie within the nucleon (experimentally, the proton's 
charge radius $r_p = 0.87\,$fm), a point also emphasised by the scalar 
diquark's charge radius, calculated as described in Ref.~[\ref{RjacquesA}]: 
$r_{0^+}^2 = (0.55\,{\rm fm})^2$.  Moreover, defining $\omega_{f_{1,2}}$ by 
requiring a least-squares fit of ${\cal F}(\ell^2/\omega_{f_{1,2}})$ to 
$f_{1,2}(\ell^2)$, matched in magnitude at $\ell^2\simeq 0$, one obtains a 
scale characterising the quark-diquark separation: 
\begin{equation} 
l_{q(qq)_{f_1}}:=1/\omega_{f_1} = 0.40 \, {\rm fm} > 0.15\,{\rm fm} = 
\frac{1}{2}\,l_{0^+}\,. 
\end{equation} 
These scales reveal a significant spatial separation between the dormant 
quark and the diquark-participant quarks while permitting both the quark and 
diquark to remain within the baryon's volume, and thereby provide an 
intuitively appealing picture of confined constituents.  For the pseudovector 
analogue 
\begin{equation} 
l_{q(qq)_{a_1}} = 0.36\,{\rm fm} > \frac{1}{2}\,l_{1^+}\,. 
\end{equation} 
Finally, the ratio 
\begin{equation} 
\label{scR} \mbox{\sc r} = f_2(\ell^2=0)/f_1(\ell^2=0) 
\end{equation} 
is a measure of the importance of the lower component of the positive energy 
nucleon's spinor.  It is not small, a fact that emphasises the need to treat 
baryons using a Poincar\'e covariant framework.  As a point of comparison, an 
analogue in the MIT bag model is $\mbox{\sc r}_{\rm MIT}:= {\rm max}_{x\in 
[0,1]}\,j_1(2.04\, x )/j_0(0) = 0.43$. 
 
\subsection{Nucleon mass and pion loops} 
\addtocounter{subsection}{1} \noindent 
We have illustrated that an internally consistent and accurate description of
the nucleon and $\Delta$ masses is readily obtained using a Poincar\'e
covariant Faddeev equation based on confined diquarks and quarks.  However,
the $\pi N N$ and $\pi N \Delta$ couplings are large so it is important to
estimate the shift in the masses owing to $\pi$-dressing.  That was the
primary goal of Ref.\ [\ref{RNpiN}] but in seeking the estimate a number of
additional important results were confirmed or established.
 
\subsubsection{One-loop mass shift} 
\addtocounter{subsubsection}{1} 
\noindent 
As a first step, the shift in the mass of a positive-energy nucleon, $\delta 
M_+$, was evaluated perturbatively; i.e., the effect of just one pion loop on 
the nucleon's mass was calculated.  Assuming a pseudoscalar coupling: $g \bar 
N i \gamma_5 \vec{\tau}\cdot \vec{\pi} N$, $g=M/f_\pi$ with $M$ the mass of 
the nucleon in the loop, this can be separated into a sum of three terms: 
\begin{equation} 
\label{deltaMplus} 
\delta M_+ = \delta_F M_+^+ + \delta_F M_+^- + \delta_H M_+\,, 
\end{equation} 
where $\delta_F M_+^+$ represents the Fock diagram with a positive-energy 
nucleon in the loop, $\delta_F M_+^-$ is the $Z$-diagram; i.e., the Fock 
diagram with a negative-energy nucleon in the loop, and $\delta_H M_+$ is the 
tadpole (or Hartree) contribution arising from the contact term: 
$[M/(2f_\pi^2)] \, \bar N \vec{\pi}\cdot\vec{\pi} N$. 
 
The loop integrals in Eq.\ (\ref{deltaMplus}) must be defined and that was 
achieved in Ref.\ [\ref{RNpiN}] by implementing a Poincar\'e invariant 
Pauli-Villars regularisation.  It was immediately apparent that this is 
equivalent to employing a monopole form factor at each $\pi N N$ vertex: 
$g/(1+k^2/\lambda^2)$, and since the procedure modifies the pion propagator it 
may be interpreted as expressing compositeness of the pion and regularising its 
off-shell contribution.  A related effect was identified in Refs.\ 
[\ref{RMitchell:1996dn},\ref{Rreglaws}]. 
 
\begin{figure}[t] 
\centerline{\includegraphics[height=18em]{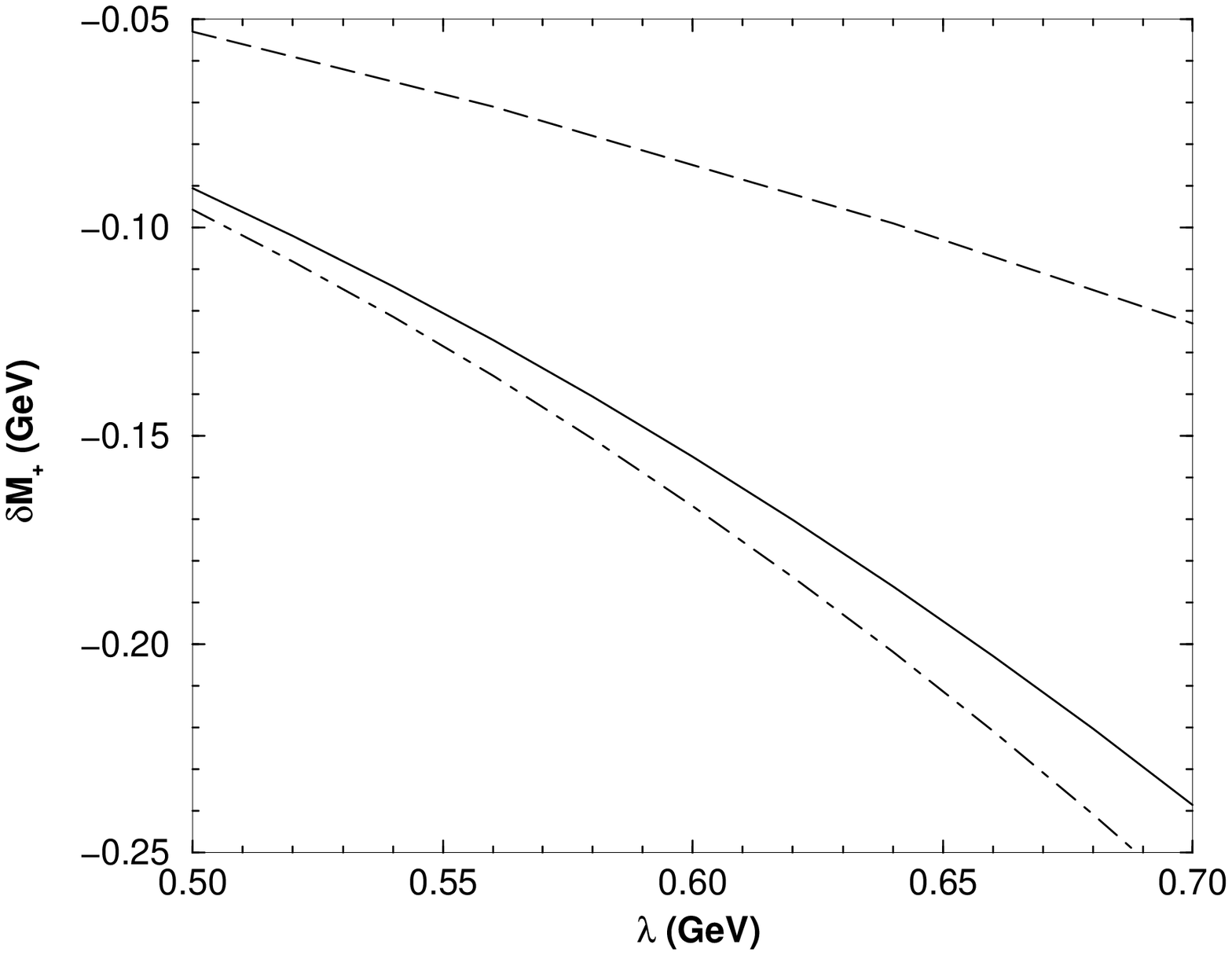}} 
\fcaption{\label{deltaMlambda} Solid line: Shift in a positive-energy nucleon's 
mass due to the O$(g^2)$ $\pi$-contribution to the self energy. ($M=0.94\,$GeV, 
$m_\pi=0.14\,$GeV.)  $\delta M_+(\lambda= 0.6\,{\rm GeV})= -0.15\,$GeV. Dashed 
line: $\delta_F M_+^+$, Eq.\ (\protect\ref{LNA}); dot-dashed line: $\delta^A 
M_+^+$, Eq.\ (\protect\ref{ALNA}). (Adapted from Ref.\ [\protect\ref{RNpiN}].)} 
\end{figure}

Each term in Eq.\ (\ref{deltaMplus}) is finite after regularisation and 
straightforward to evaluate numerically.  However, some insight can be gained 
algebraically by supposing that the mass of the nucleon in the loop is much 
greater than any other scale present.  In that case,\cite{NpiN} 
\begin{eqnarray} 
\label{LNA} \delta_F M_+^+ & = & -\,\frac{3}{32\pi} \frac{1}{f_\pi^2} m_\pi^3 - 
m_\pi^2 \, f^+_{(1)}(\lambda) - f^+_{(0)}(\lambda)\,,\\ 
\label{LNAZ} \delta_F M_+^- & = & \frac{3}{32\pi^2} \frac{M}{f_\pi^2}\, 
m_\pi^2\, (\ln [m_\pi^2/\lambda^2] - 1) + m_\pi^2\, f^-_{(1)}(\lambda) + 
f^-_{(0)}(\lambda) 
\,,\\ 
\label{LNAH} \delta_H M_+ &= & -\, \frac{3}{32 \pi^2} \frac{M}{f_\pi^2} \, 
\left(  m_\pi^2 \left( \ln\left[m_\pi^2/\lambda^2\right] - 1 \right) + 
\lambda^2 \right)\,, 
\end{eqnarray} 
where all the regularisation-scheme-dependence resides in
$f^\pm_{(0,1)}(\lambda)>0$, which are only functions of the regularisation
mass-scale.  One sees immediately that, for sufficiently small
$m_\pi>\lambda$, $\delta_F M_+^+ +\delta_F M_+^->0$, and it is only when the
Hartree term is included that $\delta M_+<0$.  Furthermore, the leading
behaviour of the nucleon's mass shift that is nonanalytic in the
current-quark mass, $m\propto \sqrt m_\pi$, is only given by Eq.\ (\ref{LNA})
because DCSB constrains the couplings and ensures a precise cancellation
between the $\ln m_\pi$-terms in Eqs.\ (\ref{LNAZ}), (\ref{LNAH}).  In
general, with the pseudoscalar coupling, $\delta M_+(\lambda)<0$ and
decreases monotonically from zero with increasing $\lambda$ only because of
destructive interference between the tadpole- and $Z$-diagrams.\cite{NpiN} At
realistic values of $m_\pi$ the actual value of $\delta M_+(\lambda)$ is
determined by the regularisation-scale-dependent terms, as visible in Fig.\
\ref{deltaMlambda}. 
 
There is no tadpole contribution to the nucleon's mass shift if one elects to 
use an axial-vector $\pi N$ coupling:\cite{weinberg} $[1/(2 f_\pi)] \, \bar N 
\gamma_5 \gamma^\mu \,\vec{\tau} \cdot \partial_\mu\vec{\pi} \,N$.  In this case 
$\delta^A M_+ = \delta^A M_+^+ + \delta^A M_+^-$; i.e., there are two 
contributions, one with a positive-energy nucleon in the loop and the other 
with a negative-energy nucleon.  Some insight is again obtained by 
considering the large-$M$ limit, wherein\cite{NpiN} 
\begin{equation} 
\label{ALNA} \delta^A M_+^+ = -\,\frac{3}{32\pi} \frac{1}{f_\pi^2} m_\pi^3 + 
m_\pi^2 \, f^+_{(1^A)}(\lambda) + 
f^+_{(0^A)}(\lambda); 
\end{equation} 
i.e., the same contribution, nonanalytic in the current-quark mass, as in 
Eq.~(\ref{LNA}), but with different regularisation-dependent terms, and 
$\delta^A M_+^- \propto 1/M$ because the pseudovector coupling suppresses the 
$Z$-diagram.  The leading nonanalytic contribution to the nucleon's mass is 
now unambiguous: it is given by the positive-energy nucleon Fock diagram, 
irrespective of the coupling's character.  Implementing the Pauli-Villars 
regularisation, a numerical evaluation of $\delta^A M_+ $ is straightforward 
with the result depicted in Fig.\ 
\ref{deltaMlambda}.  It is evident that $\delta^A M_+^+ \neq \delta_F M_+^+$, 
which illustrates the difference between the regularisation-dependent terms 
in Eqs.~(\ref{LNA}) and (\ref{ALNA}).  In addition, although it may not be 
immediately obvious, 
\begin{equation} 
\label{ONSHELL} \delta^A M_+ \equiv \delta M_+\,, 
\end{equation} 
which is why there is only one solid curve in the figure.  In this outcome
one has a quantitative verification of the on-shell equivalence of the
pseudoscalar and pseudovector interactions, in perturbation theory, as long
as the interactions are treated in a manner consistent with chiral
symmetry.\cite{sakurai}
 
\subsubsection{Mass shift, nonperturbatively} 
\addtocounter{subsubsection}{1} 
\noindent 
The effect of infinitely many pion loops can be calculated using a DSE for the 
nucleon's self energy: 
\begin{equation} 
 \Sigma(P) =   3 \int\!\frac{d^4 k}{(2\pi)^4}\, 
g_{PV}^2(P,k) \,\Delta_\pi((P-k)^2)\, \gamma\cdot (P-k)\gamma_5\, G(k)\, 
\gamma\cdot (P-k)\gamma_5\,, \label{EDSE} 
\end{equation} 
with $ G^{-1}(k) =i \gamma\cdot k + M + \Sigma(P) = i \gamma\cdot k \, {\cal 
A}(k^2) + M + {\cal B}(k^2)$, where $M$ is the nucleon's bare mass, which is 
obtained, e.g., by solving the Faddeev equation.  In Eq.~(\ref{EDSE}), 
$\Delta_\pi(k^2)=1/[k^2+m_\pi^2]$ is the pion propagator, and $g_{PV}(P,k)$ is 
a form factor that must describe the composite nature of {\it both} the pion 
and the nucleon.  The self-consistent solution of Eq.~(\ref{EDSE}) yields 
${\cal A}(k^2)$ and ${\cal B}(k^2)$, and therefrom the nonperturbative mass 
shift. 
 
The $\pi N$ vertex function can be calculated using a Poincar\'e covariant 
model of the nucleon, however, a calculation of the mass shift may again be 
expedited by employing an algebraic parametrisation that is constrained by such 
studies; e.g.,\cite{NpiN} 
\begin{equation} 
\label{gNoff} g_{PV}(P,k) = \frac{g}{2 M} \, g_\pi((P-k)^2) \, 
g_N(P^2)\,g_N(k^2)\,, 
\end{equation} 
$g_\pi(x) = {\rm e}^{-x/\Lambda^2}\!$, $g_N(x)= {\rm 
e}^{-(x+M^2)/\Lambda_N^2}\!$. This model provides for suppression of the 
coupling when either or both the nucleon and pion are off-shell, and in this 
it represents the compositeness of both.  The exponential form facilitates an 
algebraic evaluation of many necessary integrals and each term in the product 
is phenomenologically equivalent to a monopole form factor 
$1/[1+x/\lambda^2]$ if the mass scales are related via $\Lambda = \surd 2\, 
\lambda$.  Another advantage of this algebraic form is that it enables an 
elucidation of the precise equivalence between the Minkowski and Euclidean 
space calculations of the mass shift. 
 
A key aspect of a nonperturbative evaluation of $\delta M_+$ is that the 
position of the pole in the nucleon's propagator is not known \textit{a 
priori}: locating it is the goal, and this precludes an algebraic evaluation 
of the energy-integral that was straightforward in the one-loop calculation. 
In this case one must proceed by first evaluating the angular integrals in 
Eq.~(\ref{EDSE}), which are independent of $G(k)$.  That can be illustrated 
with the kernel of the equation for ${\cal B}$ 
\begin{eqnarray} 
{\cal K}_{\cal B}(P^2,k^2) & = & \int\! d\Omega_k\, g_{PV}^2(P,k)\bigg[1 - 
\frac{2 m_\pi^2}{(P-k)^2 + m_\pi^2} \bigg], \label{KBang} 
\end{eqnarray} 
with $d\Omega_k$ the usual angular measure.  It is evident that ${\cal K}_{\cal 
B}$ can be considered as a sum of two terms. The first is proportional to the 
angular average of $g_{PV}^2(P,k)$, and using Eq.~(\ref{gNoff}) that integral 
can be evaluated exactly: 
\begin{eqnarray} 
\nonumber 
\bar g_{PV}^2(P^2,k^2) & := & \int d\Omega_k \,g_{PV}^2((P-k)^2)\\ 
 & = &  \frac{g^2}{4 M^2}\, {\rm e}^{-\sfrac{ 2 (P^2+k^2)}{\Lambda^2}} 
\,g_N^2(P^2)\, g_N^2(k^2) \frac{\Lambda^2}{2 P k} \, {\rm I}_1(\sfrac{4 P 
k}{\Lambda^2}), 
\end{eqnarray} 
where ${\rm I}_1(x)$ is a modified Bessel function and $P = \sqrt{P^2}$, 
$k=\sqrt{k^2}$.  The second term is proportional to 
\begin{eqnarray} 
\omega_{g^2}(P^2,k^2) & := &\int d\Omega_k  \, \frac{g_{PV}^2(P,k)} {(P-k)^2 + 
m_\pi^2}\,, 
\end{eqnarray} 
which, in general, cannot be expressed as a finite sum of known functions. 
However, if $g_{PV}$ is regular at $P=k$ and its analytic structure is not a 
key influence on the solution, then the approximation 
\begin{eqnarray} 
\nonumber \omega_{g^2}(P^2,k^2) & \approx & \frac{g^2}{2 
M^2}\,g_{\pi}^2(|P^2-k^2|) g_N^2(P^2)\, g_N^2(k^2)\, 
\int\! d\Omega_k\, \frac{1}{(P-k)^2 + m_\pi^2} \\ 
\label{omegaB} &=& \frac{g^2}{2 M^2}\, g_{\pi}^2(|P^2-k^2|)\, g_N^2(P^2)\, 
g_N^2(k^2)\, 
\frac{1}{a+\sqrt{a^2-b^2}}\\ 
& =: & \tilde g_{PV}^2(P^2,k^2) \,\frac{1}{a+\sqrt{a^2-b^2}}\,, 
\end{eqnarray} 
where $a= P^2 + k^2 +m_\pi^2$, $b= 2 P k$, is a reliable 
tool.\cite{angleapprox} These preconditions are obviously satisfied in this 
application because the dominant physical effect in $\pi N$ physics is the 
pion pole and that appears at a mass-scale much lower than those present in 
$g_{PV}$.  Qualitatively identical considerations apply to ${\cal K}_{\cal 
A}$. 
 
The nucleon's mass appears at $P^2<0$ and hence to complete the specification 
of Eq.\ (\ref{EDSE}) one must define the continuation of the kernels into the 
timelike region.  The kernels' primary nonanalyticity is a square-root branch 
point associated with the simple pole in the pion propagator, and in continuing 
to $P^2<0$ it is necessary to include the discontinuity across the associated 
cut.  That must not be forgotten and is readily accomplished\cite{fukuda76} so 
that the nucleon's DSE is expressed by two coupled integral equations, which we 
illustrate with that for ${\cal B}$ 
\begin{eqnarray} 
\nonumber {\cal B}(x) & = &- \frac{3}{16 \pi^2}\int_{x_b}^0\!\! dy\, y\, \tilde 
g^2(x,y)\,\Delta\tilde{\cal K}_{\cal B}(x,y)\, \frac{{\cal B}(y)}{y 
{\cal A}^2(y) + {\cal B}^2(y)} \\ 
&& - \,\frac{3}{16\pi^2} \int_0^\infty \!\!dy\, y\, \tilde{\cal K}_{\cal 
B}(x,y)\, \frac{{\cal B}(y)}{y {\cal A}^2(y) + {\cal B}^2(y)}\,, 
\end{eqnarray} 
where $x_b=- (\sqrt{-x}-m_\pi)^2$ is the location of the branch point and 
\begin{eqnarray} 
\Delta\tilde{\cal K}_{\cal B}(x,y) & = & m_\pi^2\,\frac{\sqrt{(x+y+m_\pi^2)^2- 
4 x y}}{x \,y}\,,\\ 
\tilde {\cal K}_{\cal B}(x,y)  &= & \bar g_{PV}^2(x,y)  - \,\tilde 
g_{PV}^2(x,y)\,\frac{ 2 m_\pi^2}{a+\sqrt{a^2-b^2}} \,. 
\end{eqnarray} 
NB.\ The $\Delta\tilde{\cal K}_{{\cal A},{\cal B}}$ terms contribute only for 
$P^2+m_\pi^2<0$. 
 
With the solutions for ${\cal A}$, ${\cal B}$ in hand, the fully-dressed 
nucleon mass, $M_D$, is obtained by solving 
\begin{equation} 
M_D^2 \, {\cal A}^2(-M_D^2) = [M + {\cal B}(-M_D^2)]^2 
\end{equation} 
and the nonperturbative mass shift is $\delta M_+ = M_D - M$.  The widths 
$\Lambda$, $\Lambda_N$ in Eq.\ (\ref{gNoff}) are constrained by model and 
phenomenological analyses:\cite{oettel2,jacquesmyriad,tonysoft,pearcea} 
\begin{equation} 
\Lambda \sim 0.9 \,{\rm GeV}\,,\; \Lambda_N/\Lambda \sim 1.5 - 2.0 
\end{equation} 
and completing the calculation,\footnote{Three ``pions in the air'' are 
sufficient to yield a self-consistent solution and one pion alone provides 
$95$\% of the mass shift.} \ Ref.\ [\ref{RNpiN}] reports a $\pi N$-loop induced 
mass shift 
\begin{equation} 
-\delta M_+ \approx (60 - 100) \, {\rm MeV}\,. 
\end{equation} 
Allowing for model-dependence it is therefore safe to say that the $\pi 
N$-loop reduces the nucleon's mass by $\sim 10\,$-$\,20$\%.  Extant 
calculations\cite{tonyCBM,tonyANU,gastaopiN} show that the contribution from 
the analogous $\pi \Delta$-loop is of the same sign and no greater in 
magnitude so that the total reduction is $ \lsim 20\,$-$\,40$\%.  These same 
calculations indicate that the $\Delta$ mass is also reduced by $\pi$ loops 
but by a smaller amount ($\sim 50\,$-$\,100\,$MeV less). 
 
\subsubsection{Pions, quarks and diquarks} 
\addtocounter{subsubsection}{1} 
\noindent 
The question now is does this contribution materially affect the quark-diquark 
picture of baryons?  That may be addressed by solving the Faddeev equations 
again but this time requiring that the quark-diquark component yield higher 
masses for the $N$ and $\Delta$: $M_N=0.94 + 0.2=1.14\,$GeV, $M_\Delta= 
1.232+0.1=1.332\,$GeV. 
 
The results\cite{NpiN} of that exercise are presented in the third and fourth 
rows of Table~\ref{tableMass} and establish that the effects are material but 
not disruptive. In this case omitting the axial-vector diquark yields 
$M_N=1.44\,$GeV, which signals a $10\,$\% increase in the importance of the 
scalar-diquark component of the nucleon. (It is an \textit{increase} because 
this component now requires less correction.  The scalar diquark's charge 
radius was found to be $r_{0^+}=0.63\,$fm; i.e., $15\,$\% larger.)  It also 
reveals a reduction in the role played by axial-vector diquark correlations 
in the nucleon, since now restoring them only reduces the nucleon's core mass 
by $21$\%, with $\pi$ self-energy corrections providing the remaining $14$\%. 
 
Requiring an exact fit to the $N$ and $\Delta$ masses using only quark and 
diquark degrees of freedom therefore leads to an overestimate of the role 
played by axial-vector diquark correlations: it forces the $1^+$ diquark to 
mimic, in part, the effect of pions since they both act to reduce the mass cf.\ 
that of a quark$+$scalar-diquark baryon.  An accurate picture would represent 
the nucleon as $\sim 60$\% quark+scalar-diquark, $\sim 20$\% 
quark+axial-vector-diquark and $\sim 20$\% pion cloud ($\pi N + \pi\Delta$). 
This result can significantly impact upon the calculation of quantities such as 
the neutron's charge form factor and the ratio $F_2^p(q^2)/F_1^p(q^2)$, and 
consequently existing Faddeev-equation-based calculations of form factors 
should be revisited. 
 
\subsection{Nucleon form factors} 
\addtocounter{subsection}{1} 
\noindent 
The Faddeev equation also provides a bound state amplitude, which is a key 
element in the impulse approximation\cite{jacquesA,mikepN} used in many studies 
of the nucleon's electromagnetic and strong form 
factors.\cite{oettel2,jacquesA,jacquesmyriad,nedm} However, those studies have 
all overlooked the pion cloud's contribution, which is certainly important at 
small $q^2$ but may also implicitly affect results at larger $q^2$ by causing 
some elements in the calculations to be overweighted in order to mimic the 
pion's effect.  The studies must be updated. Nevertheless, with these remarks 
in mind, we briefly review a topical result. 
 
References [\ref{RjacquesA},\ref{Rjacquesmyriad}] employed a product 
\textit{Ansatz} for the nucleon's Faddeev amplitude that retains only the 
scalar diquark component, $\Psi_3^{0^+}$ in Eq.\ (\ref{Psi}).  This exploratory 
model involved three parameters, which were determined in a least-squares fit 
to $G_E^p(q^2)$ on $0<q^2<3\,$GeV$^2$, and made predictions for a variety of 
other couplings and form factors.  Its description of the neutron's charge form 
factor is poor, as it should be given the omission of the axial-vector diquark 
correlation and pion cloud.  However, this is the only major quantitative 
defect; e.g., $G_M^{p,n}(q^2)$ are well described. In fact they are probably 
too well reproduced since $\pi$-loops are expected to contribute $\sim 20$\% to 
the nucleons' magnetic moments and charge radii.\cite{radiiCh} Hence, Refs.\ 
[\ref{RjacquesA},\ref{Rjacquesmyriad}] provide a Poincar\'e covariant DSE model 
of the nucleon with a systematic error of $\lsim 30$\%. The model has an 
important additional feature, absent in other approaches: calculated quantities 
evolve smoothly to their perturbative-QCD limit because the DSEs reproduce 
perturbation theory at weak coupling.  We saw this explicitly for $u_v^\pi(x)$ 
in Sec.\ 3.3 and $F_\pi(q^2)$ in Sec.\ 4.1. 
 
\begin{figure}[t] 
\centerline{\includegraphics[height=20.7em]{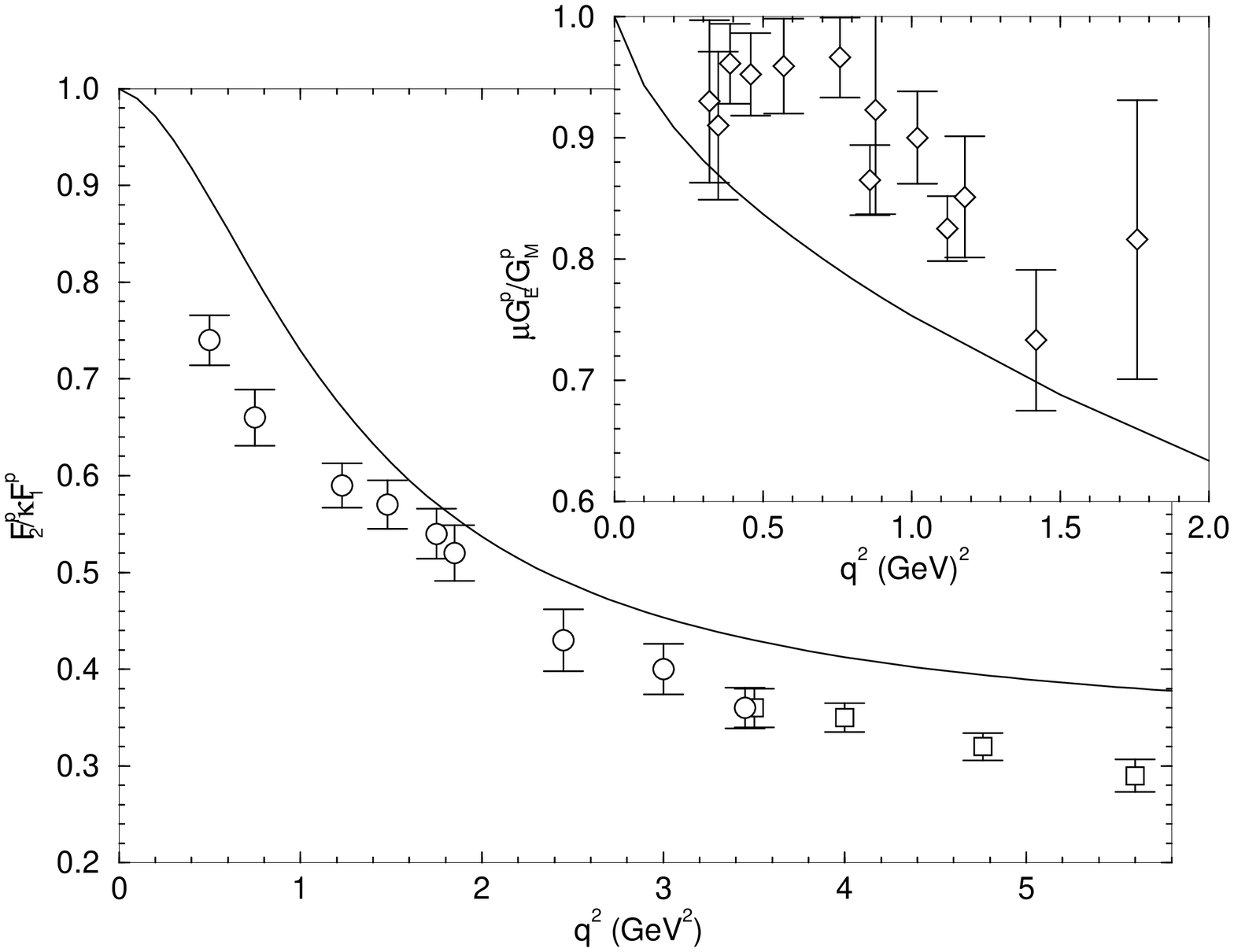}} 
\smallskip 
\fcaption{\label{figQF2F1} Main Figure -- Solid line: $F_2^p(q^2)/[\kappa 
F_1^p(q^2)]$ calculated using a model Faddeev amplitude for the nucleon that 
retains only a scalar diquark.\protect\cite{jacquesA,newjacques} Data: boxes, 
Ref.\ [\protect\ref{Rjones}]; circles, Ref.\ [\ref{Rgayou}].  Inset -- Solid 
line $\mu G_E^p/G_M^p$ calculated with the scalar-diquark 
model\protect\cite{jacquesA} cf.\ data from Ref.\ [\ref{Rroygayou}].} 
\end{figure}

We plot results for $F_2^p(q^2)/[\kappa F_1^p(q^2)]$, $\kappa=F_2(0)$, and $\mu 
G_E^p(q^2)/G_M^p(q^2)$ in Fig.\ \ref{figQF2F1}. It was remarked in Ref.\ 
[\ref{Rjacquesmyriad}] that the model agreed semiquantitatively with the 
then-current JLab data\cite{jones} but this is the first 
illustration.\cite{newjacques} The data has excited some 
interest\cite{jerryF2F1} but we judge that here the situation is just as with 
$F_\pi(q^2)$ [Fig.\ \ref{fig:fpi}].  On the domain hitherto explored the form 
factors are evolving through the region on which infrared phenomena, such as 
the length-scales defined by bound state amplitudes and strongly-dressed quark 
propagators, are dominant.  Thereupon a quantitative agreement with data is 
sensitive to model details.  Eventually, the perturbative 
behaviour:\cite{Lepage:1980fj} $q^2 F_2^p(q^2) / F_1^p(q^2) = \,$constant, will 
become evident but that is unlikely until significantly larger $q^2$. For 
instance, with $F_\pi(q^2)$ the perturbative behaviour is not unambiguously 
evident until $q^2 \gsim 15\,$GeV$^2$.  It is a challenge for quantum field 
theoretical DSE models of the nucleon, with their unique capacity for 
interpolating between the soft and hard domains, to locate the onset of 
perturbative behaviour in nucleon form factors.  Given the experience with 
other elastic form factors it would have been a surprise to see that at 
$q^2<10\,$GeV$^2$. 
 
\pagebreak

%% file: sect6.tex
 
\vspace*{1pt}\textlineskip  
\section{Epilogue}      
\addtocounter{section}{1} 
\vspace*{-0.5pt} 
\noindent 
We \label{epiloguelabel} have provided a perspective on the contemporary 
application of Dyson-Schwin\-ger equations (DSEs) to hadron physics.  The 
keystone of this approach's success has always been an appreciation and 
expression of the momentum-dependence of dressed-parton propagators at infrared 
length-scales.  That is responsible for the magnitude of constituent-quark and 
-gluon masses, and the length-scale characterising confinement in bound states. 
It is now recognised as a fact.  Modern hadron physics experiments are probing 
a domain on which this phenomenon underpins observable behaviour.  The next 
generation is likely to advance to a region wherein the transition to 
perturbative behaviour takes place.  DSE methods, with their unique capacity to 
connect phenomena dominated by soft scales with their perturbative limits, will 
necessarily become increasingly valuable. 
 
In recent years it has become clear why the simple rainbow-ladder DSE 
truncation has been successful for light vector and flavour nonsinglet 
pseudoscalar mesons.  It is the first term in a systematic and nonperturbative 
scheme that preserves all the Ward-Takahashi identities that express 
conservation laws at an hadronic level.  Studies in these channels showed that 
resumming subclasses of diagrams to infinite order provides a correction to 
bound state masses of $\lsim 10$\%.  Indeed, with just the first order 
correction to the rainbow-ladder kernel, the calculated masses are accurate to 
$99$\%.  This analysis also explains why the truncation should, and does, fail 
for scalar mesons and points the way toward a systematically improved hadron 
phenomenology. 
 
The existence of a systematic, nonperturbative, symmetry preserving truncation 
scheme has enabled the proof of exact results in QCD.  It provides for a 
straightforward explanation of the dichotomy of the pion as both a Goldstone 
mode and a bound state of effectively very massive quarks.  In arriving at that 
understanding, a mass formula for flavour nonsinglet pseudoscalar mesons was 
exposed which unifies the light- and heavy-quark regimes of QCD and provides a 
qualitative understanding of lattice simulations and their extrapolation to the 
chiral limit. 
 
There have been numerous applications of well-constrained DSE models to 
hadronic phenomena. Among them, a calculation of the pion's valence-quark 
mo\-men\-tum-fraction probability distribution, $u_v^\pi(x)$, highlights the 
framework's ability to provide a description that unifies the soft and hard 
domains of a given phenomenon.  In this case that has been crucial in 
exposing a serious discrepancy between theory and experiment.  In agreement 
with perturbative QCD, the DSE study predicts $u_v^\pi(x)\propto (1-x)^2$ in 
the valence-quark domain.  However, extant experiments are consistent with 
$u_v^\pi(x)\propto (1-x)$.  That is profoundly disturbing because a 
verification of the experimental result would even challenge the assumed 
vector-exchange nature of the force underlying the strong interaction. 
 
The widespread success of a renormalisation-group-improved rainbow-ladder 
model is certainly one of the most significant achievements of the last five 
years.  The model is defined by one parameter, which is an analogue for 
light-quarks of the string tension.  That parameter and the two light-quark 
masses $m_u=m_d$, $m_s$ are fitted to $m_\pi$, $f_\pi$, $m_K$, $f_K$ and 
everything calculated subsequently is a parameter-free prediction.  These 
\textit{ab initio} calculations have provided a unified description of many 
phenomena, among them: the spectrum of light-vector mesons; $\pi$ and $K$ 
electroweak form factors; vector meson transition form factors; and even 
$\pi$-$\pi$ scattering.  And this has not merely been a quantitative success. 
The studies have provided important information on how and where the 
predictions of perturbative QCD become apparent in exclusive processes and, 
for the pion, tied that to features of low-energy $\pi$-$\pi$ scattering and 
current-algebra constraints.  This body of work is unique in providing a 
systematically improvable, nonperturbative, Poincar\'e covariant, symmetry 
preserving approach to hadron physics. 
 
A significant challenge is to emulate this success with baryons. Studies 
conducted hitherto, while preserving these important features, have been 
exploratory.  They can and must be improved; e.g., by properly including 
$\pi$-cloud effects. To understand the data obtained at current and future 
hadron physics facilities a Poincar\'e covariant framework for baryons that 
naturally expresses the transition from the nonperturbative to perturbative 
domains will be necessary.  The early studies suggest that a Faddeev equation 
built upon the DSE framework can fill that need.